\documentclass[zpreprint,zbstnp]{zeus_paper}
\usepackage[english]{babel}
\usepackage[T1]{fontenc}
\usepackage{supertabular}
\newcommand{\ZcoosysA}{%
The ZEUS coordinate system is a right-handed Cartesian system, with the $Z$
axis pointing in the proton beam direction, referred to as the ``forward
direction'', and the $X$ axis pointing left towards the center of HERA.
The coordinate origin is at the nominal interaction point.\xspace}

\newcommand{\Zdetdesc}{%
A detailed description of the ZEUS detector can be found 
elsewhere~\cite{bluebook}. A brief outline of the 
components which are most relevant for this analysis is given
below.\xspace}
\newcommand{\Zctddesc}[1]{%
Charged particles were tracked in the central tracking detector (CTD)~\cite{CTD},
which operated in a magnetic field of $1.43\Tesla$ provided by a thin 
superconducting coil. The CTD consisted of 72~cylindrical drift chamber 
layers, organized in 9~superlayers covering the polar angle\footnote{\ZcoosysA} region 
\mbox{$15^\circ<\theta<164^\circ$}. The transverse-momentum resolution for
full-length tracks was $\sigma(p_T)/p_T=0.0058p_T\oplus0.0065\oplus0.0014/p_T$,
with $p_T$ in $\Gev$.}
\newcommand{\Zcaldesc}{%
The high-resolution uranium--scintillator calorimeter (CAL)~\cite{CAL} consisted 
of three parts: the forward (FCAL), the barrel (BCAL) and the rear (RCAL)
calorimeters. Each part was subdivided transversely into towers and
longitudinally into one electromagnetic section (EMC) and either one (in RCAL)
or two (in BCAL and FCAL) hadronic sections (HAC). The smallest subdivision of
the calorimeter is called a cell.  The CAL energy resolutions, as measured under
test-beam conditions, were $\sigma(E)/E=0.18/\sqrt{E}$ for electrons and
$\sigma(E)/E=0.35/\sqrt{E}$ for hadrons, with $E$ in $\Gev$.}

\chardef\usc=95
\chardef\til=126
\catcode`\@=11 
\DeclareRobustCommand\xdotspace{\futurelet\@let@token\@xdotspace}
\def\@xdotspace{%
  \ifx\@let@token.\else
  \ifx\@let@token\bgroup.\else
  \ifx\@let@token\egroup.\else
  \ifx\@let@token\/.\else
  \ifx\@let@token\ .\else
  \ifx\@let@token~.\else
  \ifx\@let@token!.\else
  \ifx\@let@token,.\else
  \ifx\@let@token:.\else
  \ifx\@let@token;.\else
  \ifx\@let@token?.\else
  \ifx\@let@token/.\else
  \ifx\@let@token'.\else
  \ifx\@let@token).\else
  \ifx\@let@token-.\else
  \ifx\@let@token\@xobeysp.\else
  \ifx\@let@token\space.\else
  \ifx\@let@token\@sptoken.\else
   .\space
   \fi\fi\fi\fi\fi\fi\fi\fi\fi\fi\fi\fi\fi\fi\fi\fi\fi\fi}
\catcode`\@=12 

\newcommand{\stru}[2]{%
   \relax\ifmmode\hbox{\vrule height#1 depth#2 width0pt}%
   \else\vrule height#1 depth#2 width0pt\fi}

\newcommand{\Ronum}[1]{\uppercase\expandafter{\romannumeral#1}}
\newcommand{\ronum}[1]{\expandafter{\romannumeral#1}}
\DeclareRobustCommand{\LaTeXZ}{%
  \LaTeX\kern-.05em4\kern-.1em
  {\raisebox{-0.2ex}{$\scriptstyle\text{ZEUS}$}}\xspace}

\newcommand{\tab}[1]{Table~\ref{tab-#1}}

\DeclareMathAlphabet{\mathbf}{OT1}{cmr}{bx}{sl}
\newcommand{\eVdist}{\kern-0.06667em}

\newcommand{\Gev}{{\text{Ge}\eVdist\text{V\/}}}

\newcommand{\met}{\,\text{m}}

\newcommand{\Tesla}{\,\text{T}}

\newcommand{\slashfrac}[2]{%
  \raisebox{0.5ex}{\ensuremath #1}\kern-0.12em/\kern-0.08em
  \raisebox{-.8ex}{\ensuremath #2}}

\newcommand{\sqr}[3]{%
    {\vcenter{\hrule height.#3ex\hbox{\vrule width.#2ex height#1ex
     \kern#1ex\vrule width.#3ex}\hrule height.#2ex}}}

\catcode`\@=11 
\newcommand{\parenbar}{\mathpalette\p@renb@r}
\def\p@renb@r#1#2{\vbox{%
  \ifx#1\scriptscriptstyle \dimen@.7em\dimen@ii.2em\else
  \ifx#1\scriptstyle \dimen@.8em\dimen@ii.25em\else
  \dimen@1em\dimen@ii.4em\fi\fi \offinterlineskip
  \ialign{\hfill##\hfill\cr
    \vbox{\hrule width\dimen@ii}\cr
    \noalign{\vskip-.3ex}%
    \hbox to\dimen@{$\mathchar300\hfil\mathchar301$}\cr
    \noalign{\vskip-.3ex}%
    $#1#2$\cr}}}
\catcode`\@=12 

\newcommand{\IP}{{\rm I$\kern-0.01667em$P}\xspace}

\mathchardef\qsm=63
\mathchardef\pls=43
\mathchardef\mns=512
\mathchardef\plm=518
\mathchardef\eql=61
\mathchardef\smallleft=300
\mathchardef\smallright=301
\mathchardef\les=316
\mathchardef\gre=318
\mathchardef\leq=532
\mathchardef\grq=533

\catcode`\@=11 
\newcounter{pict@width}
\newcounter{pict@height}
\newlength{\pict@scale}
\newcommand{\psfigadd}[4]{%
\setcounter{pict@width}{1*\ratio{#2+\pict@scale/2}{\pict@scale}}
\setcounter{pict@height}{1*\ratio{#3+\pict@scale/2}{\pict@scale}}
\setlength{\unitlength}{\pict@scale}
\hbox to #2{\hspace{-\fill}\begin{picture}(\thepict@width,\thepict@height)
\put(0,0){\psfig{figure=#1,width=#2,height=#3,clip=}}
\SetScale{0.283466457}
\SetWidth{1.763889}
{#4}
\end{picture}}
}
\newcounter{pict@widthfst}
\newcounter{pict@widthscd}
\newcounter{pict@widthtot}
\newcommand{\psfigaddtwo}[7]{%
\setcounter{pict@widthfst}{1*\ratio{#2+\pict@scale/2}{\pict@scale}}
\setcounter{pict@widthscd}{1*\ratio{#2+#4+\pict@scale/2}{\pict@scale}}
\setcounter{pict@widthtot}{1*\ratio{#2+#4+#6+\pict@scale/2}{\pict@scale}}
\setcounter{pict@height}{1*\ratio{#3+\pict@scale/2}{\pict@scale}}
\setlength{\unitlength}{\pict@scale}
\hbox{\hspace{-\fill}\begin{picture}(\thepict@widthtot,\thepict@height)
\put(0,0){\psfig{figure=#1,width=#2,height=#3,clip=}}
\put(\thepict@widthscd,0){\psfig{figure=#5,width=#6,height=#3,clip=}}
\SetScale{0.283466457}
\SetWidth{1.763889}
{#7}
\end{picture}}
}
\newcommand{\psfigror}[4]{%
\setcounter{pict@width}{1*\ratio{#2+\pict@scale/2}{\pict@scale}}
\setcounter{pict@height}{1*\ratio{#3+\pict@scale/2}{\pict@scale}}
\setlength{\unitlength}{\pict@scale}
\hbox{\begin{picture}(\thepict@width,\thepict@height)
\put(0,\thepict@height){\psfig{figure=#1,width=#3,height=#2,clip=,angle=270}}
\SetScale{0.283466457}
\SetWidth{1.763889}
{#4}
\end{picture}}
}
\newcommand{\psfigrol}[4]{%
\setcounter{pict@width}{1*\ratio{#2+\pict@scale/2}{\pict@scale}}
\setcounter{pict@height}{1*\ratio{#3+\pict@scale/2}{\pict@scale}}
\setlength{\unitlength}{\pict@scale}
\hbox{\begin{picture}(\thepict@width,\thepict@height)
\put(0,0){\psfig{figure=#1,width=#3,height=#2,clip=,angle=90}}
\SetScale{0.283466457}
\SetWidth{1.763889}
{#4}
\end{picture}}
}
\catcode`\@=12 
\newlength\listtextwidth

\catcode`\@=11 
\newlength{\@tabfninsert}
\newlength{\@tabfnwidth}
\newcommand{\tabfootnote}[2]{%
  \setlength{\@tabfninsert}{0.8em}
  \setlength{\@tabfnwidth}{\textwidth}
  \addtolength{\@tabfnwidth}{-\@tabfninsert}
  \addtolength{\@tabfnwidth}{-0.4em}
  \noindent\makebox[\@tabfninsert][r]{\footnotesize$^{#1}$\hfil}\hfill%
  \parbox[t]{\@tabfnwidth}{\footnotesize #2\hfill}}
\catcode`\@=12 

\newcommand{\CPC}[3]{Comp. Phys.\ Comm.\ {\bf#1}, #3 (#2)}

\newcommand{\NP}[3]{Nucl.\ Phys.\ {\bf#1}, #3 (#2)}
\newcommand{\NPPS}[3]{Nucl.\ Phys.\ Proc.\ Suppl.\ {\bf#1}, #3 (#2)}
\newcommand{\PL}[3]{Phys.\ Lett.\ {\bf#1}, #3 (#2)}

\newcommand{\PR}[3]{Phys.\ Rev.\ {\bf#1}, #3 (#2)}
\newcommand{\PRL}[3]{Phys.\ Rev.\ Lett.\ {\bf#1}, #3 (#2)}

\newcommand{\NIM}[3]{Nucl.\ Instr.\ and Meth.\ {\bf#1}, #3 (#2)}
\newcommand{\ZP}[3]{Z.\ Phys.\ {\bf#1}, #3 (#2)}
\newcommand{\EP}[3]{Eur.\ Phys.\ J.\ {\bf#1}, #3 (#2)}

\newcommand{\APP}[3]{Acta\ Phys.\ Pol.\ {\bf#1}, #3 (#2)}

\def\mx2{${ M_{x}^{2} } $}

\def\Q2{${ Q^{2} }  $}
\def\q2{${ Q^{2} }  $}

\def\pt2{${ p_{T}^2}$ }
\def\pt{${ p_{T}\/\/ }$ }
\def\e+p{$ e^+-p$ }

\newcommand{\ptsq}{$p_T^2$}

\begin{document}

\prepnum{DESY--08--176}

\title{Leading proton production in deep inelastic scattering at HERA}

\author{ZEUS Collaboration}

\draftversion{7.01 desy version }

\date{10 December 2008}

\abstract{
The semi-inclusive reaction $e^+p\to e^+Xp$  was studied with the ZEUS
detector at HERA using an integrated luminosity of 12.8 pb$^{-1}$.
The final-state proton, which was detected with the ZEUS
leading proton spectrometer, carried a large fraction of the 
incoming proton energy, $x_L>0.32$, and its  
transverse momentum squared  satisfied $p_T^2<0.5$ GeV$^2$;
the exchanged photon  virtuality, $Q^2$, was greater than 3 GeV$^2$
and the range of the masses of the photon-proton system was $45<W<225$ GeV.
The leading proton production cross section and  rates are presented
as a function of $x_L$, $p_T^2$, $Q^2$ and the Bjorken scaling
variable, $x$.
}

\makezeustitle 

\pagenumbering{Roman}                                                                              
                                                   %
\begin{center}                                                                                     
{                      \Large  The ZEUS Collaboration              }                               
\end{center}                                                                                       
  S.~Chekanov,                                                                                     
  M.~Derrick,                                                                                      
  S.~Magill,                                                                                       
  B.~Musgrave,                                                                                     
  D.~Nicholass$^{   1}$,                                                                           
  \mbox{J.~Repond},                                                                                
  R.~Yoshida\\                                                                                     
 {\it Argonne National Laboratory, Argonne, Illinois 60439-4815, USA}~$^{n}$                       
\par \filbreak                                                                                     
  M.C.K.~Mattingly \\                                                                              
 {\it Andrews University, Berrien Springs, Michigan 49104-0380, USA}                               
\par \filbreak                                                                                     
  P.~Antonioli,                                                                                    
  G.~Bari,                                                                                         
  L.~Bellagamba,                                                                                   
  D.~Boscherini,                                                                                   
  A.~Bruni,                                                                                        
  G.~Bruni,                                                                                        
  G.~Cara~Romeo                                                                                    
  F.~Cindolo,                                                                                      
  M.~Corradi,                                                                                      
  P.~Giusti,                                                                                       
\mbox{G.~Iacobucci},                                                                               
  A.~Margotti,                                                                                     
  T.~Massam,                                                                                       
  R.~Nania,                                                                                        
  A.~Polini\\                                                                                      
  {\it INFN Bologna, Bologna, Italy}~$^{e}$
% ...
\newpage                                                                                           
                                                           %
$^{\    1}$ also affiliated with University College London,                                        
United Kingdom\\                                                                                   
$^{\    2}$ now at University of Salerno, Italy \\                                                 
% ...
%
\newpage   
                                                           %
                                                           %
\begin{tabular}[h]{rp{14cm}}                                                                       
$^{a}$ &  supported by the Natural Sciences and Engineering Research Council of Canada (NSERC) \\  
$^{b}$ &  supported by the German Federal Ministry for Education and Research (BMBF), under        
          contract numbers 05 HZ6PDA, 05 HZ6GUA, 05 HZ6VFA and 05 HZ4KHA\\                         
$^{c}$ &  supported in part by the MINERVA Gesellschaft f\"ur Forschung GmbH, the Israel Science   
          Foundation (grant no. 293/02-11.2) and the U.S.-Israel Binational Science Foundation \\  
$^{d}$ &  supported by the Israel Science Foundation\\                                             
$^{e}$ &  supported by the Italian National Institute for Nuclear Physics (INFN) \\                
$^{f}$ &  supported by the Japanese Ministry of Education, Culture, Sports, Science and Technology 
          (MEXT) and its grants for Scientific Research\\                                          
$^{g}$ &  supported by the Korean Ministry of Education and Korea Science and Engineering          
          Foundation\\                                                                             
$^{h}$ &  supported by the Netherlands Foundation for Research on Matter (FOM)\\                   
$^{i}$ &  supported by the Polish State Committee for Scientific Research, project no.             
          DESY/256/2006 - 154/DES/2006/03\\                                                        
$^{j}$ &  partially supported by the German Federal Ministry for Education and Research (BMBF)\\   
$^{k}$ &  supported by RF Presidential grant N 1456.2008.2 for the leading                         
          scientific schools and by the Russian Ministry of Education and Science through its      
          grant for Scientific Research on High Energy Physics\\                                   
$^{l}$ &  supported by the Spanish Ministry of Education and Science through funds provided by     
          CICYT\\                                                                                  
$^{m}$ &  supported by the Science and Technology Facilities Council, UK\\                         
$^{n}$ &  supported by the US Department of Energy\\                                               
$^{o}$ &  supported by the US National Science Foundation. Any opinion,                            
findings and conclusions or recommendations expressed in this material                             
are those of the authors and do not necessarily reflect the views of the                           
National Science Foundation.\\                                                                     
$^{p}$ &  supported by the Polish Ministry of Science and Higher Education                         
as a scientific project (2006-2008)\\                                                              
$^{q}$ &  supported by FNRS and its associated funds (IISN and FRIA) and by an Inter-University    
          Attraction Poles Programme subsidised by the Belgian Federal Science Policy Office\\     
$^{r}$ &  supported by an FRGS grant from the Malaysian government\\                               
\end{tabular}                                                                                      
                                                           %
                                                           %

\pagenumbering{arabic}  
\pagestyle{plain}

\section{Introduction}

Hadron-hadron collisions predominantly give rise to leading particles of  
the same type as those in the incoming beams and carrying a large fraction of 
the momentum of incoming particles.
The spectrum of leading particles approximately scales with
the centre-of-mass energy, a property  known as limiting 
fragmentation~\cite{review1}. 
The properties of the accompanying final state are also  universal 
when studied as a function of the centre-of-mass energy available 
after excluding the leading particles~\cite{review2,review3}.

Events with a final-state proton carrying a large fraction of the
available energy, $x_L$, but a small transverse momentum, $p_T$, 
have been studied in detail in high-energy hadron-proton 
collisions~\cite{review4,review5,review6}.
More recently, the HERA experiments reported measurements of 
the production of leading protons in $ep$ collisions~\cite{lp95,lph1}.  
Several
mechanisms have been suggested to explain the production of leading
protons. None of them are, as yet, amenable to calculations based on
perturbative quantum chromodynamics (pQCD).  This is, in part, a
consequence of the small values of  $p_T$ of
the leading proton which necessitates a non-perturbative approach. Some
models~\cite{SULLIVAN,ZOLLER,HOLTMANN,KOPE,SNS} are based on the Regge
formalism, with leading proton production occurring through
$t$-channel exchanges, both isoscalar and isovector, notably of the Pomeron,
pion and Reggeon trajectories. These exchanges mediate the interaction between
the proton and the hadronic fluctuations of the virtual photon.  Other
models retain quarks and gluons as fundamental entities,
but add non-perturbative elements, such as soft-colour
interactions (SCI)~\cite{sci}. Alternatively, 
the concept of fracture functions~\cite{FF} 
offers a QCD framework in which to describe the leading baryon momentum
spectra.

This paper presents measurements of leading proton production in
$e^+p$ collisions, $e^+p \rightarrow e^+Xp$, with a four-fold
increase in statistics compared to an earlier measurement~\cite{lp95}.
High-energy protons with low transverse momentum
carrying at least a fraction $x_L$=0.32 of the incoming-proton
momentum were measured in the ZEUS leading proton spectrometer
(LPS)~\cite{lpsrho}.  The cross sections are presented
 as a function of the proton variables $x_L$ and $p_T^2$.
The dependence on the Bjorken variable, $x$, and on the photon
virtuality, $Q^2$, was also studied and compared to that of the
inclusive deep inelastic scattering (DIS) reaction $e^+p\to e^+X$. 
The measurements
cover  the kinematic range $Q^2>3$ GeV$^2$ and $45<W<225$
GeV, where $W$ is the total mass of the photon-proton system. 
The  data for $x_L > 0.93$ were used in an earlier study~\cite{lp97-diff}
to extract the diffractive structure function of the proton.

The leading proton structure function, $F_2^{\rm LP}$, defined in 
Section~\ref{kincross},
which can be identified with a fracture function, is also presented. 
The latter parameterises the momentum 
spectra of leading particles through parton
distribution functions in the proton. 
This approach can be incorporated in Monte Carlo (MC) programs simulating
hadronic final states in $p p$ interactions at the LHC~\cite{lhc-eee} 
and extended cosmic-ray showers~\cite{sybill,qgsjet}.

\section{Experimental set-up}
\label{detector}

The measurements were performed with data collected in 1997 at the
$ep$ collider HERA using the ZEUS detector, when HERA operated with a 
proton beam energy $E_p= 820$~GeV and a positron beam energy $E_e=27.5$~GeV.

\Zdetdesc

\Zctddesc

\Zcaldesc

The position of the scattered positron was determined by combining 
information from the CAL, the small-angle rear tracking detector~\cite{SRTD} 
and the hadron-electron separator~\cite{HES}.

The LPS~\cite{lpsrho} was used during the data-taking period 1994$-$2000
to detect positively charged particles scattered at very 
small angles and carrying a large fraction of the longitudinal  momentum 
of the incoming proton. It consisted of 54 planes of silicon microstrip
 detectors grouped into six stations, S1 to S6, and located along the 
outgoing proton beam line between $Z=20$ m and $Z=90$ m. 

The stations S1, S2, S3 and S4, S5, S6 
can be considered as two independent spectrometers, called s123 and s456, 
respectively, due to their different 
phase-space coverage, in particular in azimuth.
During data taking, the stations were inserted very close to the proton 
beam (typically a few mm). Charged particles inside the beampipe were 
deflected by the magnetic field of the proton-beamline magnets and measured 
in the LPS with a resolution better than 1\% on the longitudinal momentum 
and of 5 MeV on the transverse momentum. The beam transverse momentum 
spread at the interaction point was $\approx$ 40 MeV in the horizontal 
plane and $\approx$ 90 MeV in the vertical plane and dominated the 
transverse-momentum resolution.

A forward neutron calorimeter (FNC)~\cite{FNC} was installed in the HERA 
tunnel at \mbox{$\theta = 0^{\circ}$} and at $Z = 106\met$ from the 
interaction point in the proton-beam direction. The FNC, a 
lead--scintillator calorimeter, had an energy resolution for hadrons
 $\sigma(E)/E = 0.70/\sqrt{E}$, with $E$ in GeV, as measured in a test beam. 
The calorimeter was segmented vertically into 14 towers. Three planes of 
veto counters were located in front of the FNC to reject events in which 
a particle showered in inactive material along the beamline upstream
of the FNC.

The luminosity was measured from the rate of the bremsstrahlung process 
$ep \to e\gamma p$. The resulting small-angle energetic photons were 
measured by the luminosity monitor~\cite{lumi}, a lead--scintillator 
calorimeter placed in the HERA tunnel at $Z=-107$ m.

\section{Kinematics and cross sections}
\label{kincross}

\begin{figure}[htb]
\begin{center}
\leavevmode
\hbox{%
\epsfxsize = 10cm
\epsffile{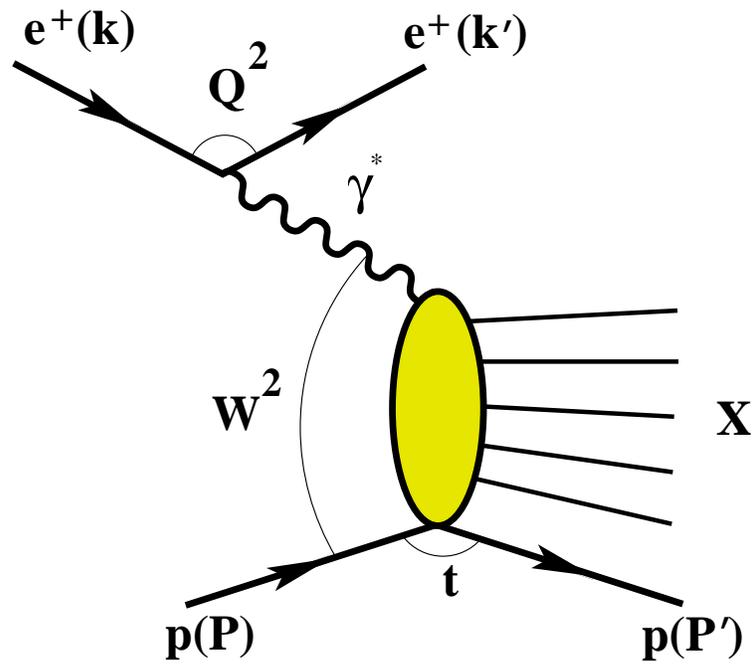}}
\end{center}
\caption{{\protect{
Schematic diagram of the reaction $e^+p \rightarrow
e^+Xp$.}}}
\label{fig:diag}
\end{figure}

Figure~\ref{fig:diag} illustrates semi-inclusive leading proton production 
in $ep$ collisions.
Four kinematic variables are needed to describe the reaction
$e^+ p\rightarrow e^+ X p$.
They are defined in terms of the four-momenta of the incoming and
outgoing positron, $k$ and $k'$, and of the incoming and outgoing
proton, $P$ and $P'$, respectively.

The Lorentz-invariant kinematic variables used in inclusive studies
are $Q^2= -q^2 = -(k-k')^2$, the virtuality of the exchanged photon;
$x=Q^2/(2P\cdot q)$ and the inelasticity,  
$y=q\cdot P/(k\cdot P)\simeq Q^2/(sx)$;
$W^2=(P+k-k')^2=m^2_p+Q^2(1-x)/x$, the square of the photon-proton
centre-of-mass energy.  In these equations, $m_p$ is the mass of the
proton and $\sqrt{s}=$ 300~GeV is the $e^+p$ centre-of-mass energy.
Among these variables, only two are independent.

Two additional variables are required to describe the leading
proton. They are chosen as the momentum fraction carried by the
outgoing proton

\[
x_L =\frac{P' \cdot k}{P\cdot k}
\]

and its transverse momentum with respect to the direction of the
incoming proton, $p_T$. In terms of these variables, the square of the
four-momentum transfer from the target proton is given by

\begin{equation}
t=(P-P')^2 \simeq -\frac{p_T^2}{x_L} - \frac{(1-x_L )^2}{x_L 
}m^2_p, \nonumber
\end{equation}

where the second term is the minimum kinematically-allowed value of
$|t|$ for a given $x_L $. The variable $t$ is the
square of the four-momentum of the exchanged particle.

The differential cross section for inclusive $e^+p\rightarrow e^+ X$
scattering, in the $Q^2$ region of this analysis, 
is written in terms of the proton structure function,
$F_2(x,Q^2)$, as

\begin{equation}
\frac{d^2\sigma_{e^+p\rightarrow e^+ X} }{dx dQ^2} =
 \frac{4\pi \alpha^2}{x Q^4}\left( 1-y+\frac{y^2}{2} \right)
  F_2(x , Q^2)(1+\Delta), 
  \label{eq:dis:xsec}
\end{equation}

where $\Delta$ is a correction that takes into account the contribution of 
the longitudinal structure function, $F_L$, and of electroweak
radiative effects.
Similarly, the differential cross section for semi-inclusive
leading proton (LP) production can be written
in terms of the leading proton structure function, 
$F_2^{\rm LP(4)}(x , Q^2, x_L, p_T^2)$, as

\begin{equation}
\frac{d^4\sigma_{e^+p\rightarrow e^+ Xp}}{dx dQ^2dx_L dp_T^2} =
 \frac{4\pi \alpha^2}{x Q^4}\left( 1-y+\frac{y^2}{2} \right)
   F^{\mbox{\rm\tiny LP(4)}}_2(x , Q^2, x_L, p_T^2)
     (1+\Delta_{\rm LP}),   
\label{eq:ln}
\end{equation}

\noindent

where  $\Delta_{\rm LP}$ is the analogue of $\Delta$.

The structure function $F_2^{\rm LP(4)}(x,Q^2,x_L,p_T^2)$ corresponds to the 
proton-to-proton fracture function
\mbox{$M_2^{p/p}(x,Q^2,x_L,p_T^2)$}~\cite{FF}, i.e. the structure function of a proton
probed under the condition that the target fragmentation region contains 
a proton with a given $x_L$ and $p_T^2$.

\section{Reconstruction of the kinematic variables}
\label{recon}

In the $Q^2$ range of this analysis, DIS events are characterised
by the presence of a scattered positron, mostly in RCAL.
The scattered positron was  
reconstructed using an electron finder algorithm based on a 
neural network~\cite{sira}.

The properties of the hadronic final state in the central detector were
derived using 
the energy flow objects (EFOs)~\cite{EFO} reconstructed from CAL clusters and 
CTD tracks by combining the CTD and CAL information to optimise the
resolution of the reconstructed kinematic variables. The EFOs were 
additionally corrected for energy loss due to inactive material in front
of the CAL.

The DIS variables $x$, $y$, $Q^2$ and $W$ were obtained by using a
weighting method~\cite{fpc1}, which uses a weighted average of the
values determined from the electron~\cite{elmeth} and
double-angle~\cite{dameth} methods. The variable $y$ was also
reconstructed using the Jacquet-Blondel method~\cite{jb}, which uses
information from the hadronic final state to reconstruct the event
kinematics, and is denoted by $y_{\rm JB}$.

The momentum $p=(p_X,p_Y,p_Z)$ of the leading proton candidate was 
determined using the LPS. The variable $x_L$ was evaluated as $x_L=p_Z/E_p$
and the squared transverse momentum as $p_T^2=p_X^2+p_Y^2$.

\section{Data sample and event selection}
\label{selection}

During  1997, the ZEUS detector collected 
an integrated luminosity of 27.8 pb$^{-1}$. However,
the experimental conditions allowed the operation of the LPS only for 
an integrated luminosity of 
12.8 pb$^{-1}$. Of this sample, 4.8 pb$^{-1}$ of data were collected 
with all the LPS stations. In the remaining part, only the spectrometer 
s456 was used.

Online, a three-level trigger~\cite{trigger} 
was used. At the third level, the event variables were reconstructed with 
an accuracy close to that obtained offline. Final detector calibration and
full-event reconstruction were performed offline.

Two sets of events were selected~\cite{rinaldi}: the inclusive DIS sample 
and the LPS sample. For a fraction of the inclusive DIS candidate events 
the trigger was prescaled, thus reducing the  
effective integrated luminosity of the inclusive DIS sample to 1 pb$^{-1}$.
The selection of the LPS sample was performed with a dedicated LPS trigger.

The presence of a good positron candidate in the CAL was 
required in the trigger chain used to select the inclusive DIS sample. 
In addition, the following conditions were applied:

\begin{itemize}

\item $|Z_{\rm vtx}|<50$ cm, where $Z_{\rm vtx}$ is the $Z$ coordinate of 
the event vertex.  
This cut is needed to remove background due to proton beam-gas
interactions and cosmic rays;

\item energy of the scattered positron $E^{\prime}_e>10$ GeV. The positron 
position was required to be outside the region close to the rear beampipe 
hole, where the presence of inactive material reduced the precision of the 
energy measurement;

\item the quantity $E-P_Z$, where the energy $E$ and the longitudinal
momentum $P_Z$ are summed over all the EFOs and the scattered 
positron, in the range $38 < E-P_Z < 65$ GeV, to exclude background from 
photoproduction, proton beam-gas interactions and cosmic rays;

\item  $y_{\rm JB}>0.03$ in order to ensure hadronic activity away from 
the forward direction.

\end{itemize}

The following cuts define the kinematic region:

\begin{itemize}

\item $Q^2>3$ GeV$^2$, to select DIS events with large virtuality of the 
exchanged photon; 
\item $45<W<225$ GeV, to ensure a wide kinematic coverage of the hadronic 
system.

\end{itemize}

The number of events that passed the inclusive DIS selection cuts was 145447.

The LPS sample was selected as the DIS sample, but requiring 
in addition the LPS trigger and the following conditions:

\begin{itemize}

\item a  reconstructed track in the LPS with $p_T^2<0.5$ GeV$^2$ and 
$x_L>0.32$. 
To reduce the sensitivity of the LPS acceptance due to the uncertainties
of the location of the beampipe elements,  a cut was applied to the minimum
distance, $\Delta_{\rm pipe}$, between the track and the beampipe 
requiring $\Delta_{\rm pipe}\,>\,0.04$ (0.25) cm for s456 (s123).
The selection of
tracks with $\Delta_{\rm pot}>0.02$ cm, where $\Delta_{\rm pot}$ is the
minimum distance of the track from the edge of any LPS detector,
ensured that the tracks were well within the active regions of the
silicon detectors;

\item the sum of the energy and the longitudinal momentum of  both the 
energy deposits in the CAL and the  particle detected in the LPS, $E+P_Z$, 
was required to be smaller than 1655 GeV; this cut rejected most 
of the random overlays of DIS events with protons from the beam-halo 
or from a  proton beam-gas collision (see Section~\ref{background}).

\end{itemize}

A total of 73275 events survived the above selection criteria, of which 
6008 had a track in s123 and 67267 had a track in s456.

\section{Monte Carlo simulation}
\label{montecarlo}

To determine the acceptance of the apparatus, inclusive DIS events
with $Q^2>0.5$ GeV$^2$ were generated using {\sc Djangoh}~\cite{django}, which
is interfaced to {\sc Heracles}~\cite{heracles} for electroweak
radiative effects. 
In order to study the migration of events from low $Q^2$, a sample of
photoproduction events with $Q^2 < 0.5$ GeV$^2$ was generated with
{\sc Pythia}~\cite{pythia}.
In  the MC samples, the hadronic final state was generated using
the Matrix Element Parton Shower model (MEPS)~\cite{lepto} for QCD
radiation and {\sc Jetset}~\cite{jetset} for hadronisation. 
The  diffractive events in {\sc Djangoh} were generated using  the
soft-colour-interaction mechanism (SCI)~\cite{sci}.

All  MC events were passed through the standard simulation of the
ZEUS detector, based on {\sc Geant} 3.13~\cite{geant}, and trigger
and through the same reconstruction and analysis programs as the data. The
simulation included the geometry of the beampipe apertures, the
magnetic field along the leading proton trajectory and the proton-beam
emittance.

To obtain a good description of the data, it was necessary to reweight
the leading proton $x_L$ and $p_T^2$ distributions generated by the
MC~\cite{reweighting,rinaldi}; the fraction of diffractive events with respect
to the total was also reweighted in bins of $x_L$. In particular, the
slope of the exponential $p_T^2$ distribution, 
ranging from 2.5 to 4.5 GeV$^{-2}$, was increased by a
constant value $\Delta b=3.4$ GeV$^{-2}$ and the $x_L$ spectrum was
reweighted to a flat distribution below the diffractive peak. The
reweighting parameters were chosen according to previous
measurements~\cite{lp95}. The reweighting preserved the total MC cross section.

\begin{figure}[htb]
\begin{center}
\leavevmode
\hbox{%
\epsfxsize = 15cm
\epsffile{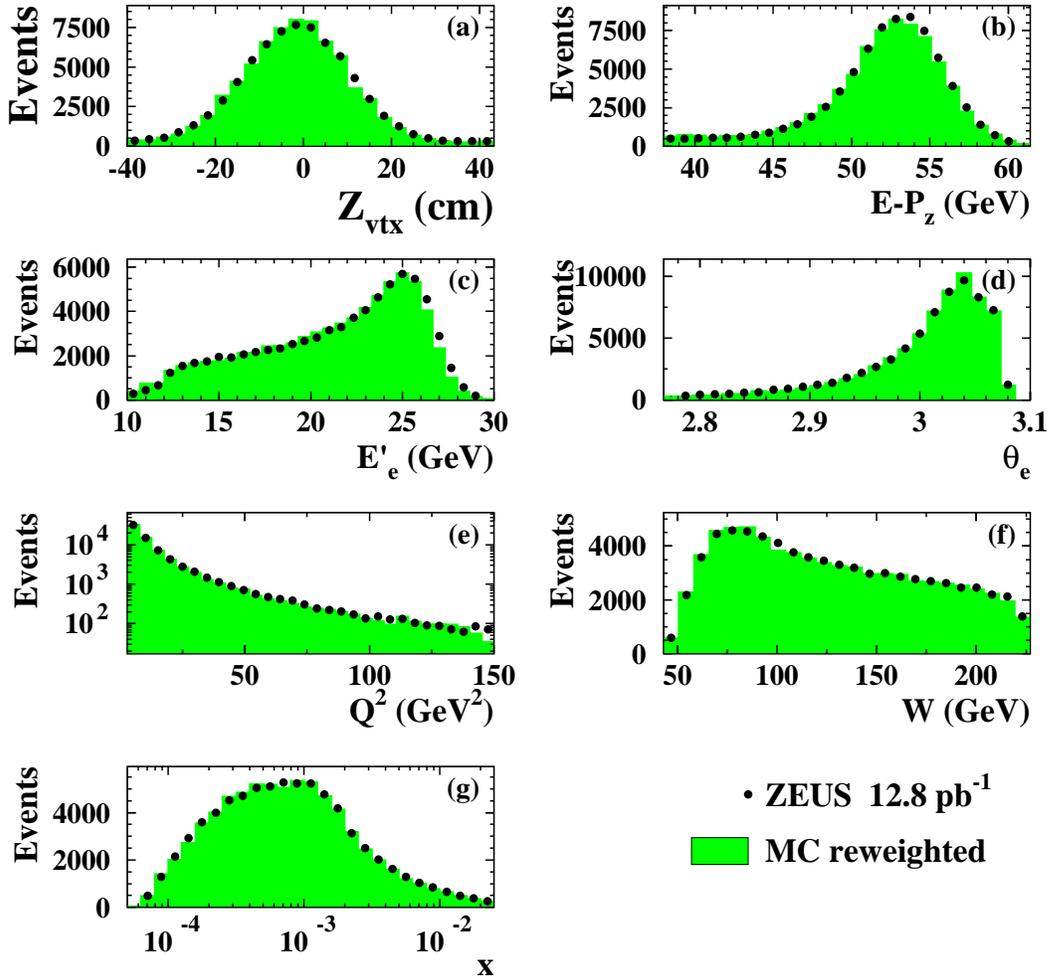}}
\end{center}
\caption{{\protect{
Comparison between data (dots) and reweighted MC (shaded histograms) 
for DIS quantities for the LPS sample: 
(a) $Z$ coordinate of the vertex, 
(b) $E-P_{Z}$ distribution, 
(c) energy of the scattered positron, $E^{\rm \prime}_e$, 
(d) polar angle of the scattered positron, $\theta_e$, 
(e) virtuality of the exchanged photon, \q2, 
(f) invariant mass of the hadronic system, $W$, 
and (g) Bjorken scaling variable, $x$.}}}
\label{fig:datamc:dis}
\end{figure}

\begin{figure}[htb]
\begin{center}
\leavevmode
\hbox{%
\epsfxsize = 15cm
\epsffile{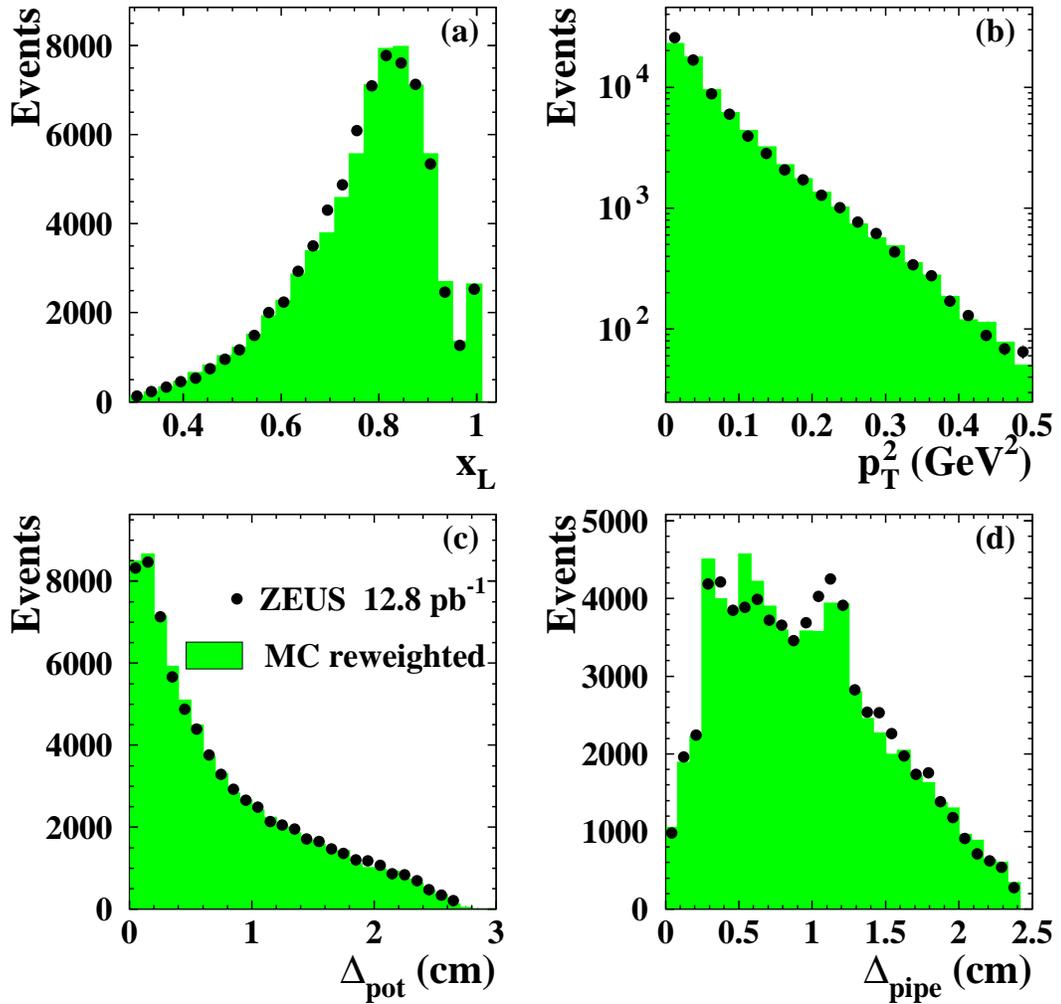}}
\end{center}
\caption{{\protect{
Comparison between data (dots) and reweighted MC (shaded histograms) for
LPS specific quantities: 
(a) fractional longitudinal momentum, $x_L$, 
(b) squared transverse momentum, $p_T^2$, 
(c) minimum track distance from the edge of the pot, $\Delta_{\rm pot}$, 
and (d) minimum track distance from the beampipe, $\Delta_{\rm pipe}$.  
}}}
\label{fig:datamc:lps}
\end{figure}

For the LPS sample, the comparison between the data and the sum of the
reweighted MC samples ({\sc Djangoh} and {\sc Pythia}), for the DIS 
variables and the LPS specific
variables, is shown in Figs.~\ref{fig:datamc:dis} and~\ref{fig:datamc:lps}.
The agreement is generally good. The LPS
variable $\Delta_{\rm pipe}$ is not perfectly reproduced by the reweighted
MC, but the selection cut applied is far from the region in which the
disagreement is observed.

\section{Acceptance}
\label{acceptance}

\begin{figure}[htb]
\begin{center}
\leavevmode
\hbox{%
\epsfxsize = 13cm
\epsffile{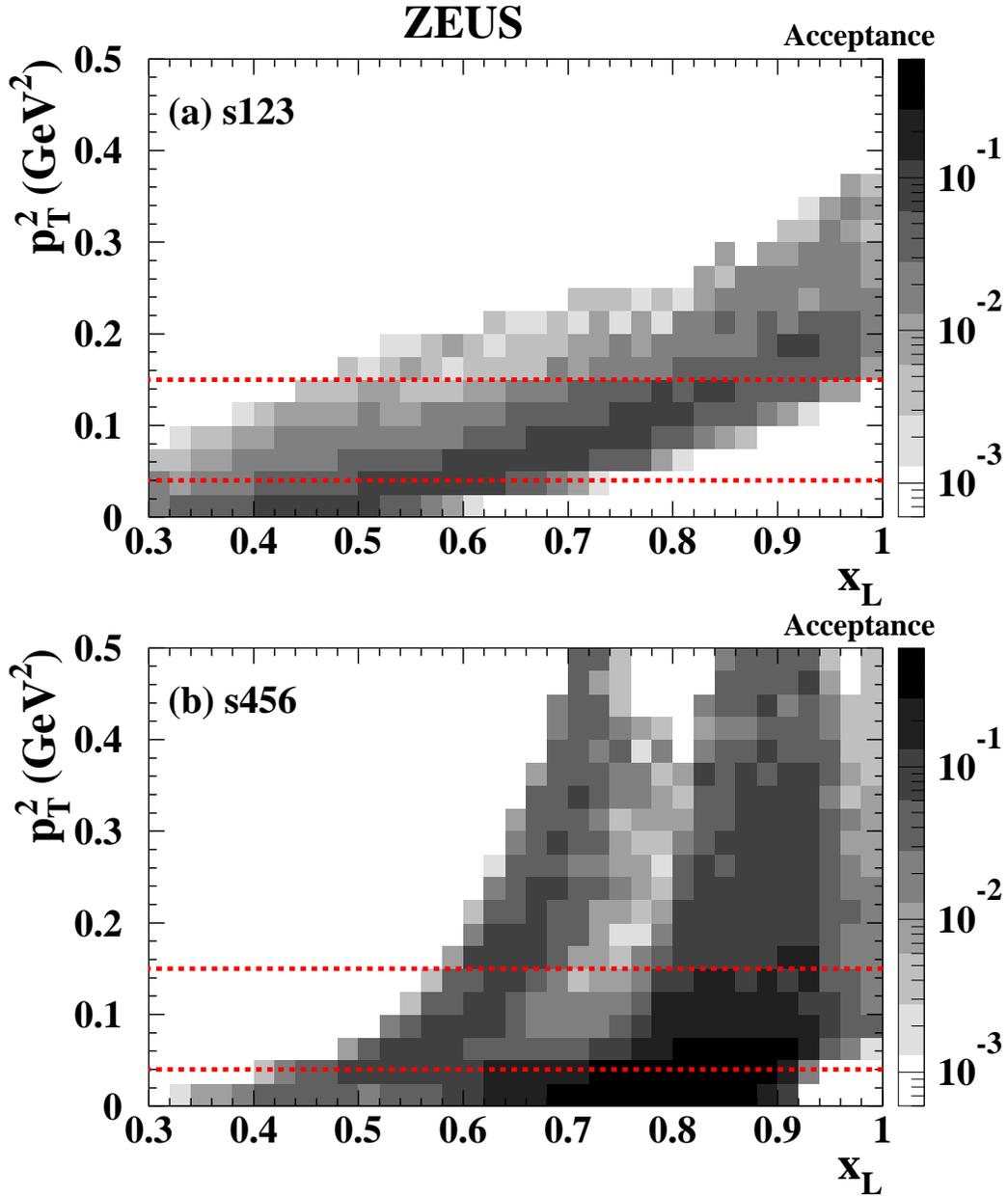}}
\end{center}
\caption{{\protect{
Acceptance of the LPS spectrometers (a) s123 and (b)
s456  in the $x_L$, $p_T^2$ plane. The dashed lines delimit the
three $p_T^2$ integration ranges.}}}
\label{fig:acceptance}
\end{figure}

The acceptance was defined as the ratio of the number of reconstructed
events in a bin to the number of generated events in that bin. This
definition includes the effects of the geometrical acceptance of the
apparatus, its efficiency and resolution, and the event selection
efficiency. Figure~\ref{fig:acceptance} shows the acceptances of the LPS
station combinations s123 and s456  as a function of
$x_L$ and $p_T^2$. The maximum acceptance  is 10\% in the region
$0.63<x_L<0.65$, $0.05<p_T^2<0.1$ GeV$^2$ for the spectrometer s123
and 52\% in the region $0.77<x_L<0.8$, $p_T^2<0.05$ GeV$^2$ for the
spectrometer s456.

The analysis bins were chosen according to the LPS acceptance, resolution
(see Section~\ref{detector}) and available statistics. 

For completeness, also shown in Fig.~\ref{fig:acceptance} are the three
regions of $p_T^2$ used for the cross-section measurements.

\section{Background studies}
\label{background}

The LPS data sample contains  three  background contributions,

\begin{itemize}

\item non-baryon contributions;

\item overlay events;

\item misidentified low-$Q^2$ background.

\end{itemize}

\begin{figure}[htb]
\begin{center}
\leavevmode
\hbox{%
\epsfxsize = 14cm
\epsffile{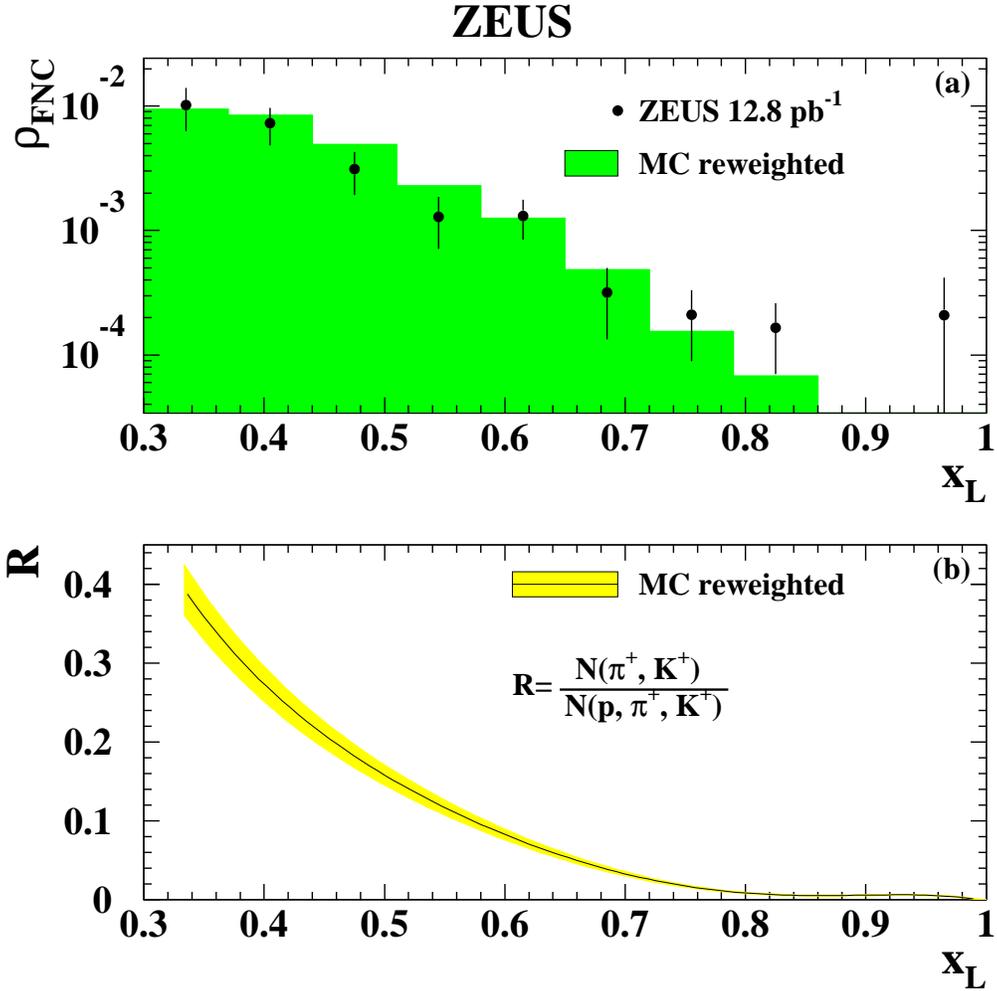}}
\end{center}
\caption{{\protect{
(a) Ratio $\rho_{FNC}$ of the number of events with a track in
the LPS and a neutron candidate in the FNC to the number of events 
with a track in the LPS. The dots represent the data and the shaded 
histogram the reweighted MC;
(b) expected fraction $R$ of positive pions and kaons reconstructed in the LPS
as a function of $x_L$. The error band reflects the statistical
uncertainty derived from the LPS-FNC data sample.
}}}
\label{fig:pik}
\end{figure}

The LPS had no particle identification capability, but MC studies indicate 
that most high-$x_L$ particles in the LPS are protons. The MC expectations 
were tested with a subsample of LPS-tagged events where a neutron candidate 
was found in the FNC~\cite{ln}. The neutron candidate was required to have 
a minimum energy deposit of 50 GeV and the total $E+P_Z$ of the event, 
including the neutron, was required to be below 1750 GeV. A total of 47 
events were found. For this class of events the track in the LPS is most 
likely either a $\pi^+$ or a $K^+$. Figure~\ref{fig:pik}a shows the ratio 
$\rho_{\rm FNC}$ of the number of events with a track in the LPS and a
neutron candidate in the FNC to the number of events with a track in
the LPS. The reweighted MC describes the data well and 
therefore can be used to subtract the background. The fraction $R$ of events 
with a positive meson reconstructed in the LPS, evaluated with the reweighted 
MC, is shown in Fig.~\ref{fig:pik}b as a function of $x_L$.
The fraction $R$ in the MC was found to be independent of $p_T^2$. 
It is substantial at low $x_L$ and falls below 10\% above $x_L\approx 0.6$. 
This contribution was subtracted.

The $E+P_Z$ spectrum for the beam-halo events was constructed as
a combination of generic DIS events and a beam-halo track
reconstructed in the LPS in randomly triggered events. The $E+P_Z$ 
distribution was normalised to the data for $E+P_Z> 1685$ GeV, which
mainly contain beam-halo events. The background remaining after the
$E+P_Z< 1655$ GeV cut was negligible for $x_L<0.9$, and reached
$(8\pm3)$\% for $x_L>0.98$. The expected fraction of overlay events was
subtracted.

The acceptance corrections were calculated using the reweighted 
MC generated with 
$Q^2>2$  GeV$^2$. The contribution of events which migrate from the region
$Q^2<2$ GeV$^2$ was found to be independent of $x_L$ and $p_T^2$ and 
equal to ($7.3\pm 0.5$)\%; it was subtracted.

\section{Systematic studies}
\label{systematics}

The systematic uncertainties were calculated by varying the cuts and
by modifying the analysis procedure.
The stability of the DIS selection was checked by varying the selection cuts,

\begin{itemize}

\item the $|Z_{\rm vtx}|$ cut was varied by $\pm10$ cm;

\item the cut on the scattered positron energy was varied by $\pm$ 2 GeV  and 
the width of the fiducial region in the rear calorimeter was varied by
0.5 cm in the $X$ and $Y$ directions;

\item the $E-P_Z$ cut was changed to $35<E-P_Z<68$ GeV and $41<E-P_Z<62$ GeV;

\item the $y_{\rm JB}$ cut was raised to 0.04.

\end{itemize}

The observed changes in the cross section were below 1\% and  neglected.
The variation in the LPS selection of the $\Delta_{\rm pipe}$ threshold by 
$\pm$ 0.03 cm and the $\Delta_{\rm pot}$ threshold by $\pm$ 0.01 led to 
negligible changes in the cross section (<1\%).

The following checks resulted in non-negligible systematic uncertainties of 
the cross section (the mean value is given in brackets):

\begin{itemize}

\item the reweighting parameter $\Delta b$ was varied by $\pm 0.9$ GeV$^{-2}$, 
compatible with the spread of $b$ versus $x_L$ ($^{+6.8\%}_{-5.9\%}$);

\item the subtracted fraction of background from $\pi^+$ and  $K^+$ 
reconstructed in the LPS was varied by the statistical uncertainty derived
from the LPS-FNC data (see Fig.~\ref{fig:pik}) ($^{+1.7\%}_{-1.7\%}$
for $x_L<0.7$);

\item the fraction of overlay events to be subtracted from the data
was increased and decreased by its statistical uncertainty
($^{+2.0\%}_{-2.0\%}$ for $x_L>0.9$);

\item the uncertainty on the beam optics was evaluated by varying 
the transverse momentum spread of the proton beam in the MC by  
$\pm 10\%$~\cite{lp97-diff} and led to a change of typically 
$\pm 1.4\%$. In addition, the positions of the LPS stations  were varied 
 to reflect the actual position of the stations during the data-taking
period. This was done because, in the simulation, the MC assumes 
only one average position  ($^{+2.5\%}_{-1.8\%}$). In the diffractive region
the uncertainty related to the beam optics increased to $\pm$10\%.

\end{itemize}

The resulting total systematic uncertainty, obtained by adding in quadrature
all the individual systematic uncertainties (combining positive and negative
contributions separately), excluding an 
overall normalisation uncertainty of 2\% from the luminosity
measurement, is shown in the figures as an error band, that includes both a 
correlated and an uncorrelated component.

\section{Results}
\label{results}

All measurements were performed separately with the s123 and s456 
spectrometers,
and were then combined in a weighted average, using only statistical
uncertainties. This procedure was repeated for every systematic check.
Some measurements are presented as normalised to the inclusive DIS 
cross section, 
$\sigma_{\rm inc}$, determined from the inclusive DIS sample described in
Section~\ref{selection}.
All cross-section measurements are averaged over a given bin and quoted
at the mean value of the variable in that bin.
The measurements for $x_L>0.93$ are presented here in a different kinematic 
domain than those previously published~\cite{lp97-diff}.

\subsection{Transverse-momentum spectra and \boldmath{$p_T^2$} slopes}

\begin{figure}[htb]
\begin{center}
\leavevmode
\hbox{%
\epsfxsize = 15cm
\epsffile{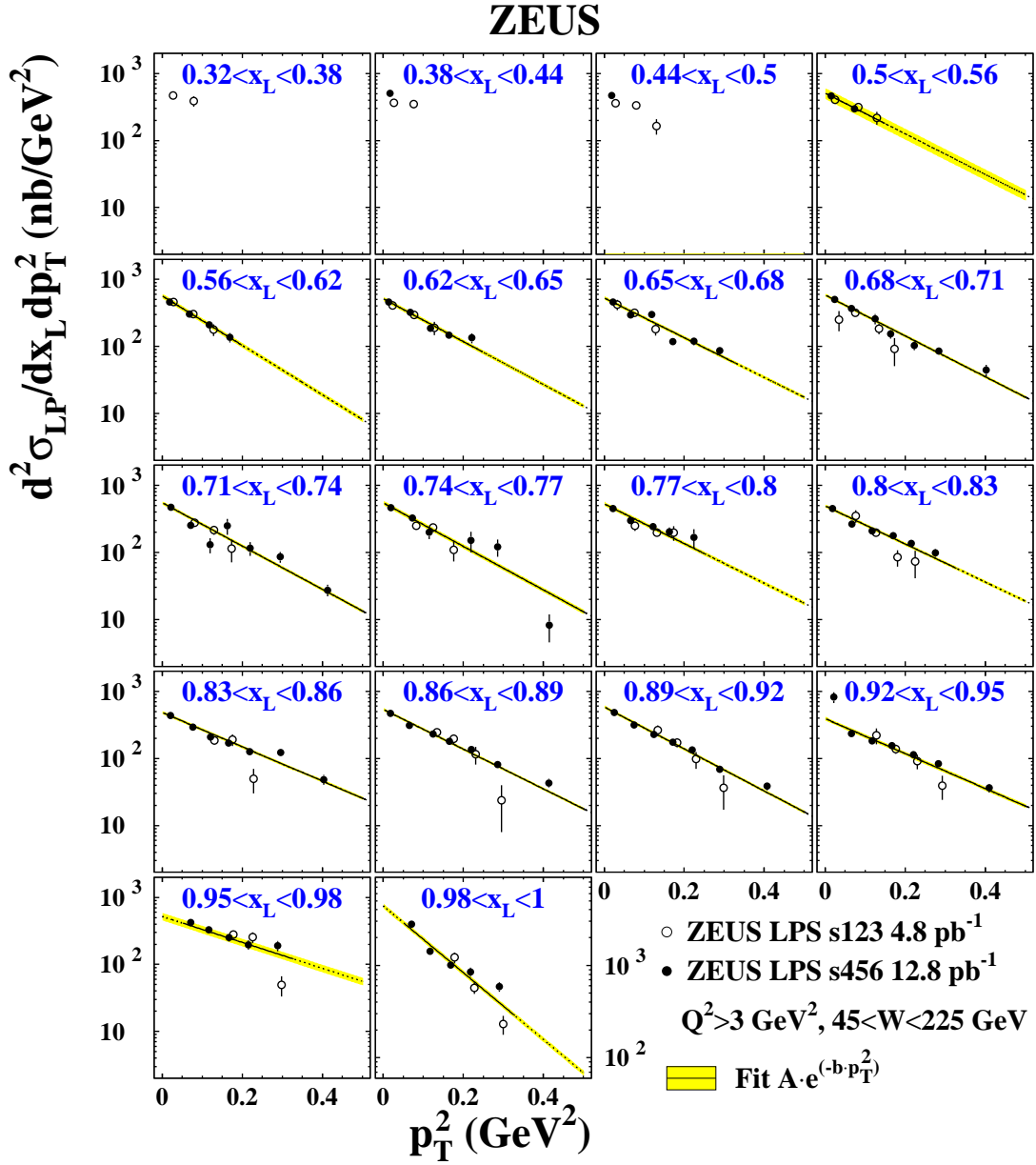}}
\end{center}
\caption{{\protect{
The double-differential cross-section
$d^2\sigma_{\rm LP}/dx_Ldp_T^2$ for $Q^2>3$
$GeV^2$ and $45<W<225$ GeV  as a function of $p_T^2$ in
bins of $x_L$. The circles and the dots are the ZEUS data measured
with the spectrometers s123 and s456, respectively. 
For  clarity, only the statistical uncertainties are
shown. The systematic uncertainties are listed in~\tab{xlpt2}.
The lines are the result of a fit to a function $A\cdot
e^{-b\cdot p_T^2}$, as described in the text. The solid lines indicate 
the range in which the fit was performed.
The bands show the statistical uncertainty of the fit. 
}}}
\label{fig:sigma:xlpt2}
\end{figure}

\begin{figure}[htb]
\begin{center}
\leavevmode
\hbox{%
\epsfxsize = 14cm
\epsffile{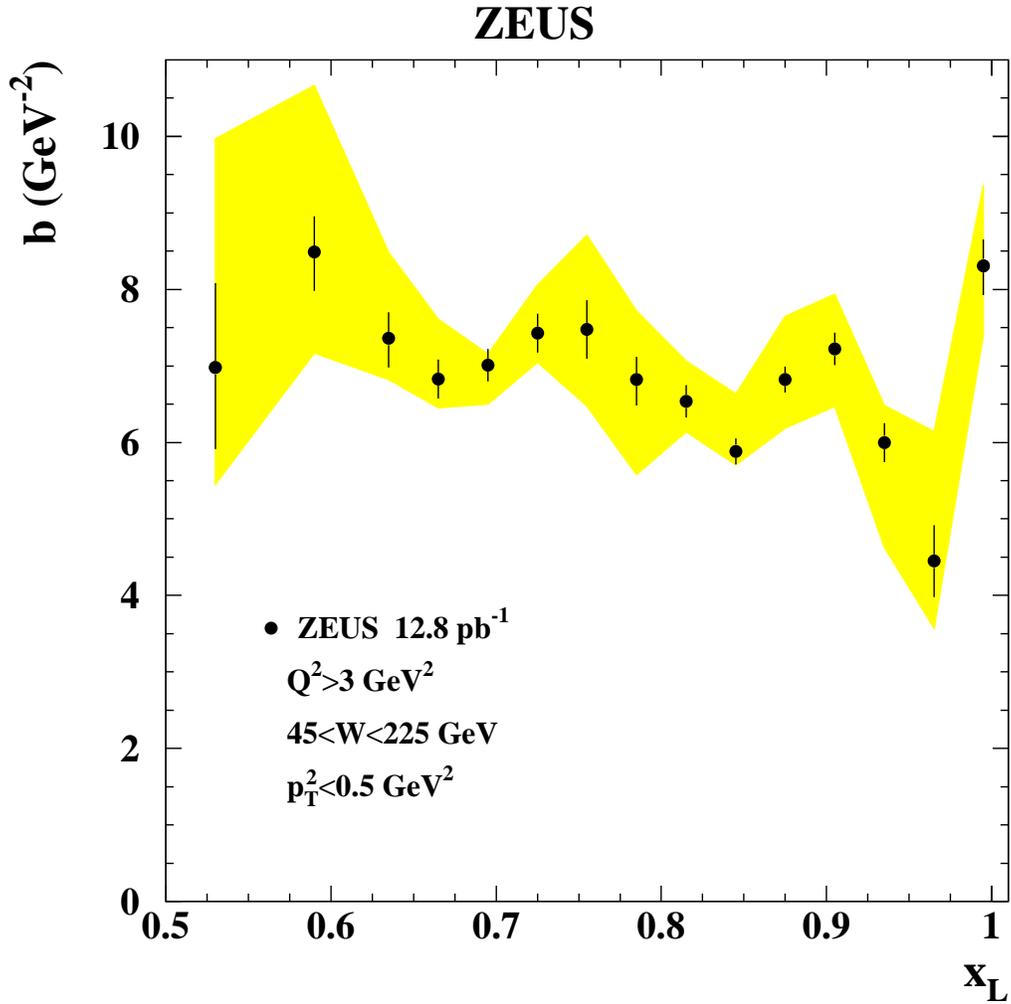}}
\end{center}
\caption{{\protect{
The $p_T^2$-slope, $b$, of the cross-section
$d^2\sigma_{\rm LP}/dx_Ldp_T^2$, as defined by the parameterisation $A
\cdot e^{-b\cdot p_T^2}$ and obtained from a fit to the data in bins of
$x_L$, in the kinematic range indicated in the figure. 
The bands represent the systematic uncertainty. 
}}}
\label{fig:sigma:slopes}
\end{figure}

\begin{figure}[htb]
\begin{center}
\leavevmode
\hbox{%
\epsfxsize = 14cm
\epsffile{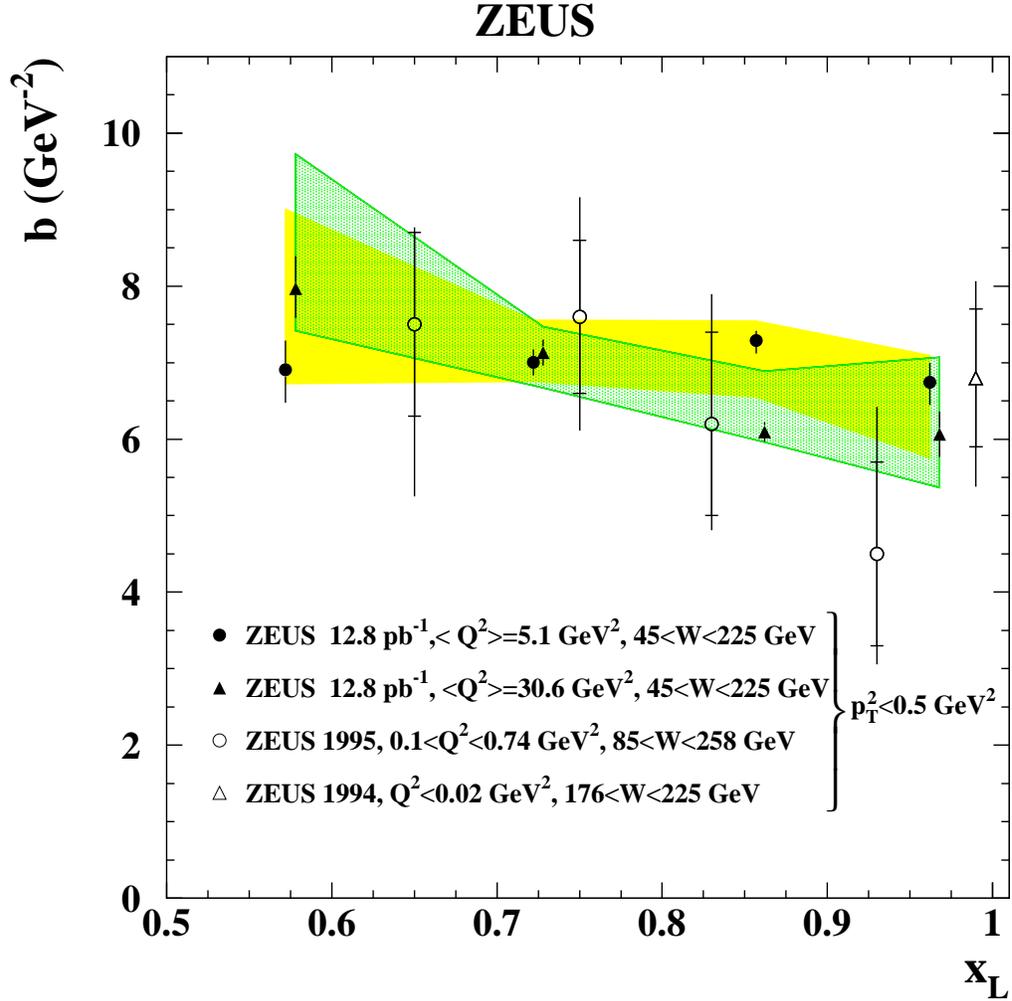}}
\end{center}
\caption{{\protect{
The $p_T^2$-slope, $b$, of the cross-section
$d^2\sigma_{\rm LP}/dx_Ldp_T^2$, as defined by the parameterisation $A
\cdot e^{-b\cdot p_T^2}$ and obtained from a fit to the data in bins of
$x_L$, in different kinematic ranges as indicated in the figure. 
The bands represent the systematic uncertainty. 
For the ZEUS 1995 data~\cite{lp95} and the ZEUS 1994
data~\cite{lpsrho}, the inner vertical bars represent the statistical
uncertainties, the outer bars the statistical and systematic
uncertainties added in quadrature.}}} 
\label{fig:sigma:slopes:q2}
\end{figure}

The double-differential cross-section $d^2\sigma_{\rm LP}/dx_Ldp_T^2$ as a
function of $p_T^2$ in bins of $x_L$ is presented in
Fig.~\ref{fig:sigma:xlpt2} and given in~\tab{xlpt2}. The results
obtained with the s123 and s456 spectrometers are shown separately. 
Within the uncorrelated uncertainties, 
the two samples lead to consistent results. 
The lines shown in Fig.~\ref{fig:sigma:xlpt2} represent the results of
a fit of an exponential function $A\cdot e^{-bp_T^2}$ to the combined
cross-section $d^2\sigma_{\rm LP}/dx_Ldp_T^2$. The band shows the statistical
uncertainty of the fit. 
The slopes are presented as a function of $x_L$ in
Fig.~\ref{fig:sigma:slopes} and given in~\tab{slopes}.
The slopes show no strong dependence on $x_L$. 
The mean value of the slopes is $\langle b \rangle =6.76\pm 0.07
({\rm stat.})^{+0.63}_{-0.52} ({\rm syst.})$ GeV$^{-2}$.
The measurements of the
$p_T^2$ slopes at $\langle Q^2 \rangle =5.1$ GeV$^2$ and $\langle Q^2
\rangle=30.1$ GeV$^2$, in the
range $45<W<225$ GeV, where $x_L$ bins were combined, are presented in
Fig.~\ref{fig:sigma:slopes:q2} and given in~\tab{slopesq2}.
Also shown are the ZEUS 1994 data~\cite{lpsrho} in the range 
$Q^2<0.02$ GeV$^2$ and $176<W<225$ GeV
and the ZEUS 1995~\cite{lp95} data in the range $0.1<Q^2<0.74$ GeV$^2$ and
$85<W<258$ GeV. The $p_T^2$ slopes are independent of the
virtuality of the exchanged photon.

\subsection{Longitudinal momentum spectra}

\begin{figure}[htb]
\begin{center}
\leavevmode
\hbox{%
\epsfxsize = 14cm
\epsffile{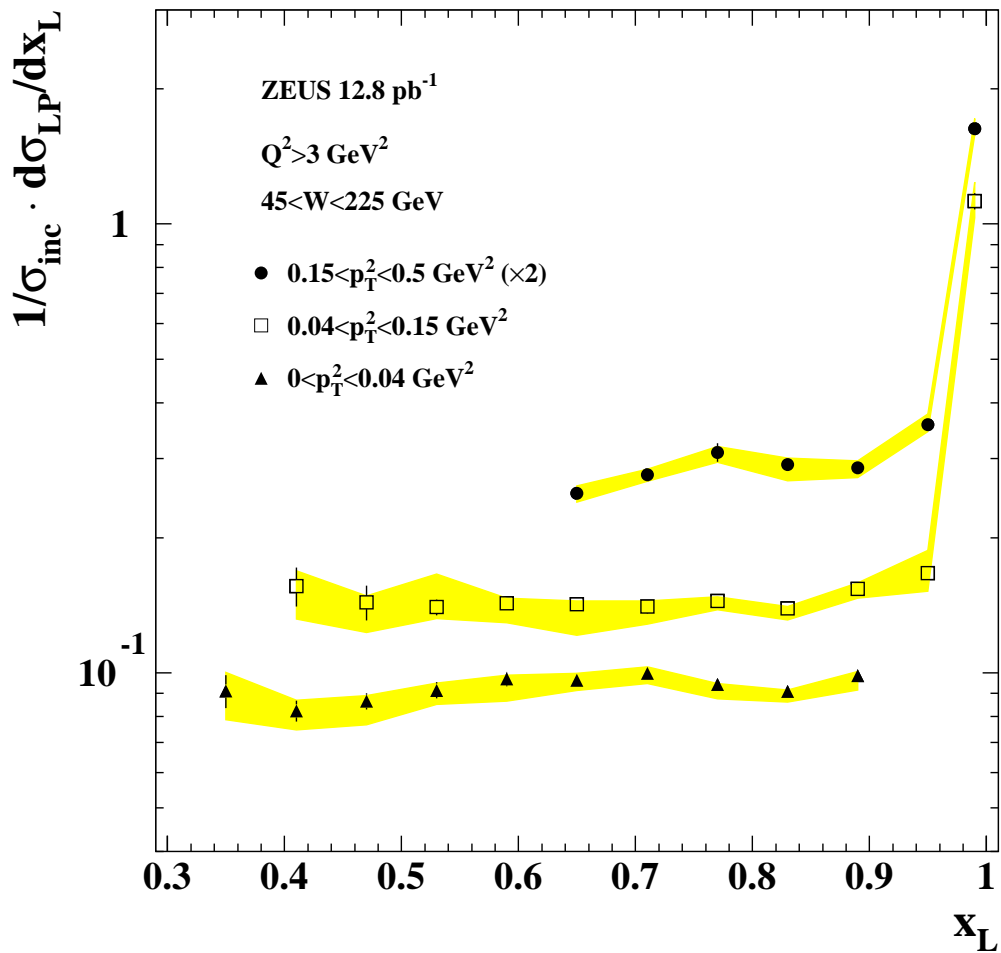}}
\end{center}
\caption{{\protect{
The leading proton production rate, $1/\sigma_{\rm inc}\cdot \sigma_{\rm LP}/dx_L$, as a function of
$x_L$ in three $p_T^2$ ranges as indicated in the figure. The measurements
in the higher $p_T^2$ range are multiplied by a factor two for
visibility. The bands represent the systematic uncertainty. 
}}}
\label{fig:sigma:xl3}
\end{figure}

The cross section as a function of $x_L$ has been measured in three bins of 
$p_T^2$: $0<p_T^2<0.04$, $0.04<p_T^2<0.15$ and $0.15<p_T^2<0.5$ GeV$^2$. The
leading proton production rate, $1/\sigma_{\rm inc}\cdot d\sigma_{\rm
LP}/dx_L$, for the three
$p_T^2$ ranges is shown in Fig.~\ref{fig:sigma:xl3} and listed in \tab{xl3}.
Due to the LPS acceptance, the accessible $x_L$ range changes as a function of
$p_T^2$, as seen in the figure.
The rate as a function of $x_L$ is 
approximately flat up to the diffractive peak, where it increases by
a factor of about six. This behaviour of the cross section as a function of
$x_L$ is essentially independent of $p_T^2$. 

\begin{figure}[htb]
\begin{center}
\leavevmode
\hbox{%
\epsfxsize = 14cm
\epsffile{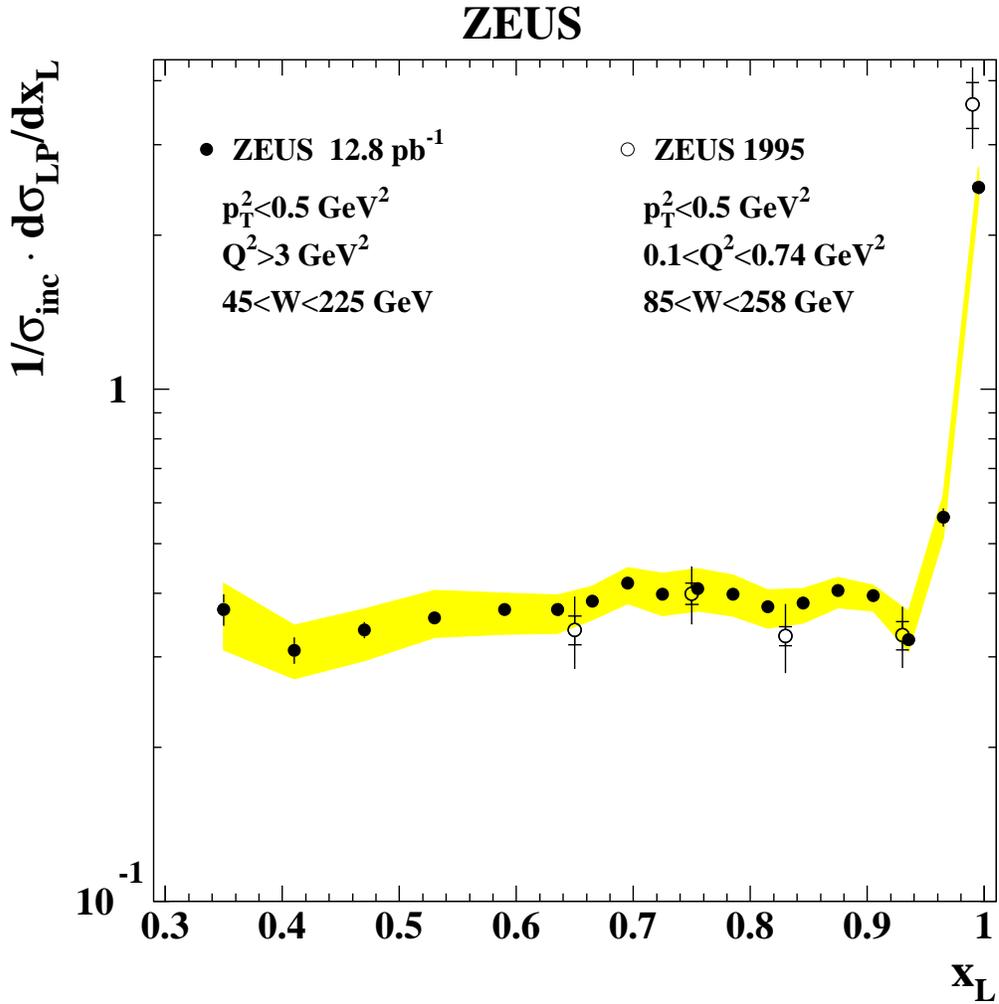}}
\end{center}
\caption{{\protect{
The leading proton production rate,
$1/\sigma_{\rm inc}\cdot d\sigma_{\rm LP}/dx_L$, for
two ranges of \q2 as indicated in the figure. 
The bands represent the systematic
uncertainty.  For the ZEUS 1995 data~\cite{lp95} 
the inner vertical bars represent the statistical uncertainties, 
the outer bars the statistical and systematic
uncertainties added in quadrature.
}}}
\label{fig:sigma:xldis}
\end{figure}

Since, as discussed in the previous section, also the $p_T^2$ slopes are
independent of $x_L$, the cross section as a function of $x_L$ can be 
extrapolated to the full $x_L>0.32$ and $p_T^2<0.5$ GeV$^2$ range.
The measurement of $1/\sigma_{\rm inc}\cdot d\sigma_{\rm LP}/dx_L$ as a 
function of $x_L$, extrapolated to the full $p_T^2<0.5$ GeV$^2$ 
range is shown in Fig.~\ref{fig:sigma:xldis} and given in~\tab{xldis}.
For comparison, the ZEUS 1995 data~\cite{lp95} with lower $Q^2$ are also 
shown. The two measurements are consistent.

\begin{figure}[htb]
\begin{center}
\leavevmode
\hbox{%
\epsfxsize = 14cm
\epsffile{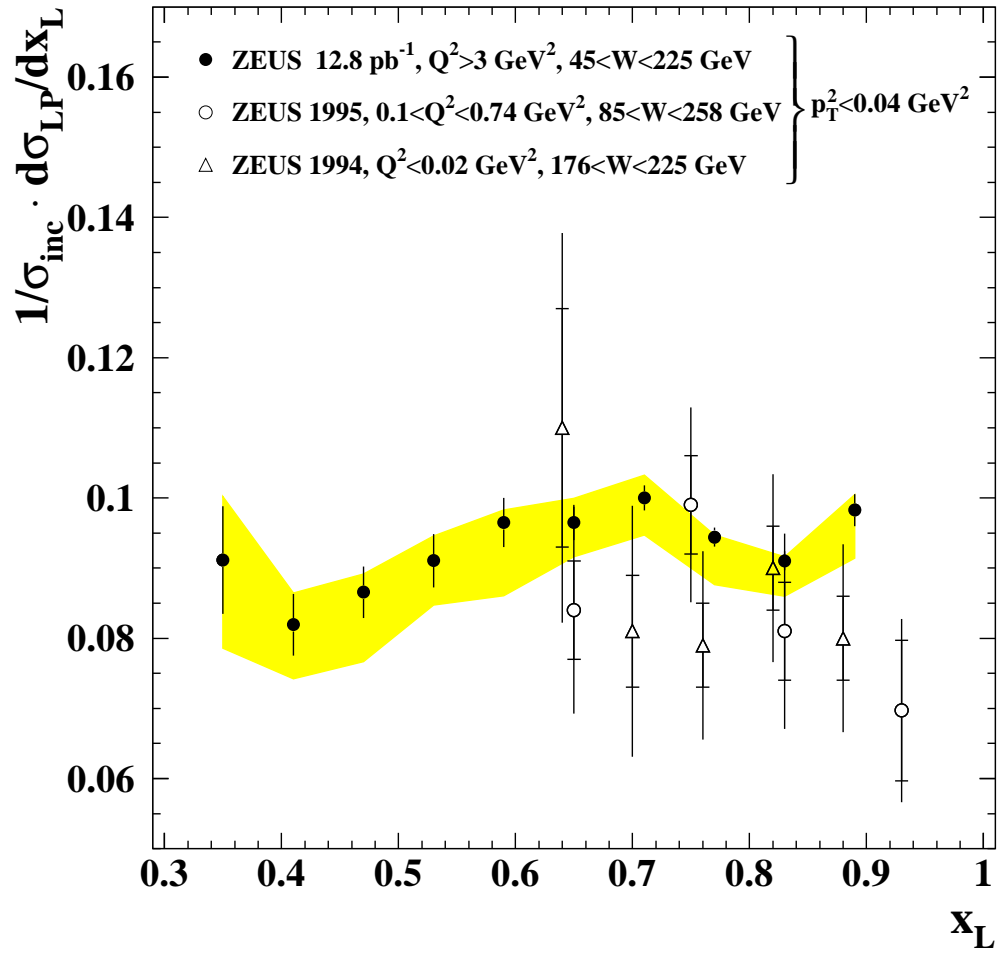}}
\end{center}
\caption{{\protect{
The  leading proton production rate, 
$1/\sigma_{\rm inc}\cdot d\sigma_{\rm LP}/dx_L$,
for $p_T^2<0.04$ $GeV^2$ in the kinematic ranges indicated in the
figure. Other details as in Fig.~\ref{fig:sigma:slopes:q2}.
}}}
\label{fig:sigma:xl004}
\end{figure}

For $p_T^2<0.04$ GeV$^2$, the measurement of 
$1/\sigma_{\rm inc}\cdot d\sigma_{\rm LP}/dx_L$  can also be compared  
to previous measurements in the photoproduction regime 
($Q^2<0.02$ GeV$^2$)~\cite{lp95}. The comparison is shown 
in Fig.~\ref{fig:sigma:xl004}. Due to the low $p_T^2$ values, the diffractive
peak is not accessible (see Fig.~\ref{fig:acceptance}). 
The photoproduction data tend to lie
systematically below the higher-$Q^2$ measurement, though within
uncertainties the results are consistent.

\subsection{Ratios of leading proton production to inclusive DIS yields}

The rate of leading proton production, $r^{\rm LP(3)}(x,Q^2,x_L)$, in $e^+p$ scattering  was determined according to

\begin{equation}\label{rlp3:meas}
r^{\rm LP(3)}(x,Q^2,x_L)=\frac{N^{\rm LPS}(x,Q^2,x_L)}{N^{\rm DIS}(x,Q^2)}\frac{A_{\rm DIS}}
{A_{\rm LPS}}\frac{\mathcal{L}^{\rm DIS}}{\mathcal{L}^{\rm LPS}}\frac{1}{\Delta
x_L},
\end{equation}

where $N^{\rm LPS}(x,Q^2,x_L)$ is the number of events
corresponding to an integrated luminosity, $\mathcal{L}^{\rm LPS}$, 
 with a proton
candidate in the LPS in a given ($x,Q^2,x_L$) bin and integrated
over $0<p_T^2<0.5$ GeV$^2$,
and $N^{\rm DIS}(x,Q^2)$ is the number of DIS events 
corresponding to an integrated luminosity, $\mathcal{L}^{\rm DIS}$,
in that ($x,Q^2$) bin. 
The acceptance $A_{\rm DIS}$ was estimated by applying
only the DIS selection cuts and $A_{\rm LPS}$ is the acceptance of the 
LPS sample.
The variable $\Delta x_L$ is the size of the $x_L$ bin.

\begin{figure}[htb]
\begin{center}
\leavevmode
\hbox{%
\epsfxsize = 15cm
\epsffile{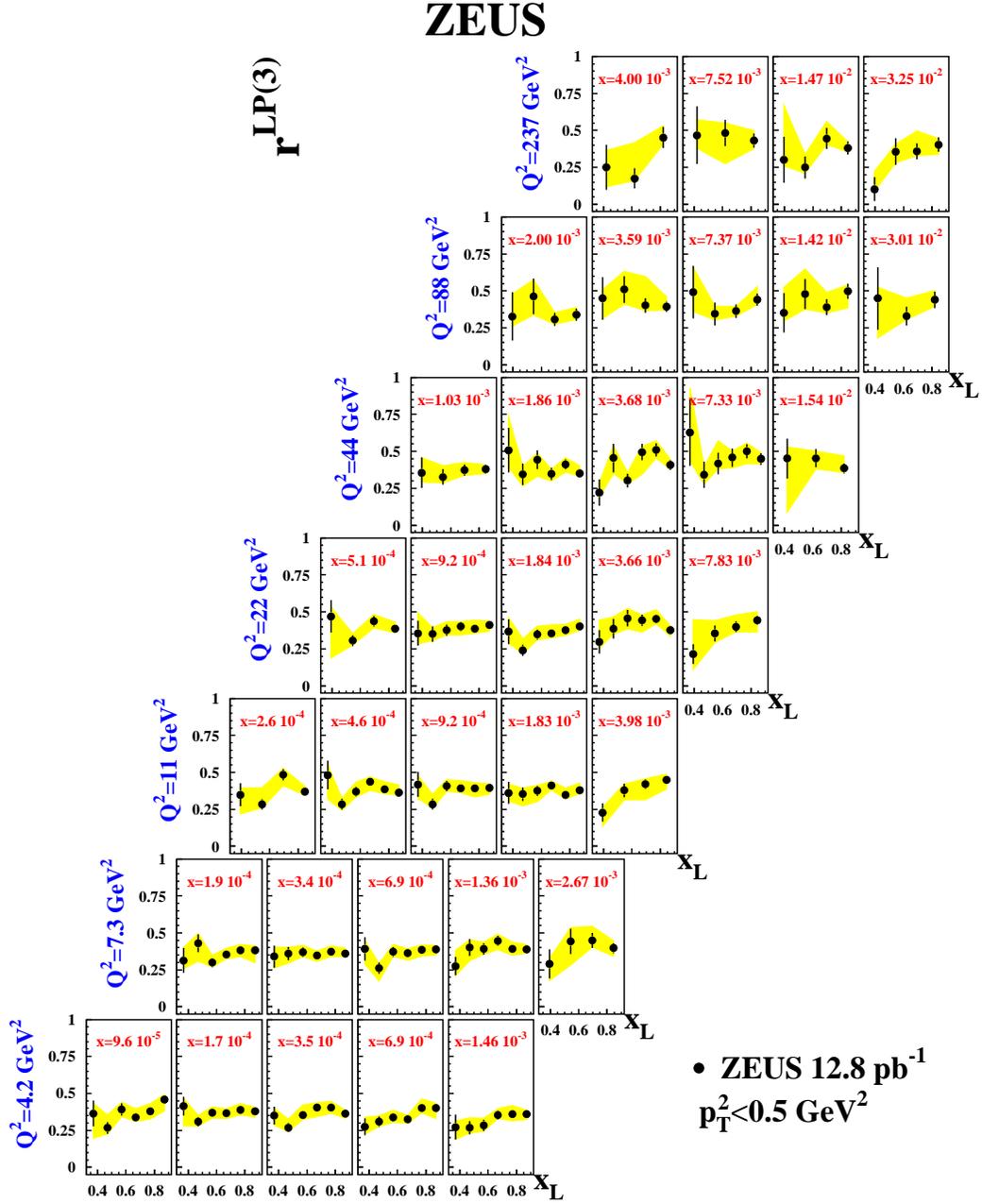}}
\end{center}
\caption{{\protect{
The leading proton production rate, $r^{\rm LP(3)}$,
as a function of $x_L$ in bins of $x$ and $Q^2$, for $p_T^2<0.5$ $GeV^2$.
The bands represent the systematic uncertainty.
}}}
\label{fig:rlp3}
\end{figure}

The ratio $r^{\rm LP(3)}$ as a function of $x_L$ in bins of $x$ and $Q^2$
is shown in Fig.~\ref{fig:rlp3} and given in~\tab{rlp3full}.
The $x_L$ range of the measurement is
limited to $0.32<x_L<0.92$, as detailed studies of the diffraction region were 
presented elsewhere~\cite{lp97-diff}. 
The ratio  $r^{\rm LP(3)}$ has also been  measured in the three ranges of  
$p_T^2$ and the values are given in Tables~\ref{tab-rlp3_0.04},
~\ref{tab-rlp3_0.15} and~\ref{tab-rlp3_0.5}. 
The $r^{\rm LP(3)}$ values are approximately constant over the kinematic 
range of this analysis, independent of the $p_T^2$ range.

\begin{figure}[htb]
\begin{center}
\leavevmode
\hbox{%
\epsfxsize = 14cm
\epsffile{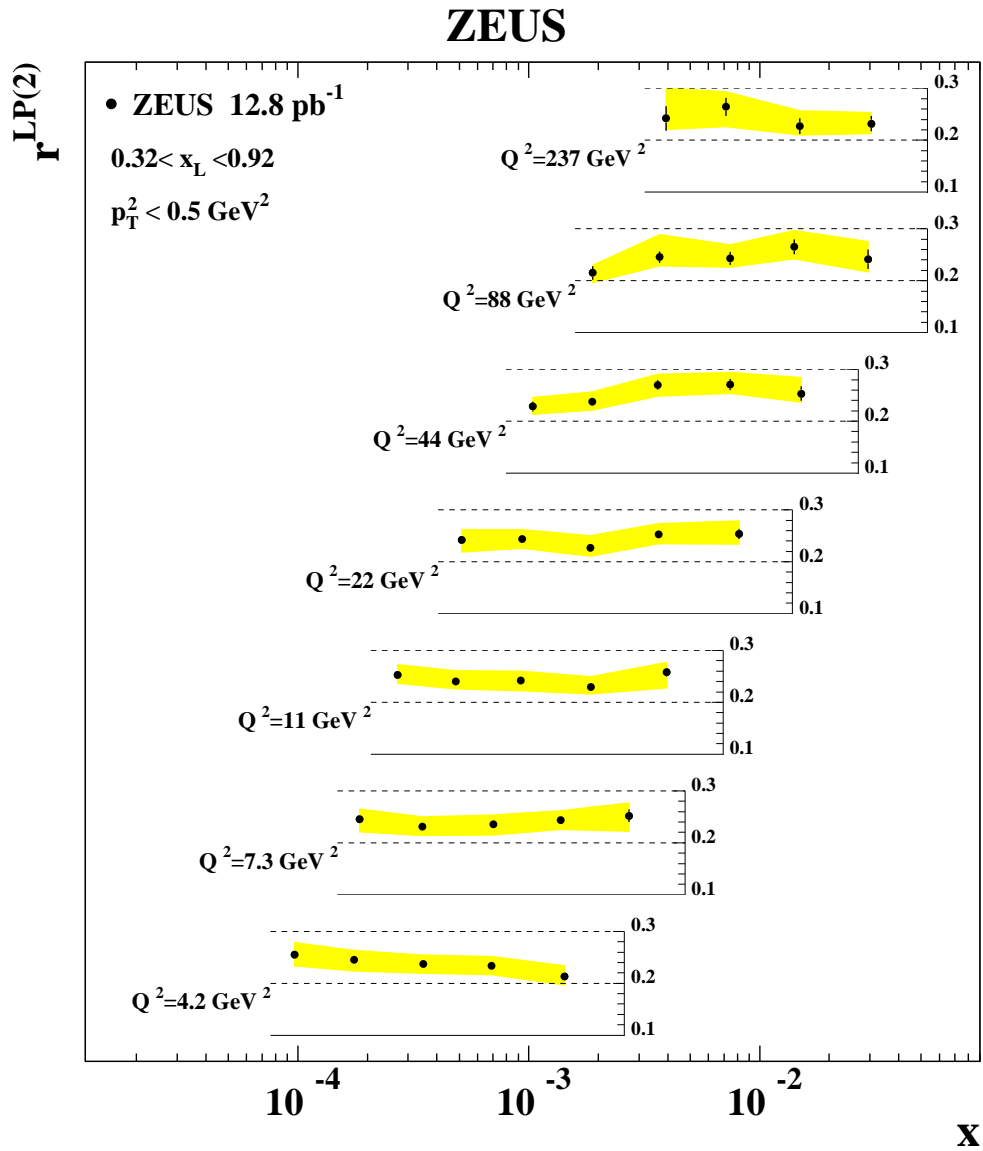}}
\end{center}
\caption{{\protect{
The leading proton production rate, $r^{\rm LP(2)}$,
as a function of $x$ in bins $Q^2$, in the kinematic range indicated
in the figure.  The bands represent the correlated systematic uncertainty.}}}
\label{fig:rlp2}
\end{figure}

The rate of leading proton production as a function of $x$ and $Q^2$,
integrated over $0.32<x_L<0.92$ and $p_T^2<0.5$ GeV$^2$, 
$r^{\rm LP(2)}(x,Q^2)$, 
is shown in Fig.~\ref{fig:rlp2} and given in~\tab{rlp2}.
The ratio $r^{\rm LP(2)}$ is
approximately constant as a function of $x$ and of $Q^2$. The mean
value $\langle r^{\rm LP(2)} \rangle = 0.240 \pm 0.001 ({\rm stat.})
^{+0.020}_{-0.018} ({\rm syst.})$ means that 
approximately 24\% of inclusive DIS events have a leading proton in the 
range $0.32<x_L<0.92$ with $p_T^2<0.5$ GeV$^2$.

\begin{figure}[htb]
\begin{center}
\leavevmode
\hbox{%
\epsfxsize = 14cm
\epsffile{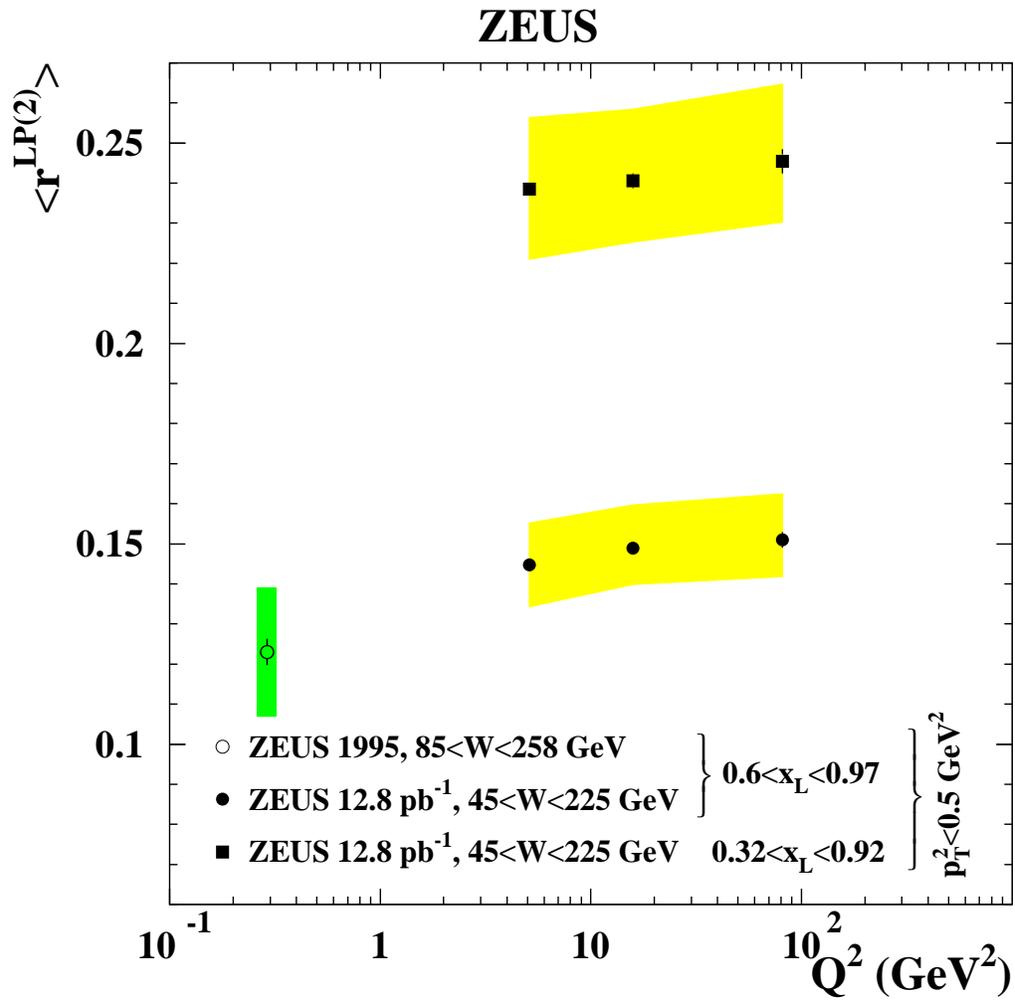}}
\end{center}
\caption{{\protect{
The leading proton production rate averaged over $x$, 
$\langle r^{\rm LP(2)} \rangle$, 
as a function of $Q^2$ in two different ranges of $x_L$, 
in the kinematic range indicated in the figure. 
The bands represent the correlated systematic uncertainty.
}}} 
\label{fig:rlp95}
\end{figure}

The ratio $\langle r^{\rm LP(2)} \rangle$ averaged over $x$ as a function
of $Q^2$ is shown in Fig.~\ref{fig:rlp95} and given in~\tab{rlpmean},
in the range $0.32<x_L<0.92$, $p_T^2<0.5$ GeV$^2$ and $45<W<225$ GeV. 
A mild increase with $Q^2$ cannot be excluded.
To further investigate the $Q^2$ dependence, the rates integrated over
$0.6<x_L<0.97$ can be compared to the equivalent rates  for 
$\langle Q^2 \rangle=0.29$ GeV$^2$ measured by the ZEUS collaboration
 previously~\cite{lp95}.
The result is shown in Fig.~\ref{fig:rlp95} and included in
~\tab{rlpmean}.
 Assuming that the systematic uncertainties have a similar origin, 
dominated by the LPS
acceptance, a steady rise with $Q^2$ is observed. A similar effect was observed
in leading neutron production~\cite{ln}, where it was attributed to absorption
and rescattering effects~\cite{abs}, which disappear when the transverse size 
of the virtual photon decreases with increasing $Q^2$.

\subsection{The leading proton structure functions}

The ratio $r^{\rm LP(2)}$ can be expressed as the ratio of $F_2^{\rm LP(2)}$
to  $F_2$,

\begin{equation}\label{f2lp}
r^{\rm LP(2)}(x,Q^2)=\frac{F_2^{\rm LP(2)}(x,Q^2)}{F_2(x,Q^2)},
\end{equation}

where $F_2^{\rm LP(2)}$ is obtained from $F_2^{\rm LP(4)}$ by integrating over
$x_L$ and $p_T^2$. Therefore, the values of $F_2^{\rm LP(2)}$ can be obtained
from the measured $r^{\rm LP(2)}$ and $F_2$.
The values of $F_2$ were obtained from the NLO ZEUS-S fit
parameterisation of the parton distribution functions of 
the proton~\cite{zeus-s}.

\begin{figure}[htb]
\begin{center}
\leavevmode
\hbox{%
\epsfxsize = 15cm
\epsffile{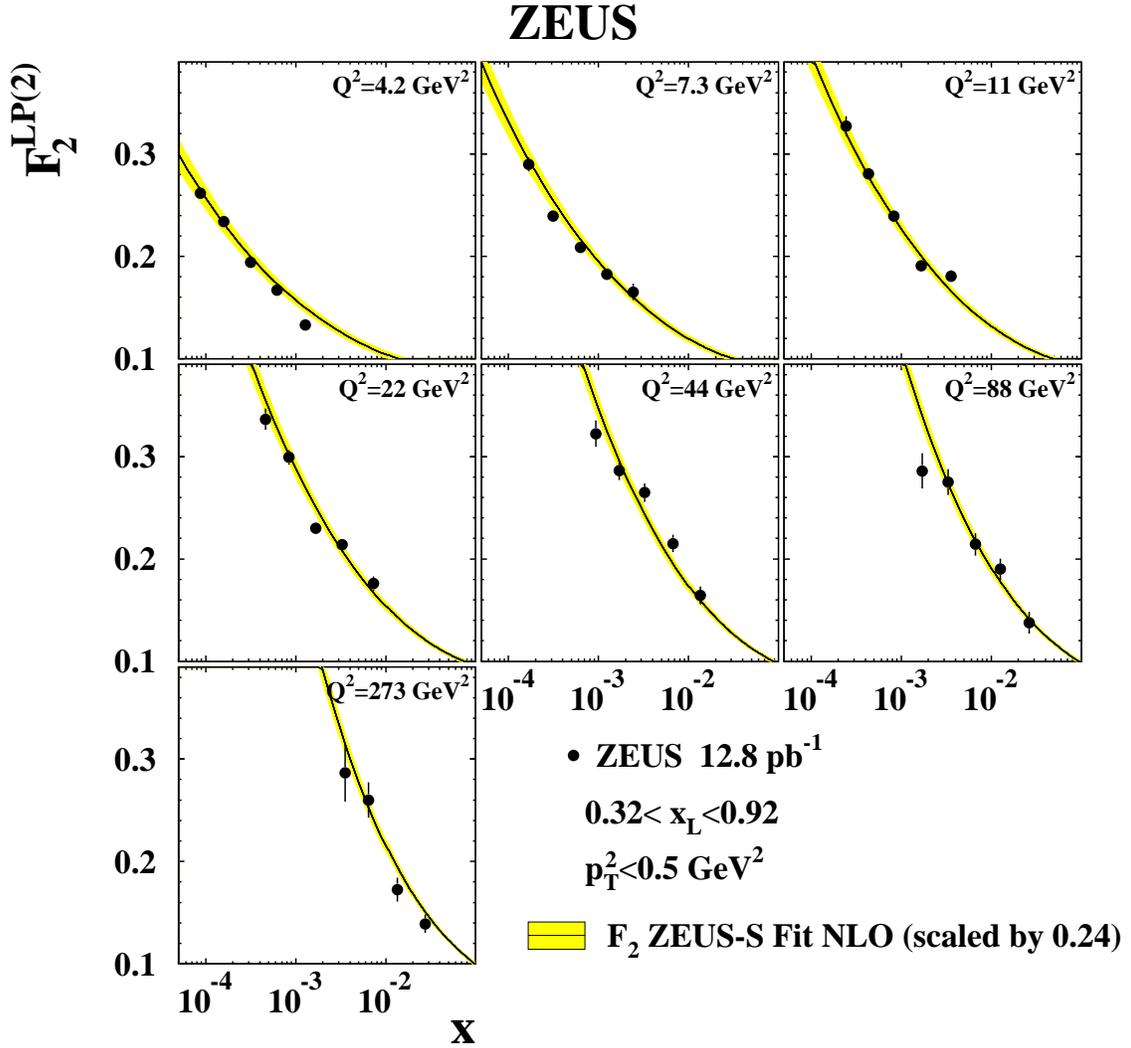}}
\end{center}
\caption{{\protect{
The leading proton structure function, $F^{\rm LP(2)}$,
as a function of $x$ in bins of $Q^2$, in the kinematic range
indicated in the figure. 
For clarity, only the statistical uncertainties are
shown. The systematic uncertainties are listed in~\tab{flp2}.
The lines show the $F_2$ parameterisation  scaled by  0.24,
the approximate average value of $r^{\rm LP(2)}$.
The bands show the one-standard deviation limits of the $F_2$ parameterisation.
}}}
\label{fig:flp2}
\end{figure}

The structure-function  $F_2^{\rm LP(2)}$ is presented in Fig.~\ref{fig:flp2},
plotted as a function of $x$ at fixed $Q^2$ values for $0.32<x_L<0.92$, 
$p_T^2<0.5$ GeV$^2$ and given in~\tab{flp2}.
 The curves in the plot show the $F_2$ parameterisation
scaled by the average value $\langle r^{\rm LP(2)} \rangle =0.24 $ and 
the bands represent the one-standard-deviation limits of the NLO
ZEUS-S parameterisation.
A very good description of $F_2^{\rm LP(2)}$ is obtained, as expected from 
the approximate $x$ and $Q^2$ independence of $r^{\rm LP(2)}$.

\subsection{Comparison to leading neutrons}

\begin{figure}[htb]
\begin{center}
\leavevmode
\hbox{%
\epsfxsize = 14cm
\epsffile{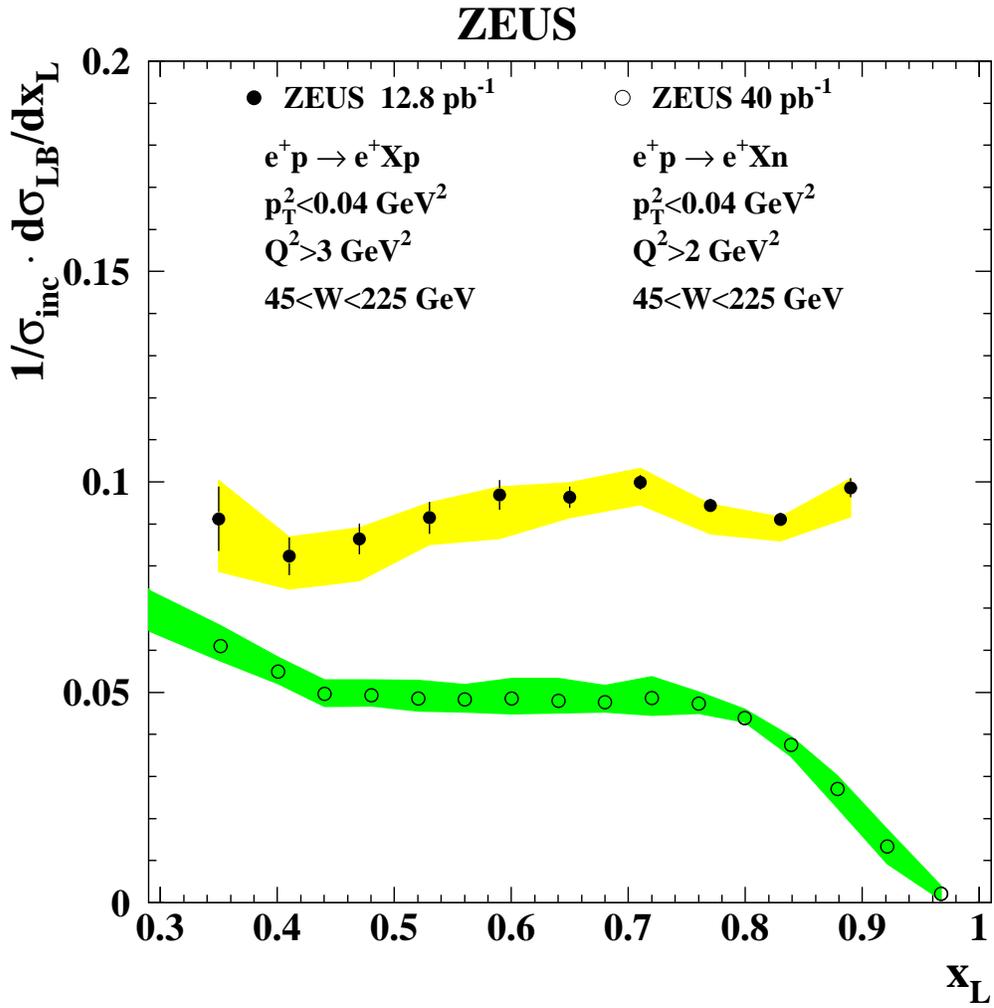}}
\end{center}
\caption{{\protect{
The  rate
$1/\sigma_{\rm inc}\cdot d\sigma_{\rm LB}/dx_L$ for leading proton
(dots) and leading neutron production (circles). The bands show the 
systematic uncertainties.
}}}
\label{fig:xlln}
\end{figure}

The rate of leading proton production for $p_T^2<$0.04 GeV$^2$ 
can be compared to the recent ZEUS measurement of leading neutrons~\cite{ln}.
The comparison is shown in Fig.~\ref{fig:xlln}. 
In the range $0.32<x_L<0.92$, there are approximately twice as many
protons as neutrons. This is consistent with the additive quark
model~\cite{addquark}, in which the probabilities to have a proton or
a neutron in the final state are 2/3 and 1/3, respectively. In a
particle exchange model, the exchange of isovector particles would
result in half as many protons as neutrons. Thus, exchange of
isoscalars must be invoked to account for the observed proton
rate~\cite{SNS}. This contribution is likely also to explain the
different behaviour of the rates at large $x_L$.

\begin{figure}[htb]
\begin{center}
\leavevmode
\hbox{%
\epsfxsize = 14cm
\epsffile{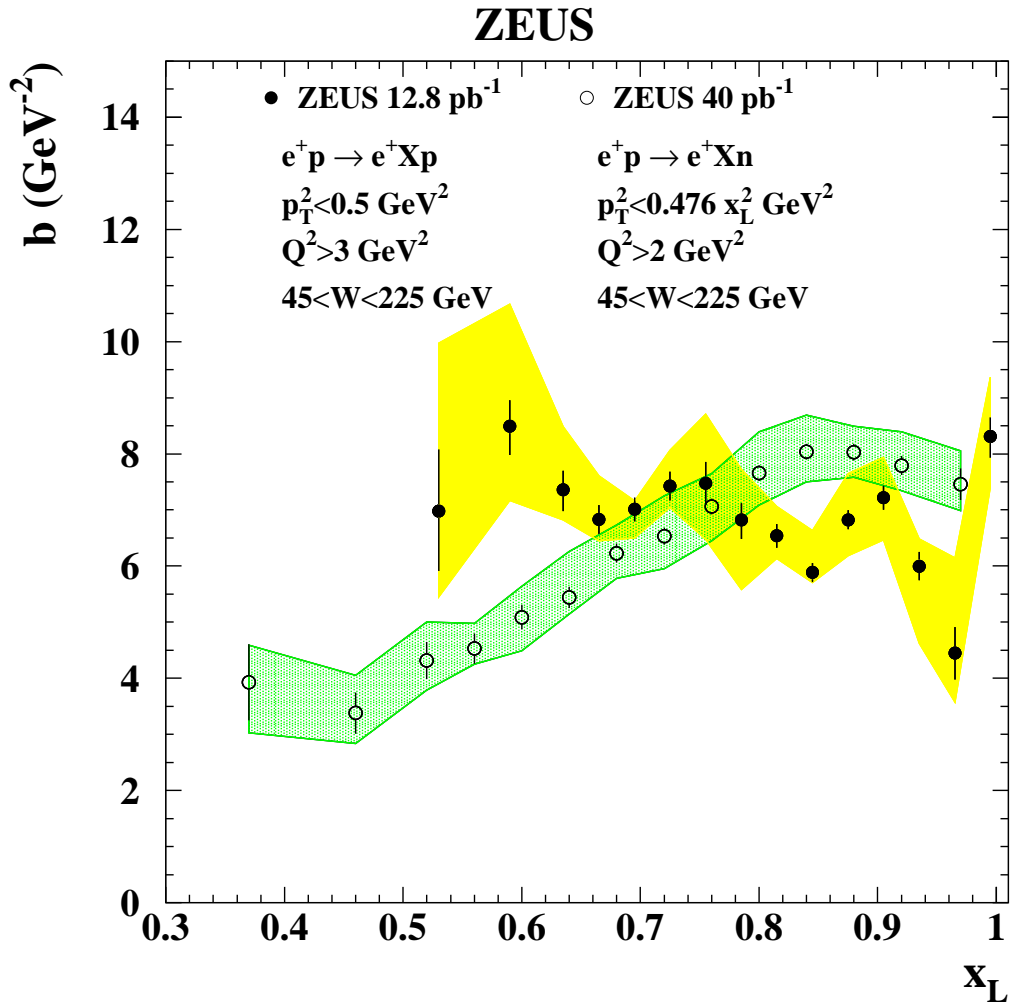}}
\end{center}
\caption{{\protect{
The $p_T^2$-slope, $b$, as a function of $x_L$ for
leading proton (dots) and leading neutron (circles) production. 
The bands show the systematic uncertainties.
}}}
\label{fig:slopesln}
\end{figure}

The slopes of the $p_T^2$ distributions for leading protons and
neutrons are shown in Fig.~\ref{fig:slopesln}. Although the $p_T^2$
and $Q^2$ ranges are different, the two
samples have similar values of $b$ in the region $0.65<x_L<0.8$, where pion
exchange is expected to dominate the production mechanism~\cite{SNS}.

\subsection{Comparison to models}

\begin{figure}[htb]
\begin{center}
\leavevmode
\hbox{%
\epsfxsize = 14cm
\epsffile{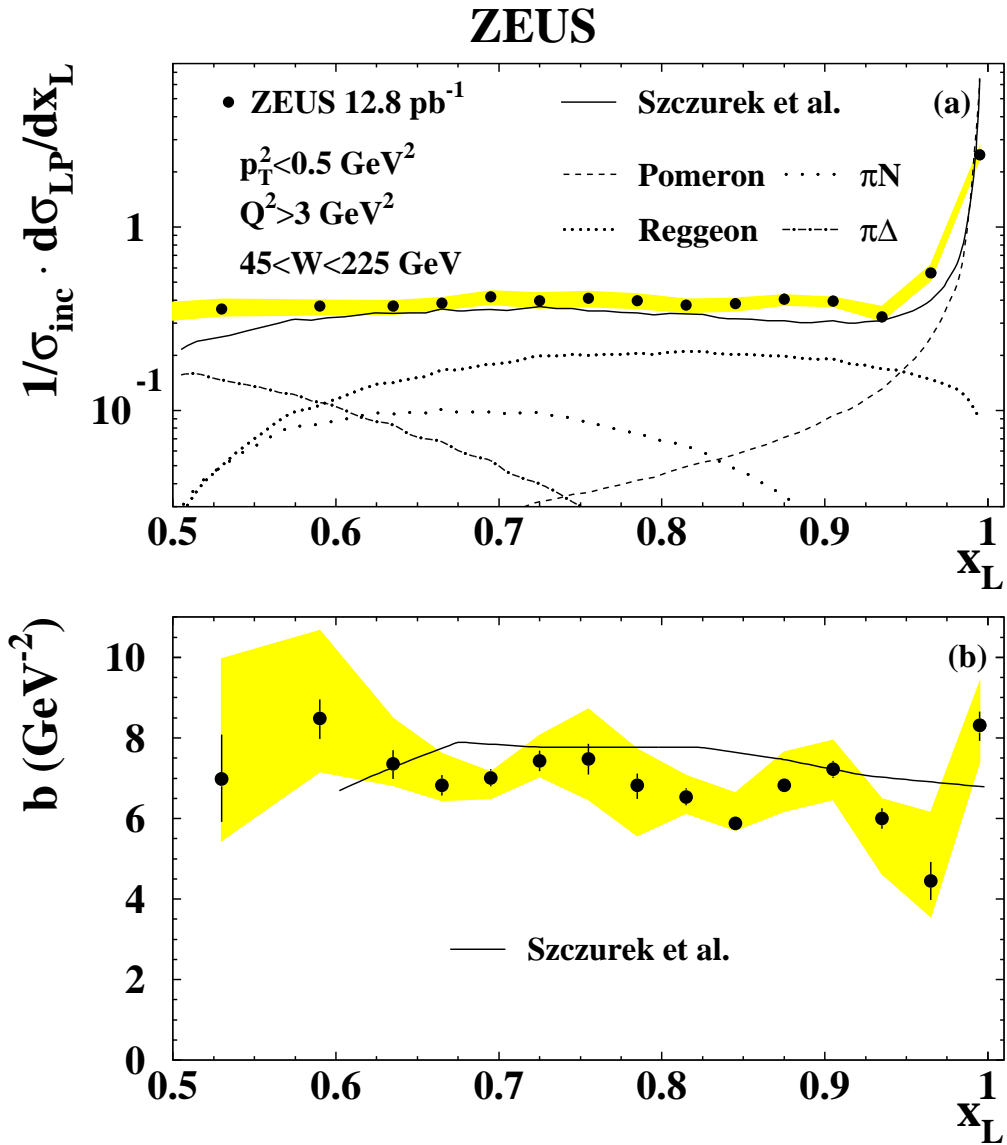}}
\end{center}
\caption{{\protect{
A Regge-based model~\cite{SNS} compared to (a) the
measured leading proton production rate, $1/\sigma_{\rm inc}\cdot
d\sigma_{\rm LP}/dx_L$, and (b) the $p_T^2$-slope, $b$.
The bands show the systematic uncertainties.
}}}
\label{fig:regge}
\end{figure}

The predictions of the model of Szczurek et 
al.~\cite{SNS} are compared to the leading proton production rate
$1/\sigma_{\rm inc}\cdot d\sigma_{\rm LP}/dx_L$ and the $p_T^2$ slopes
in Fig.~\ref{fig:regge}.
In this model, leading proton production for $0.6<x_L<0.9$ is dominated by
isoscalar Reggeon exchange. Diffractive processes due to Pomeron
exchange become increasingly important as $x_L$ approaches unity. The
contribution of pion exchange plays an important role in the medium $x_L$ 
range. The model describes  the shape of the longitudinal momentum
spectrum and of the $p_T^2$ slopes reasonably well. 
The normalisation of the Reggeon contribution, which has a
large theoretical uncertainty~\cite{SNS}, may be constrained by this 
measurement. 
The model does not include absorptive corrections and rescattering 
effects~\cite{abs}, since they are expected to be small in DIS regime.

\begin{figure}[htb]
\begin{center}
\leavevmode
\hbox{%
\epsfxsize = 14cm
\epsffile{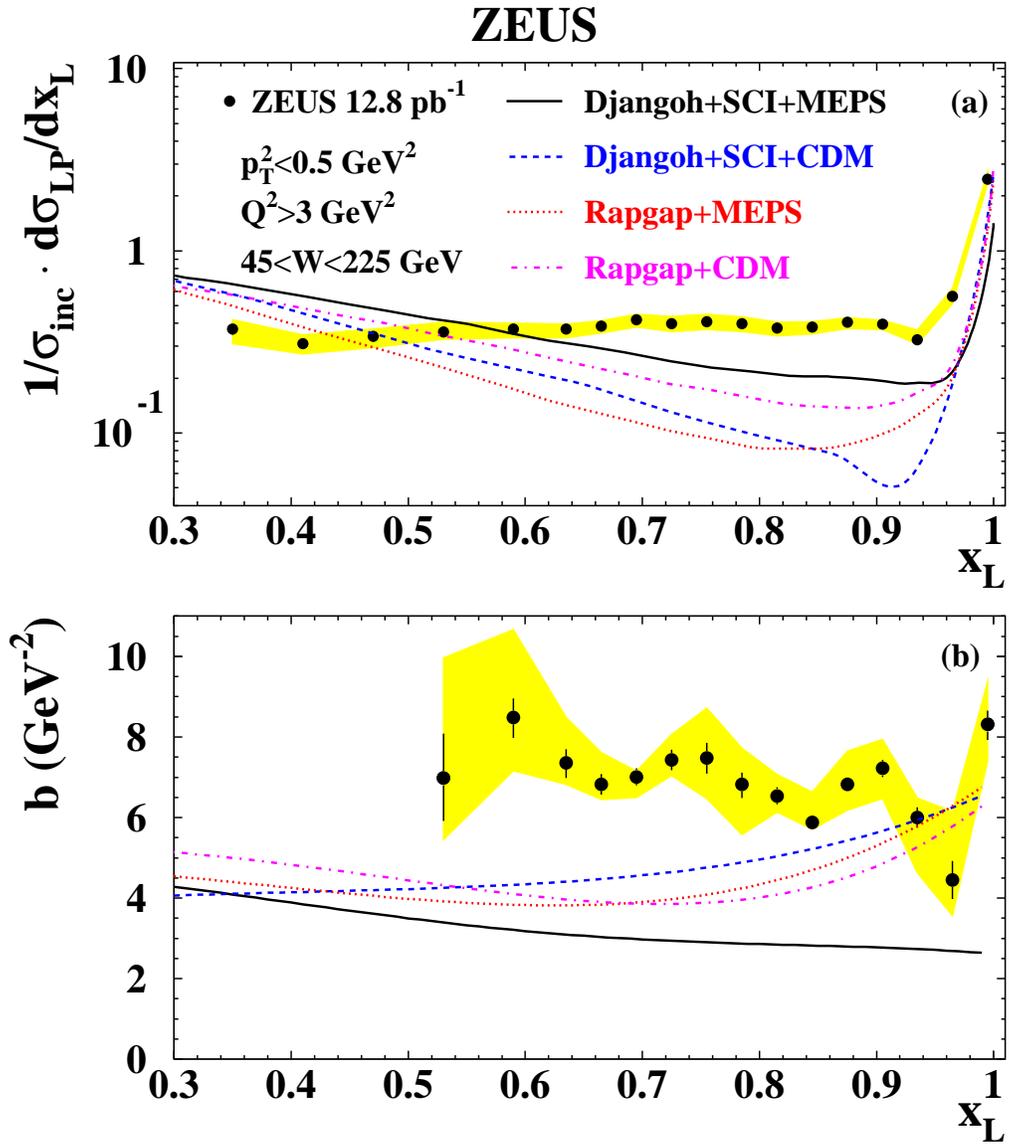}}
\end{center}
\caption{{\protect{  
Expectations of various Monte Carlo models of DIS, as described in the
figure, compared to (a) the
leading proton production rate, $1/\sigma_{\rm inc}\cdot d\sigma_{\rm
LP}/dx_L$, and (b) the $p_T^2$-slope, $b$. The bands show the systematic
uncertainties. 
}}}
\label{fig:mc}
\end{figure}

In Fig.~\ref{fig:mc}, various DIS Monte Carlo models are compared to the 
experimental data. The prediction of {\sc Djangoh}~\cite{django} with
SCI~\cite{lepto} and {\sc Rapgap}~\cite{rapgap}  are compared to the
leading proton production rate $1/\sigma_{\rm inc}\cdot
d\sigma_{\rm LP}/dx_L$ and to the $p_T^2$ slopes. In both MC 
models, the QCD radiation was  performed either by the parton
shower~\cite{lepto} or colour dipole (CDM)~\cite{ariadne} models.  
None of the DIS Monte Carlo models can reproduce the flat dependence
of $x_L$ below the diffractive peak.
The MC generator {\sc Djangoh}, with SCI and MEPS, reproduces quite well 
the dependence of $b$ on $x_L$, although the mean values of
the slope are lower than those measured. In the other MC
models, the value of the slope is consistent with the measurements
only at high $x_L$.

\section{Summary}

The cross section of leading proton production for $x_L>0.32$ and
$p_T^2<0.5$ GeV$^2$ and its ratio to the
inclusive DIS cross section have been measured in the range $Q^2>3$
GeV$^2$ and $45<W<225$ GeV, with 12.8 pb$^{-1}$ collected during 1997
with the ZEUS leading proton spectrometer.
The leading proton production cross section as a function of $p_T^2$
falls exponentially with a mean slope $\langle b \rangle = 6.76 \pm
0.07({\rm stat.})^{+0.63}_{-0.52} ({\rm syst.})$ GeV$^{-2}$, 
approximately independent of $x_L$ and of the photon virtuality, $Q^2$.
Below the diffractive peak, the $x_L$ distribution is  flat, independent 
of $p_T^2$ and $Q^2$.

The ratios of leading proton production to the inclusive DIS yields,
$r^{\rm LP(2)}$ and $r^{\rm LP(3)}$, show no strong dependence on 
$x$ or $Q^2$. In the range $0.32<x_L<0.92$ and $p_T^2<0.5$ GeV$^2$,
approximately 24\%  of DIS events have a leading proton.
The ratio $\langle r^{\rm LP(2)} \rangle$ averaged over $x$ 
rises very slowly with $Q^2$ in the DIS regime. 
This trend is further confirmed by measurements at lower $Q^2$.
The dependence of the leading proton structure-function $F_2^{\rm LP(2)}$
on $x$ and $Q^2$ is approximately the same as that of $F_2$.

The yield of leading protons in DIS is almost twice that of
leading neutrons. The $p_T^2$ slopes have a different dependence
on $x_L$, but are compatible in the range $0.65<x_L<0.8$, where
pion exchange is dominant.

The main features of the experimental data are reproduced by 
a Regge-inspired model. The results of this study are an important 
ingredient for modelling leading particle production in 
hadron-hadron interactions, which is not properly reproduced by existing 
generators.

\section*{Acknowledgments}

We thank the DESY Directorate for their encouragement, and
gratefully acknowledge
the support of the DESY computing and network services. We are specially
grateful to the HERA machine group: collaboration with them was crucial for
the successful installation and operation of the leading proton
spectrometer.
The design, construction  and installation of the ZEUS
detector have been made possible by the ingenuity and effort of many
people
from DESY and home institutes who are not listed as authors. Finally,
it is a pleasure to thank F. A. Ceccopieri and L. Trentadue
for many useful discussions.

\vfill\eject


\begin{small}


\begin{center}

\tablefirsthead{
\hline \hline 
$x_L$ range & $p_T^2$ range  & $d^2\sigma_{\rm LP}/dx_Ldp_T^2$  &
$d^2\sigma_{\rm LP}/dx_Ldp_T^2$  & $d^2\sigma_{\rm LP}/dx_Ldp_T^2$   \\
            &    (GeV$^2$)    & s123 (nb/GeV$^2$)       &  s456 (nb/GeV$^2$)   & Combined (nb/GeV$^2$) \\
\hline \hline 
}

\tablehead{
\multicolumn{4}{l}{\tab{xlpt2} (cont.)} \\
\hline  \hline
$x_L$ range & $p_T^2$ range  & $d^2\sigma_{\rm LP}/dx_Ldp_T^2$  &
$d^2\sigma_{\rm LP}/dx_Ldp_T^2$  & $d^2\sigma_{\rm LP}/dx_Ldp_T^2$   \\
            &      (GeV$^2$)    & s123 (nb/GeV$^2$)       &  s456 (nb/GeV$^2$)   & Combined (nb/GeV$^2$) \\
\hline \hline
}

\tabletail{\hline}

\bottomcaption{
The double-differential cross-section $ d^2 \sigma_{\rm LP}/dx_L dp_T^2$ as a function of $x_L$ and $p_T^2$,
separately measured with the spectrometers s123 and s456 and the combined result. 
Statistical uncertainties are listed first, followed by systematic uncertainties.
}

\begin{supertabular}{|c|c|c|c|c|}

0.32-0.38 &0.00-0.05& $  467\pm  40^{+  49}_{-  53}$& $                             $& $  467\pm  40^{+  49}_{-  53}$\\
          &0.05-0.10& $  387\pm  68^{+  45}_{- 161}$& $                             $& $  387\pm  68^{+  45}_{- 161}$\\
 \hline
0.38-0.44 &0.00-0.05& $  367\pm  28^{+  32}_{-  22}$& $  502\pm  41^{+  42}_{-  83}$& $  409\pm  23^{+  28}_{-  31}$\\
          &0.05-0.10& $  349\pm  46^{+  35}_{-  57}$& $                             $& $  349\pm  46^{+  35}_{-  57}$\\
 \hline
0.44-0.50 &0.00-0.05& $  362\pm  25^{+  18}_{-  40}$& $  469\pm  25^{+  46}_{-  49}$& $  415\pm  18^{+  20}_{-  36}$\\
          &0.05-0.10& $  332\pm  40^{+  19}_{-  38}$& $                             $& $  332\pm  40^{+  19}_{-  38}$\\
          &0.10-0.15& $  165\pm  44^{+  22}_{-  51}$& $                             $& $  165\pm  44^{+  22}_{-  51}$\\
 \hline
0.50-0.56&0.00-0.05& $  403\pm  26^{+  21}_{-  26}$& $  463\pm  22^{+ 128}_{-  14}$& $  437\pm  17^{+  44}_{-  12}$\\
        &0.05-0.10& $  317\pm  33^{+  51}_{-  17}$& $  298\pm  19^{+  52}_{-  16}$& $  303\pm  17^{+  41}_{-  12}$\\
        &0.10-0.15& $  219\pm  47^{+  40}_{-  13}$& $                             $& $  219\pm  47^{+  40}_{-  13}$\\
 \hline
0.56-0.62&0.00-0.05& $  460\pm  33^{+  22}_{-  70}$& $  463\pm  19^{+  30}_{-  26}$& $  462\pm  16^{+  20}_{-  31}$\\
        &0.05-0.10& $  303\pm  27^{+  29}_{-  25}$& $  306\pm  16^{+  16}_{-  30}$& $  305\pm  13^{+  16}_{-  23}$\\
        &0.10-0.15& $  178\pm  35^{+  42}_{-  14}$& $  210\pm  17^{+   8}_{-  46}$& $  204\pm  15^{+   7}_{-  36}$\\
        &0.15-0.20& $                             $& $  136\pm  22^{+  40}_{-  11}$& $  136\pm  22^{+  40}_{-  11}$\\
 \hline
0.62-0.65&0.00-0.05& $  410\pm  53^{+  28}_{-  56}$& $  463\pm  20^{+  19}_{-  17}$& $  456\pm  19^{+  17}_{-  19}$\\
        &0.05-0.10& $  296\pm  29^{+  18}_{-  12}$& $  324\pm  31^{+  16}_{-  51}$& $  309\pm  21^{+  12}_{-  22}$\\
        &0.10-0.15& $  189\pm  43^{+  19}_{-  51}$& $  187\pm  16^{+  13}_{-  16}$& $  187\pm  15^{+  12}_{-  18}$\\
        &0.15-0.20& $                             $& $  148\pm  17^{+   8}_{-  27}$& $  148\pm  17^{+   8}_{-  27}$\\
        &0.20-0.25& $                             $& $  134\pm  22^{+   8}_{-  42}$& $  134\pm  22^{+   8}_{-  42}$\\
 \hline
0.65-0.68&0.00-0.05& $  421\pm  74^{+  64}_{-  71}$& $  464\pm  16^{+  27}_{-  16}$& $  462\pm  16^{+  27}_{-  17}$\\
        &0.05-0.10& $  320\pm  29^{+  17}_{-  43}$& $  297\pm  33^{+  24}_{-  66}$& $  310\pm  22^{+  15}_{-  52}$\\
        &0.10-0.15& $  181\pm  36^{+  81}_{-  19}$& $  298\pm  37^{+  20}_{- 131}$& $  238\pm  26^{+  11}_{-  28}$\\
        &0.15-0.20& $                             $& $  118\pm  13^{+  42}_{-   6}$& $  118\pm  13^{+  42}_{-   6}$\\
        &0.20-0.25& $                             $& $  119\pm  16^{+  18}_{-   6}$& $  119\pm  16^{+  18}_{-   6}$\\
        &0.25-0.35& $                             $& $   87\pm  12^{+   7}_{-  15}$& $   87\pm  12^{+   7}_{-  15}$\\

\hline 

\end{supertabular}
\label{tab-xlpt2}
\end{center}

\clearpage

\begin{center}

\tablefirsthead{
\multicolumn{4}{l}{\tab{xlpt2} (cont.)} \\
\hline 
$x_L$ range & $p_T^2$ range  & $d^2\sigma_{\rm LP}/dx_Ldp_T^2$  & $d^2\sigma_{\rm LP}/dx_Ldp_T^2$  & $d^2\sigma_{\rm LP}/dx_Ldp_T^2$   \\
            &      (GeV$^2$)   & s123 (nb/GeV$^2$)       &  s456 (nb/GeV$^2$)   & Combined (nb/GeV$^2$) \\
\hline \hline 
}

\tablehead{
\multicolumn{4}{l}{\tab{xlpt2} (cont.)} \\
\hline  
$x_L$ range & $p_T^2$ range  & $d^2\sigma_{\rm LP}/dx_Ldp_T^2$  & $d^2\sigma_{\rm LP}/dx_Ldp_T^2$  & $d^2\sigma_{\rm LP}/dx_Ldp_T^2$   \\
            &       (GeV$^2$)    & s123 (nb/GeV$^2$)       &  s456 (nb/GeV$^2$)   & Combined (nb/GeV$^2$) \\
\hline \hline
}

\tabletail{\hline}

\begin{supertabular}{|c|c|c|c|c|}

0.68-0.71&0.00-0.05& $  253\pm  85^{+  27}_{- 134}$& $  500\pm  15^{+  17}_{-  17}$& $  493\pm  15^{+  16}_{-  28}$\\
        &0.05-0.10& $  320\pm  29^{+  22}_{-  42}$& $  366\pm  41^{+  24}_{- 138}$& $  335\pm  24^{+  17}_{-  74}$\\
        &0.10-0.15& $  184\pm  31^{+  57}_{-  17}$& $  259\pm  47^{+  39}_{-  55}$& $  207\pm  26^{+  47}_{-  15}$\\
        &0.15-0.20& $   92\pm  41^{+  40}_{-  74}$& $  155\pm  24^{+  40}_{-  11}$& $  139\pm  21^{+  33}_{-  45}$\\
        &0.20-0.25& $                             $& $  103\pm  16^{+  32}_{-   7}$& $  103\pm  16^{+  32}_{-   7}$\\
        &0.25-0.35& $                             $& $   85\pm  10^{+   6}_{-   4}$& $   85\pm  10^{+   6}_{-   4}$\\
        &0.35-0.50& $                             $& $   44\pm   9^{+   4}_{-   9}$& $   44\pm   9^{+   4}_{-   9}$\\
 \hline
0.71-0.74&0.00-0.05& $                             $& $  472\pm  12^{+  28}_{-  13}$& $  472\pm  12^{+  28}_{-  13}$\\
        &0.05-0.10& $  279\pm  27^{+  59}_{-  16}$& $  255\pm  24^{+  51}_{-  42}$& $  266\pm  18^{+  49}_{-  24}$\\
        &0.10-0.15& $  217\pm  33^{+  24}_{-  38}$& $  130\pm  33^{+  85}_{-  16}$& $  173\pm  23^{+  52}_{-  15}$\\
        &0.15-0.20& $  114\pm  43^{+  34}_{-  58}$& $  250\pm  66^{+  51}_{- 133}$& $  154\pm  36^{+  25}_{-  69}$\\
        &0.20-0.25& $                            $& $  116\pm  27^{+  55}_{-  15}$& $   62\pm  19^{+  55}_{-  15}$\\
        &0.25-0.35& $                             $& $   86\pm  16^{+  11}_{-  25}$& $   86\pm  16^{+  11}_{-  25}$\\
        &0.35-0.50& $                             $& $   27\pm   5^{+   2}_{-   2}$& $   27\pm   5^{+   2}_{-   2}$\\
 \hline
0.74-0.77&0.00-0.05& $                             $& $  467\pm  10^{+   9}_{-  28}$& $  467\pm  10^{+   9}_{-  28}$\\
        &0.05-0.10& $  250\pm  30^{+  51}_{-  32}$& $  325\pm  21^{+  42}_{-  14}$& $  300\pm  17^{+  42}_{-  15}$\\
        &0.10-0.15& $  234\pm  30^{+  53}_{-   6}$& $  199\pm  40^{+  33}_{-  19}$& $  222\pm  24^{+  40}_{-   7}$\\
        &0.15-0.20& $  108\pm  35^{+  38}_{-  31}$& $                             $& $  108\pm  35^{+  38}_{-  31}$\\
        &0.20-0.25& $                             $& $  152\pm  52^{+  18}_{-  84}$& $  152\pm  52^{+  18}_{-  84}$\\
        &0.25-0.35& $                             $& $  121\pm  34^{+   9}_{-  80}$& $  121\pm  34^{+   9}_{-  80}$\\
        &0.35-0.50& $                             $& $    8\pm   4^{+  11}_{-   1}$& $    8\pm   4^{+  11}_{-   1}$\\
 \hline
0.77-0.80&0.00-0.05& $                             $& $  452\pm  10^{+  10}_{-  17}$& $  452\pm  10^{+  10}_{-  17}$\\
        &0.05-0.10& $  248\pm  39^{+  30}_{-  27}$& $  297\pm  14^{+  13}_{-  38}$& $  292\pm  13^{+  13}_{-  35}$\\
        &0.10-0.15& $  196\pm  23^{+  97}_{-   7}$& $  241\pm  20^{+  50}_{-   9}$& $  222\pm  15^{+  62}_{-   5}$\\
        &0.15-0.20& $  198\pm  48^{+  40}_{-  39}$& $  203\pm  32^{+  97}_{-   9}$& $  201\pm  27^{+  56}_{-  13}$\\
        &0.20-0.25& $                             $& $  168\pm  54^{+  58}_{-  46}$& $  168\pm  54^{+  58}_{-  46}$\\
 \hline
0.80-0.83&0.00-0.05& $                             $& $  451\pm  10^{+   9}_{-  19}$& $  451\pm  10^{+   9}_{-  19}$\\
        &0.05-0.10& $  353\pm  74^{+  16}_{- 117}$& $  266\pm  11^{+  18}_{-   8}$& $  268\pm  11^{+  18}_{-   9}$\\
        &0.10-0.15& $  197\pm  25^{+  33}_{-  14}$& $  210\pm  14^{+   7}_{-  16}$& $  207\pm  12^{+   5}_{-  13}$\\
        &0.15-0.20& $   84\pm  23^{+  25}_{-  13}$& $  178\pm  17^{+   5}_{-  75}$& $  147\pm  14^{+   9}_{-  33}$\\
        &0.20-0.25& $   73\pm  32^{+  68}_{-  52}$& $  136\pm  16^{+   9}_{-  33}$& $  123\pm  15^{+  22}_{-  45}$\\
        &0.25-0.35& $                             $& $   98\pm  14^{+  22}_{-   7}$& $   98\pm  14^{+  22}_{-   7}$\\
 \hline
0.83-0.86&0.00-0.05& $                             $& $  434\pm  10^{+  13}_{-  19}$& $  434\pm  10^{+  13}_{-  19}$\\
        &0.05-0.10& $                             $& $  295\pm  12^{+   8}_{-  33}$& $  295\pm  12^{+   8}_{-  33}$\\
        &0.10-0.15& $  184\pm  24^{+  79}_{-  10}$& $  207\pm  13^{+  20}_{-  13}$& $  202\pm  11^{+  17}_{-   5}$\\
        &0.15-0.20& $  190\pm  37^{+  33}_{-  59}$& $  169\pm  15^{+   5}_{-  39}$& $  172\pm  14^{+   5}_{-  39}$\\
        &0.20-0.25& $   50\pm  20^{+  22}_{-  21}$& $  127\pm  14^{+  33}_{-   7}$& $  102\pm  11^{+  30}_{-  19}$\\
        &0.25-0.35& $                             $& $  122\pm  14^{+   7}_{-  36}$& $  122\pm  14^{+   7}_{-  36}$\\
        &0.35-0.50& $                             $& $   49\pm   8^{+  12}_{-   6}$& $   49\pm   8^{+  12}_{-   6}$\\
 \hline
0.86-0.89&0.00-0.05& $                             $& $  467\pm  13^{+  24}_{-  18}$& $  467\pm  13^{+  24}_{-  18}$\\
        &0.05-0.10& $                             $& $  309\pm  12^{+  22}_{-  20}$& $  309\pm  12^{+  22}_{-  20}$\\
        &0.10-0.15& $  247\pm  36^{+  44}_{-  29}$& $  232\pm  14^{+   8}_{-  20}$& $  234\pm  13^{+   7}_{-  19}$\\
        &0.15-0.20& $  196\pm  32^{+  23}_{-  43}$& $  180\pm  15^{+   6}_{-  35}$& $  183\pm  14^{+   6}_{-  36}$\\
        &0.20-0.25& $  115\pm  34^{+  55}_{-  24}$& $  136\pm  14^{+  21}_{-  16}$& $  133\pm  13^{+  20}_{-  16}$\\
        &0.25-0.35& $   24\pm  16^{+   8}_{-   4}$& $   81\pm   8^{+  15}_{-  11}$& $   69\pm   7^{+  24}_{-   6}$\\
        &0.35-0.50& $                             $& $   43\pm   6^{+   8}_{-   5}$& $   43\pm   6^{+   8}_{-   5}$\\
 \hline
0.89-0.92&0.00-0.05& $                             $& $  481\pm  19^{+  24}_{-  19}$& $  481\pm  19^{+  24}_{-  19}$\\
        &0.05-0.10& $                             $& $  315\pm  13^{+  16}_{-  10}$& $  315\pm  13^{+  16}_{-  10}$\\
        &0.10-0.15& $  262\pm  47^{+  68}_{-  78}$& $  228\pm  14^{+  18}_{-  21}$& $  231\pm  13^{+  15}_{-  24}$\\
        &0.15-0.20& $  172\pm  27^{+  28}_{-  44}$& $  175\pm  14^{+  22}_{-  10}$& $  174\pm  12^{+  14}_{-  11}$\\
        &0.20-0.25& $   97\pm  27^{+  58}_{-  18}$& $  133\pm  13^{+  18}_{-  19}$& $  126\pm  12^{+  15}_{-  11}$\\
        &0.25-0.35& $   37\pm  19^{+  18}_{-  11}$& $   69\pm   7^{+   7}_{-   5}$& $   65\pm   6^{+   6}_{-   5}$\\
        &0.35-0.50& $                             $& $   39\pm   5^{+   6}_{-   5}$& $   39\pm   5^{+   6}_{-   5}$\\
 \hline
0.92-0.95&0.00-0.05& $                             $& $  820\pm 145^{+ 124}_{- 432}$& $  820\pm 145^{+ 124}_{- 432}$\\
        &0.05-0.10& $                             $& $  235\pm  14^{+  54}_{-   8}$& $  235\pm  14^{+  54}_{-   8}$\\
        &0.10-0.15& $  221\pm  58^{+ 212}_{-  36}$& $  182\pm  14^{+  48}_{-  22}$& $  184\pm  13^{+  45}_{-  21}$\\
        &0.15-0.20& $  138\pm  21^{+  47}_{-  11}$& $  156\pm  14^{+  33}_{-  13}$& $  150\pm  12^{+  30}_{-   9}$\\
        &0.20-0.25& $   92\pm  22^{+  38}_{-  10}$& $  114\pm  13^{+  17}_{-  20}$& $  109\pm  11^{+  15}_{-  11}$\\
        &0.25-0.35& $   39\pm  15^{+   9}_{-  18}$& $   83\pm   9^{+  16}_{-  12}$& $   72\pm   8^{+   6}_{-  13}$\\
        &0.35-0.50& $                             $& $   36\pm   5^{+  16}_{-   4}$& $   36\pm   5^{+  16}_{-   4}$\\
 \hline
0.95-0.98&0.05-0.10& $                             $& $  423\pm  35^{+  63}_{- 116}$& $  423\pm  35^{+  63}_{- 116}$\\
        &0.10-0.15& $                             $& $  328\pm  30^{+  84}_{-  10}$& $  328\pm  30^{+  84}_{-  10}$\\
        &0.15-0.20& $  280\pm  39^{+  57}_{-  16}$& $  250\pm  32^{+ 142}_{-   5}$& $  262\pm  25^{+  90}_{-   1}$\\
        &0.20-0.25& $  254\pm  46^{+   5}_{- 116}$& $  197\pm  32^{+  60}_{-   6}$& $  215\pm  26^{+  30}_{-  26}$\\
        &0.25-0.35& $   50\pm  17^{+  18}_{-  30}$& $  190\pm  34^{+  60}_{-  72}$& $   77\pm  15^{+  19}_{-  44}$\\
 \hline
0.98-1.00&0.05-0.10& $                             $& $ 2788\pm 180^{+ 466}_{- 304}$& $ 2788\pm 180^{+ 466}_{- 304}$\\
        &0.10-0.15& $                             $& $ 1423\pm  90^{+ 188}_{- 131}$& $ 1423\pm  90^{+ 188}_{- 131}$\\
        &0.15-0.20& $ 1218\pm 160^{+ 182}_{- 214}$& $ 1012\pm  89^{+ 132}_{- 133}$& $ 1061\pm  78^{+  85}_{- 118}$\\
        &0.20-0.25& $  574\pm  81^{+ 166}_{-  35}$& $  848\pm  93^{+ 186}_{-  41}$& $  693\pm  61^{+ 174}_{-  16}$\\
        &0.25-0.35& $  231\pm  54^{+ 116}_{-  13}$& $  588\pm  69^{+  39}_{-  93}$& $  367\pm  42^{+ 147}_{-  26}$\\
 \hline \hline

\end{supertabular}
\end{center}

\clearpage


\begin{table}
\begin{center}
\begin{tabular}{|c|c|}
\hline \hline
$x_L$ range &  $b$ (GeV$^{-2}$)  \\
\hline \hline
0.50 - 0.56 & $ 6.98^{+ 1.11}_{- 1.06}$$^{+ 2.99}_{- 1.53}$ \\
0.56 - 0.62 & $ 8.49^{+ 0.47}_{- 0.51}$$^{+ 2.18}_{- 1.32}$ \\
0.62 - 0.65 & $ 7.36^{+ 0.34}_{- 0.38}$$^{+ 1.12}_{- 0.54}$ \\
0.65 - 0.68 & $ 6.83^{+ 0.26}_{- 0.26}$$^{+ 0.78}_{- 0.38}$ \\
0.68 - 0.71 & $ 7.01^{+ 0.21}_{- 0.21}$$^{+ 0.15}_{- 0.51}$ \\
0.71 - 0.74 & $ 7.43^{+ 0.26}_{- 0.26}$$^{+ 0.63}_{- 0.38}$ \\
0.74 - 0.77 & $ 7.48^{+ 0.38}_{- 0.38}$$^{+ 1.23}_{- 1.00}$ \\
0.77 - 0.80 & $ 6.82^{+ 0.30}_{- 0.34}$$^{+ 0.90}_{- 1.24}$ \\
0.80 - 0.83 & $ 6.54^{+ 0.21}_{- 0.21}$$^{+ 0.52}_{- 0.40}$ \\
0.83 - 0.86 & $ 5.88^{+ 0.17}_{- 0.17}$$^{+ 0.75}_{- 0.18}$ \\
0.86 - 0.89 & $ 6.82^{+ 0.17}_{- 0.17}$$^{+ 0.82}_{- 0.64}$ \\
0.89 - 0.92 & $ 7.22^{+ 0.21}_{- 0.21}$$^{+ 0.72}_{- 0.75}$ \\
0.92 - 0.95 & $ 6.00^{+ 0.26}_{- 0.26}$$^{+ 0.48}_{- 1.37}$ \\
0.95 - 0.98 & $ 4.45^{+ 0.47}_{- 0.47}$$^{+ 1.69}_{- 0.88}$ \\
0.98 - 1.00 & $ 8.31^{+ 0.34}_{- 0.38}$$^{+ 1.05}_{- 0.91}$ \\
\hline \hline
\end{tabular}
\caption{
The $p_T^2$-slope, $b$, of the cross-section
$d^2\sigma_{\rm LP}/dx_Ldp_T^2$, as defined by the parameterisation $A
\cdot e^{-b\cdot p_T^2}$ and obtained from a fit to the data in bins of
$x_L$. Statistical uncertainties are listed first, 
followed by systematic uncertainties.
}
\label{tab-slopes}
\end{center}
\end{table}


\begin{table}
\begin{center}
\begin{tabular}{|c|c|c|}
\hline \hline
$x_L$ range & \multicolumn{2}{c|}{$b$ (GeV$^{-2}$)}  \\
\hline
              & $\langle Q^2 \rangle=5.1$ GeV$^2$ & $\langle Q^2 \rangle=30.6$ GeV$^2$ \\
\hline \hline
0.5 - 0.65 & $ 6.90^{+ 0.38}_{- 0.43}$$^{+ 1.34}_{- 0.18}$ & $7.97^{+ 0.43}_{- 0.38}$$^{+ 1.06}_{- 0.39}$ \\
0.65 - 0.8 & $ 7.00^{+ 0.17}_{- 0.17}$$^{+ 0.20}_{- 0.21}$ & $ 7.13^{+0.17}_{- 0.17}$$^{+ 0.16}_{- 0.31}$ \\
0.8 - 0.92 & $ 7.29^{+ 0.13}_{- 0.17}$$^{+ 0.15}_{- 0.49}$ & $ 6.10^{+0.13}_{- 0.13}$$^{+ 0.64}_{- 0.04}$ \\
0.92 - 1.0 & $ 6.74^{+ 0.26}_{- 0.30}$$^{+ 0.28}_{- 0.33}$ & $6.06^{+0.30}_{- 0.30}$$^{+ 0.66}_{- 0.36}$ \\ 

\hline \hline
\end{tabular}
\caption{
The $p_T^2$-slope, $b$, of the cross-section
$d^2\sigma_{\rm LP}/dx_Ldp_T^2$, as defined by the parameterisation $A
\cdot e^{-b\cdot p_T^2}$ and obtained from a fit to the data in bins of
$x_L$, measured in two ranges of $Q^2$. Statistical uncertainties are 
listed first, followed by systematic uncertainties.
}
\label{tab-slopesq2}
\end{center}
\end{table}


\begin{table}
\begin{center}
\begin{tabular}{|c|c|c|c|}
\hline \hline
$x_L$ range & \multicolumn{3}{c|}{$1/\sigma_{\rm inc}\cdot d\sigma_{\rm LP}/dx_L$}  \\
\hline 
             & $0<p_T^2<0.04$ GeV$^2$ &  $0.04<p_T^2<0.15$ GeV$^2$ & $0.15<p_T^2<0.5$ GeV$^2$ \\
\hline \hline
0.32 - 0.38 & $0.091\pm0.008^{+0.009}_{-0.013}$ &                        &                        \\
0.38 - 0.44 & $0.082\pm0.004^{+0.005}_{-0.008}$ & $0.156\pm0.016^{+0.013}_{-0.024}$ &                        \\
0.44 - 0.50 & $0.086\pm0.004^{+0.003}_{-0.010}$ & $0.144\pm0.013^{+0.005}_{-0.021}$ &                        \\
0.50 - 0.56 & $0.091\pm0.004^{+0.004}_{-0.006}$ & $0.140\pm0.006^{+0.026}_{-0.008}$ &                        \\
0.56 - 0.62 & $0.097\pm0.004^{+0.002}_{-0.010}$ & $0.143\pm0.004^{+0.004}_{-0.014}$ &                        \\
0.62 - 0.68 & $0.096\pm0.003^{+0.003}_{-0.005}$ & $0.142\pm0.005^{+0.002}_{-0.021}$ & $0.126\pm0.007^{+0.010}_{-0.011}$ \\
0.68 - 0.74 & $0.100\pm0.002^{+0.003}_{-0.005}$ & $0.141\pm0.005^{+0.004}_{-0.012}$ & $0.138\pm0.008^{+0.008}_{-0.009}$ \\
0.74 - 0.80 & $0.094\pm0.001^{+0.000}_{-0.007}$ & $0.145\pm0.003^{+0.003}_{-0.006}$ & $0.155\pm0.015^{+0.010}_{-0.015}$ \\
0.80 - 0.86 & $0.091\pm0.001^{+0.001}_{-0.005}$ & $0.139\pm0.003^{+0.001}_{-0.008}$ & $0.146\pm0.005^{+0.010}_{-0.023}$ \\
0.86 - 0.92 & $0.099\pm0.002^{+0.002}_{-0.007}$ & $0.154\pm0.003^{+0.005}_{-0.007}$ & $0.143\pm0.004^{+0.011}_{-0.014}$ \\
0.92 - 0.98 &                        & $0.167\pm0.006^{+0.020}_{-0.015}$ & $0.179\pm0.007^{+0.021}_{-0.013}$ \\
0.97 - 1.00 &                        & $1.126\pm0.047^{+0.113}_{-0.092}$ & $0.816\pm0.036^{+0.088}_{-0.055}$ \\
\hline \hline
\end{tabular}
\caption{
The leading proton production rate, $1/\sigma_{\rm inc}\cdot d\sigma_{\rm
LP}/dx_L$, as a function of $x_L$ measured in three ranges of
\ptsq. Statistical uncertainties are listed first, followed by
systematic uncertainties.
}
\label{tab-xl3}
\end{center}
\end{table}


\begin{table}
\begin{center}
\begin{tabular}{|c|c|}
\hline \hline
$x_L$ range & $1/\sigma_{\rm inc}\cdot d\sigma_{\rm LP}/dx_L$  \\
\hline \hline
0.32 - 0.38 & $0.372\pm0.026^{+0.046}_{-0.062}$ \\
0.38 - 0.44 & $0.309\pm0.018^{+0.037}_{-0.037}$ \\
0.44 - 0.50 & $0.339\pm0.012^{+0.033}_{-0.044}$ \\
0.50 - 0.56 & $0.358\pm0.010^{+0.047}_{-0.030}$ \\
0.56 - 0.62 & $0.371\pm0.009^{+0.029}_{-0.039}$ \\
0.62 - 0.65 & $0.371\pm0.011^{+0.025}_{-0.037}$ \\
0.65 - 0.68 & $0.385\pm0.010^{+0.027}_{-0.031}$ \\
0.68 - 0.71 & $0.418\pm0.010^{+0.030}_{-0.037}$ \\
0.71 - 0.74 & $0.398\pm0.009^{+0.039}_{-0.036}$ \\
0.74 - 0.77 & $0.408\pm0.008^{+0.038}_{-0.039}$ \\
0.77 - 0.80 & $0.398\pm0.007^{+0.036}_{-0.037}$ \\
0.80 - 0.83 & $0.376\pm0.006^{+0.029}_{-0.035}$ \\
0.83 - 0.86 & $0.382\pm0.006^{+0.026}_{-0.033}$ \\
0.86 - 0.89 & $0.405\pm0.007^{+0.024}_{-0.030}$ \\
0.89 - 0.92 & $0.395\pm0.008^{+0.019}_{-0.026}$ \\
0.92 - 0.95 & $0.325\pm0.010^{+0.044}_{-0.016}$ \\
0.95 - 0.98 & $0.562\pm0.023^{+0.058}_{-0.049}$ \\
0.98 - 1.00 & $2.478\pm0.076^{+0.235}_{-0.136}$ \\
\hline \hline
\end{tabular}
\caption{
The leading proton production rate, $1/\sigma_{\rm inc}\cdot d\sigma_{\rm
LP}/dx_L$, as a function of $x_L$ measured in the region $p_T^2<0.5$
$GeV^2$. Statistical uncertainties are listed first, followed by systematic
uncertainties.
}
\label{tab-xldis}
\end{center}
\end{table}

\clearpage
   

\begin{center}

\tablefirsthead{
\hline \hline 
$\langle x \rangle$  & $\langle Q^2 \rangle$  (GeV$^2$) &  $x_L$ range & $r^{\rm LP(3)}$ \\ 
\hline \hline 
}

\tablehead{
\multicolumn{4}{l}{\tab{rlp3full} (cont.)} \\
\hline  \hline
$\langle x \rangle$  & $\langle Q^2 \rangle$  (GeV$^2$) &  $x_L$ range  & $r^{\rm LP(3)}$ \\ 
\hline \hline
}

\tabletail{\hline}

\bottomcaption{
The leading proton production rate,  $r^{\rm LP(3)}$, measured as a
function of $x_L$ for protons with $p_T^2<0.5$ $GeV^2$, in bins of 
$x$ and $Q^2$, with averages $\langle x \rangle$  and $\langle Q^2 \rangle$.
Statistical uncertainties are listed first, followed by systematic uncertainties.
}

\begin{supertabular}{|c|c|c|c|}
 $  9.6\cdot 10^{-5}$ &   4.2 & 0.32 - 0.42 & $ 0.365 \pm 0.086 ^{+ 0.066 }_{- 0.168 }$ \\
 			 &    & 0.42 - 0.52 & $ 0.268 \pm 0.040 ^{+ 0.085 }_{- 0.041 }$ \\
			 &    & 0.52 - 0.62 & $ 0.391 \pm 0.041 ^{+ 0.050 }_{- 0.048 }$ \\
			 &    & 0.62 - 0.72 & $ 0.338 \pm 0.026 ^{+ 0.058 }_{- 0.025 }$ \\
			 &    & 0.72 - 0.72 & $ 0.378 \pm 0.024 ^{+ 0.050 }_{- 0.049 }$ \\
			 &    & 0.82 - 0.72 & $ 0.458 \pm 0.027 ^{+ 0.032 }_{- 0.076 }$ \\
\hline
 $  1.7\cdot 10^{-4}$ &   4.2 & 0.32 - 0.42 & $ 0.414 \pm 0.063 ^{+ 0.050 }_{- 0.135 }$ \\
			 &    & 0.42 - 0.52 & $ 0.310 \pm 0.032 ^{+ 0.039 }_{- 0.029 }$ \\
			 &    & 0.52 - 0.62 & $ 0.370 \pm 0.026 ^{+ 0.037 }_{- 0.036 }$ \\
			 &    & 0.62 - 0.72 & $ 0.366 \pm 0.019 ^{+ 0.028 }_{- 0.035 }$ \\
			 &    & 0.72 - 0.82 & $ 0.390 \pm 0.015 ^{+ 0.039 }_{- 0.034 }$ \\
			 &    & 0.82 - 0.92 & $ 0.378 \pm 0.014 ^{+ 0.026 }_{- 0.036 }$ \\
\hline
 $  3.5\cdot 10^{-4}$ &   4.2 & 0.32 - 0.42 & $ 0.351 \pm 0.061 ^{+ 0.047 }_{- 0.062 }$ \\
 		 &  	      & 0.42 - 0.52 & $ 0.269 \pm 0.029 ^{+ 0.055 }_{- 0.028 }$ \\
		 &  	      & 0.52 - 0.62 & $ 0.354 \pm 0.025 ^{+ 0.067 }_{- 0.029 }$ \\
		 &  	      & 0.62 - 0.72 & $ 0.406 \pm 0.022 ^{+ 0.030 }_{- 0.063 }$ \\
		 &  	      & 0.72 - 0.82 & $ 0.405 \pm 0.017 ^{+ 0.040 }_{- 0.038 }$ \\
		 &   	      & 0.82 - 0.92 & $ 0.364 \pm 0.014 ^{+ 0.023 }_{- 0.029 }$ \\

\hline
 $  6.9\cdot 10^{-4}$ &   4.2 	& 0.32 - 0.42 & $ 0.275 \pm 0.057 ^{+ 0.063 }_{- 0.047 }$ \\
		 &  	 	& 0.42 - 0.52 & $ 0.308 \pm 0.035 ^{+ 0.038 }_{- 0.031 }$ \\
		 &   	 	& 0.52 - 0.62 & $ 0.339 \pm 0.027 ^{+ 0.045 }_{- 0.034 }$ \\
		 &   	 	& 0.62 - 0.72 & $ 0.326 \pm 0.020 ^{+ 0.026 }_{- 0.027 }$ \\
		 &   	 	& 0.72 - 0.82 & $ 0.400 \pm 0.018 ^{+ 0.063 }_{- 0.031 }$ \\
		 &   	 	& 0.82 - 0.92 & $ 0.403 \pm 0.017 ^{+ 0.030 }_{- 0.070 }$ \\
\hline
 $  1.46\cdot 10^{-3}$ &   4.2 	& 0.32 - 0.42 & $ 0.272 \pm 0.083 ^{+ 0.040 }_{- 0.082 }$ \\
		 &   	 	& 0.42 - 0.52 & $ 0.268 \pm 0.045 ^{+ 0.066 }_{- 0.035 }$ \\	
		 &   	 	& 0.52 - 0.62 & $ 0.285 \pm 0.038 ^{+ 0.047 }_{- 0.039 }$ \\
 		 &  	 	& 0.62 - 0.72 & $ 0.355 \pm 0.029 ^{+ 0.041 }_{- 0.023 }$ \\
		 &   	 	& 0.72 - 0.82 & $ 0.362 \pm 0.022 ^{+ 0.057 }_{- 0.046 }$ \\
		 &   	 	& 0.82 - 0.92 & $ 0.359 \pm 0.021 ^{+ 0.025 }_{- 0.034 }$ \\

\end{supertabular}
\label{tab-rlp3full}
\end{center}

\clearpage

\begin{center}

\tablefirsthead{
\multicolumn{4}{l}{\tab{rlp3full} (cont.)} \\
\hline 
$\langle x \rangle$  & $\langle Q^2 \rangle$  (GeV$^2$) &  $x_L$ range  & $r^{\rm LP(3)}$ \\ 
\hline \hline 
}

\tablehead{
\multicolumn{4}{l}{\tab{rlp3full} (cont.)} \\
\hline  
$\langle x \rangle$  & $\langle Q^2 \rangle$  (GeV$^2$) &  $x_L$ range  & $r^{\rm LP(3)}$ \\ 
\hline \hline
}

\tabletail{\hline}

\begin{supertabular}{|c|c|c|c|}
   $ 1.9\cdot 10^{-4}$ &   7.3  & 0.32 - 0.42& $ 0.314 \pm 0.084 ^{+ 0.087 }_{- 0.055 }$ \\
		 &   	 	& 0.42 - 0.52& $ 0.430 \pm 0.061 ^{+ 0.065 }_{- 0.122 }$ \\
		 &  	 	& 0.52 - 0.62& $ 0.299 \pm 0.031 ^{+ 0.049 }_{- 0.027 }$ \\
		 &    	 	& 0.62 - 0.72& $ 0.355 \pm 0.027 ^{+ 0.041 }_{- 0.029 }$ \\
		 &   	 	& 0.72 - 0.82& $ 0.382 \pm 0.024 ^{+ 0.039 }_{- 0.042 }$ \\
		 &   	 	& 0.82 - 0.92& $ 0.383 \pm 0.022 ^{+ 0.028 }_{- 0.083 }$ \\
\hline
   $ 3.4\cdot 10^{-4}$ &   7.3  & 0.32 - 0.42& $ 0.340 \pm 0.076 ^{+ 0.062 }_{- 0.075 }$ \\
		 &   	 	& 0.42 - 0.52& $ 0.360 \pm 0.044 ^{+ 0.040 }_{- 0.062 }$ \\
		 &  	 	& 0.52 - 0.62& $ 0.369 \pm 0.032 ^{+ 0.044 }_{- 0.029 }$ \\
		 &   	 	& 0.62 - 0.72& $ 0.347 \pm 0.022 ^{+ 0.030 }_{- 0.038 }$ \\
		 &   		& 0.72 - 0.82& $ 0.374 \pm 0.019 ^{+ 0.037 }_{- 0.033 }$ \\
		 &       	& 0.82 - 0.92& $ 0.359 \pm 0.016 ^{+ 0.037 }_{- 0.020 }$ \\
\hline
   $ 6.9\cdot 10^{-4}$ &   7.3  & 0.32 - 0.42& $ 0.392 \pm 0.081 ^{+ 0.056 }_{- 0.102 }$ \\
		 &  		& 0.42 - 0.52& $ 0.261 \pm 0.038 ^{+ 0.029 }_{- 0.086 }$ \\
		 &   	 	& 0.52 - 0.62& $ 0.372 \pm 0.034 ^{+ 0.048 }_{- 0.028 }$ \\
		 &   	 	& 0.62 - 0.72& $ 0.363 \pm 0.025 ^{+ 0.028 }_{- 0.047 }$ \\
		 &     		& 0.72 - 0.82& $ 0.385 \pm 0.020 ^{+ 0.034 }_{- 0.040 }$ \\
		 &  	 	& 0.82 - 0.92& $ 0.388 \pm 0.019 ^{+ 0.027 }_{- 0.037 }$ \\
\hline
   $ 1.36\cdot 10^{-3}$ &   7.3 & 0.32 - 0.42& $ 0.275 \pm 0.062 ^{+ 0.106 }_{- 0.061 }$ \\
		 &   		& 0.42 - 0.52& $ 0.401 \pm 0.056 ^{+ 0.050 }_{- 0.088 }$ \\
		 &    		& 0.52 - 0.62& $ 0.393 \pm 0.039 ^{+ 0.043 }_{- 0.035 }$ \\
		 &    		& 0.62 - 0.72& $ 0.448 \pm 0.032 ^{+ 0.042 }_{- 0.057 }$ \\
		 &    		& 0.72 - 0.82& $ 0.394 \pm 0.021 ^{+ 0.042 }_{- 0.043 }$ \\
		 &    		& 0.82 - 0.92& $ 0.389 \pm 0.020 ^{+ 0.033 }_{- 0.025 }$ \\
\hline
   $ 2.67\cdot 10^{-3}$ &   7.3 & 0.32 - 0.47 & $ 0.290 \pm 0.097 ^{+ 0.034 }_{- 0.112 }$ \\
		 &    		& 0.47 - 0.62 & $ 0.442 \pm 0.086 ^{+ 0.092 }_{- 0.159 }$ \\
		 &    		& 0.62 - 0.77 & $ 0.449 \pm 0.050 ^{+ 0.094 }_{- 0.046 }$ \\
		 &    		& 0.77 - 0.92 & $ 0.398 \pm 0.033 ^{+ 0.049 }_{- 0.054 }$ \\

\hline
$ 2.6\cdot 10^{-4}$ &  11 	& 0.32 - 0.47 & $ 0.349 \pm 0.078 ^{+ 0.042 }_{- 0.128 }$ \\
		 &   		& 0.47 - 0.62 & $ 0.285 \pm 0.034 ^{+ 0.111 }_{- 0.023 }$ \\
		 &   		& 0.62 - 0.77 & $ 0.485 \pm 0.037 ^{+ 0.041 }_{- 0.074 }$ \\
		 &   		& 0.77 - 0.92 & $ 0.371 \pm 0.021 ^{+ 0.033 }_{- 0.024 }$ \\
\hline
  $ 4.6\cdot 10^{-4}$ &  11 	& 0.32 - 0.42 & $ 0.482 \pm 0.097 ^{+ 0.067 }_{- 0.162 }$ \\
 		 &  	 	& 0.42 - 0.52 & $ 0.285 \pm 0.036 ^{+ 0.037 }_{- 0.039 }$ \\
 		 &  	 	& 0.52 - 0.62 & $ 0.370 \pm 0.031 ^{+ 0.064 }_{- 0.033 }$ \\
		  &   		& 0.62 - 0.72 & $ 0.435 \pm 0.028 ^{+ 0.034 }_{- 0.054 }$ \\
		  &   		& 0.72 - 0.82 & $ 0.386 \pm 0.019 ^{+ 0.051 }_{- 0.030 }$ \\
		  & 	 	& 0.82 - 0.92 & $ 0.365 \pm 0.016 ^{+ 0.049 }_{- 0.018 }$ \\
\hline
   $ 9.2\cdot 10^{-4}$ &  11 	& 0.32 - 0.42 & $ 0.418 \pm 0.084 ^{+ 0.056 }_{- 0.093 }$ \\
		 &   		& 0.42 - 0.52 & $ 0.286 \pm 0.037 ^{+ 0.064 }_{- 0.028 }$ \\
		 & 	  	& 0.52 - 0.62 & $ 0.408 \pm 0.034 ^{+ 0.041 }_{- 0.035 }$ \\
		 &   		& 0.62 - 0.72 & $ 0.392 \pm 0.024 ^{+ 0.052 }_{- 0.027 }$ \\
		 & 	  	& 0.72 - 0.82 & $ 0.392 \pm 0.019 ^{+ 0.039 }_{- 0.056 }$ \\
		 & 	  	& 0.82 - 0.92 & $ 0.397 \pm 0.018 ^{+ 0.024 }_{- 0.044 }$ \\
\hline
   $ 1.83\cdot 10^{-3}$ &  11 	& 0.32 - 0.42 & $ 0.361 \pm 0.075 ^{+ 0.058 }_{- 0.056 }$ \\
		 &  	 	& 0.42 - 0.52 & $ 0.353 \pm 0.047 ^{+ 0.039 }_{- 0.077 }$ \\
		 &  	 	& 0.52 - 0.62 & $ 0.375 \pm 0.033 ^{+ 0.036 }_{- 0.073 }$ \\
		 &   		& 0.62 - 0.72 & $ 0.411 \pm 0.026 ^{+ 0.030 }_{- 0.027 }$ \\
		 & 	 	& 0.72 - 0.82 & $ 0.348 \pm 0.017 ^{+ 0.043 }_{- 0.026 }$ \\
		 & 	 	& 0.82 - 0.92 & $ 0.378 \pm 0.017 ^{+ 0.051 }_{- 0.021 }$ \\
\hline
   $ 3.98\cdot 10^{-3}$ &  11 	& 0.32 - 0.47 & $ 0.228 \pm 0.063 ^{+ 0.033 }_{- 0.090 }$ \\
		 &  		& 0.47 - 0.62 & $ 0.378 \pm 0.047 ^{+ 0.043 }_{- 0.063 }$ \\
		 & 		& 0.62 - 0.77 & $ 0.421 \pm 0.033 ^{+ 0.034 }_{- 0.105 }$ \\
		 &  		& 0.77 - 0.92 & $ 0.448 \pm 0.026 ^{+ 0.038 }_{- 0.057 }$ \\
\hline
   $ 5.1\cdot 10^{-4}$ &  22 	& 0.32 - 0.47 & $ 0.470 \pm 0.111 ^{+ 0.061 }_{- 0.281 }$ \\
		 & 		& 0.47 - 0.62 & $ 0.306 \pm 0.037 ^{+ 0.045 }_{- 0.028 }$ \\
		 &  		& 0.62 - 0.77 & $ 0.436 \pm 0.035 ^{+ 0.044 }_{- 0.043 }$ \\
		 & 	 	& 0.77 - 0.92 & $ 0.388 \pm 0.023 ^{+ 0.042 }_{- 0.027 }$ \\
\hline
   $ 9.2\cdot 10^{-4}$ &  22 	& 0.32 - 0.42 & $ 0.356 \pm 0.084 ^{+ 0.137 }_{- 0.071 }$ \\
		 &	 	& 0.42 - 0.52 & $ 0.350 \pm 0.053 ^{+ 0.044 }_{- 0.044 }$ \\
		 &	 	& 0.52 - 0.62 & $ 0.376 \pm 0.039 ^{+ 0.058 }_{- 0.030 }$ \\
		 &	   	& 0.62 - 0.72 & $ 0.401 \pm 0.030 ^{+ 0.031 }_{- 0.057 }$ \\
		 &	   	& 0.72 - 0.82 & $ 0.386 \pm 0.022 ^{+ 0.056 }_{- 0.029 }$ \\
		 & 	 	& 0.82 - 0.92 & $ 0.411 \pm 0.022 ^{+ 0.026 }_{- 0.041 }$ \\
\hline
   $ 1.84\cdot 10^{-3}$ &  22 	& 0.32 - 0.42 & $ 0.367 \pm 0.084 ^{+ 0.091 }_{- 0.075 }$ \\
		 &  		& 0.42 - 0.52 & $ 0.241 \pm 0.038 ^{+ 0.075 }_{- 0.024 }$ \\
		 &  		& 0.52 - 0.62 & $ 0.347 \pm 0.034 ^{+ 0.055 }_{- 0.033 }$ \\
 		& 	 	& 0.62 - 0.72 & $ 0.353 \pm 0.025 ^{+ 0.060 }_{- 0.027 }$ \\
 		&  		& 0.72 - 0.82 & $ 0.378 \pm 0.021 ^{+ 0.036 }_{- 0.035 }$ \\
 		& 	 	& 0.82 - 0.92 & $ 0.401 \pm 0.021 ^{+ 0.048 }_{- 0.026 }$ \\
\hline
   $ 3.66\cdot 10^{-3}$ &  22 	& 0.32 - 0.42 & $ 0.297 \pm 0.078 ^{+ 0.145 }_{- 0.042 }$ \\
		 & 	 	& 0.42 - 0.52 & $ 0.386 \pm 0.066 ^{+ 0.078 }_{- 0.034 }$ \\
		 &  		& 0.52 - 0.62 & $ 0.457 \pm 0.055 ^{+ 0.066 }_{- 0.071 }$ \\
		 & 	 	& 0.62 - 0.72 & $ 0.442 \pm 0.038 ^{+ 0.035 }_{- 0.077 }$ \\
		 &  		& 0.72 - 0.82 & $ 0.453 \pm 0.029 ^{+ 0.059 }_{- 0.033 }$ \\
		 & 	 	& 0.82 - 0.92 & $ 0.377 \pm 0.023 ^{+ 0.027 }_{- 0.040 }$ \\
\hline
   $ 7.83\cdot 10^{-3}$ &  22 	& 0.32 - 0.47 & $ 0.215 \pm 0.066 ^{+ 0.232 }_{- 0.102 }$ \\
		 & 	  	& 0.47 - 0.62 & $ 0.354 \pm 0.053 ^{+ 0.087 }_{- 0.060 }$ \\
		 & 	  	& 0.62 - 0.77 & $ 0.399 \pm 0.036 ^{+ 0.079 }_{- 0.032 }$ \\
		 & 	  	& 0.77 - 0.92 & $ 0.442 \pm 0.030 ^{+ 0.060 }_{- 0.077 }$ \\
\hline
   $ 1.03\cdot 10^{-3}$ &  44 	& 0.32 - 0.47 & $ 0.356 \pm 0.104 ^{+ 0.100 }_{- 0.062 }$ \\
		 &  	 	& 0.47 - 0.62 & $ 0.327 \pm 0.052 ^{+ 0.074 }_{- 0.039 }$ \\
		 &   		& 0.62 - 0.77 & $ 0.373 \pm 0.038 ^{+ 0.051 }_{- 0.034 }$ \\
		 &  	 	& 0.77 - 0.92 & $ 0.381 \pm 0.030 ^{+ 0.027 }_{- 0.030 }$ \\
\hline
   $ 1.86\cdot 10^{-3}$ &  44 	& 0.37 - 0.42 & $ 0.508 \pm 0.150 ^{+ 0.221 }_{- 0.116 }$ \\
		 &   		& 0.47 - 0.52 & $ 0.346 \pm 0.073 ^{+ 0.049 }_{- 0.077 }$ \\
		 &  	 	& 0.57 - 0.62 & $ 0.444 \pm 0.063 ^{+ 0.043 }_{- 0.112 }$ \\
 		&   		& 0.67 - 0.72 & $ 0.348 \pm 0.041 ^{+ 0.027 }_{- 0.049 }$ \\
		 & 	  	& 0.77 - 0.82 & $ 0.413 \pm 0.033 ^{+ 0.044 }_{- 0.052 }$ \\
 		&  	 	& 0.87 - 0.92 & $ 0.352 \pm 0.026 ^{+ 0.051 }_{- 0.021 }$ \\
\hline
   $ 3.68\cdot 10^{-3}$ &  44 	& 0.37 - 0.42 & $ 0.221 \pm 0.088 ^{+ 0.065 }_{- 0.027 }$ \\
		 &  	 	& 0.47 - 0.52 & $ 0.456 \pm 0.097 ^{+ 0.079 }_{- 0.105 }$ \\
 		& 	  	& 0.57 - 0.62 & $ 0.303 \pm 0.046 ^{+ 0.052 }_{- 0.023 }$ \\
 		&  	 	& 0.67 - 0.72 & $ 0.496 \pm 0.057 ^{+ 0.036 }_{- 0.150 }$ \\
		 &   		& 0.77 - 0.82 & $ 0.511 \pm 0.044 ^{+ 0.061 }_{- 0.056 }$ \\
		 & 	  	& 0.87 - 0.92 & $ 0.408 \pm 0.033 ^{+ 0.041 }_{- 0.029 }$ \\
\hline
   $ 7.33\cdot 10^{-3}$ &  44 	& 0.37 - 0.42 & $ 0.628 \pm 0.226 ^{+ 0.297 }_{- 0.190 }$ \\
		 & 	  	& 0.47 - 0.52 & $ 0.341 \pm 0.090 ^{+ 0.053 }_{- 0.060 }$ \\
		 & 	  	& 0.57 - 0.62 & $ 0.418 \pm 0.074 ^{+ 0.153 }_{- 0.035 }$ \\
		 & 	  	& 0.67 - 0.72 & $ 0.458 \pm 0.061 ^{+ 0.044 }_{- 0.070 }$ \\
		 &   		& 0.77 - 0.82 & $ 0.501 \pm 0.048 ^{+ 0.055 }_{- 0.081 }$ \\
		 & 	  	& 0.87 - 0.92 & $ 0.448 \pm 0.039 ^{+ 0.042 }_{- 0.030 }$ \\
\hline
   $ 1.54\cdot 10^{-2}$ &  44 	& 0.32 - 0.52 & $ 0.453 \pm 0.135 ^{+ 0.078 }_{- 0.363 }$ \\
		 &   		& 0.52 - 0.72 & $ 0.454 \pm 0.062 ^{+ 0.048 }_{- 0.071 }$ \\
		 &  	 	& 0.72 - 0.92 & $ 0.385 \pm 0.033 ^{+ 0.083 }_{- 0.027 }$ \\
\hline
   $ 2.00\cdot 10^{-3}$ &  88 	& 0.32 - 0.47 & $ 0.328 \pm 0.162 ^{+ 0.151 }_{- 0.064 }$ \\
		 &  	 	& 0.47 - 0.62 & $ 0.462 \pm 0.121 ^{+ 0.111 }_{- 0.122 }$ \\
		 &   		& 0.62 - 0.77 & $ 0.307 \pm 0.045 ^{+ 0.045 }_{- 0.030 }$ \\
		 &  	 	& 0.77 - 0.92 & $ 0.339 \pm 0.040 ^{+ 0.048 }_{- 0.033 }$ \\
\hline
   $ 3.59\cdot 10^{-3}$ &  88 	& 0.32 - 0.47 & $ 0.449 \pm 0.145 ^{+ 0.058 }_{- 0.134 }$ \\
		 & 	  	& 0.47 - 0.62 & $ 0.509 \pm 0.090 ^{+ 0.120 }_{- 0.098 }$ \\
		 & 	  	& 0.62 - 0.77 & $ 0.402 \pm 0.049 ^{+ 0.194 }_{- 0.034 }$ \\
		 & 	  	& 0.77 - 0.92 & $ 0.393 \pm 0.034 ^{+ 0.068 }_{- 0.022 }$ \\
\hline
   $ 7.37\cdot 10^{-3}$ &  88 	& 0.32 - 0.47 & $ 0.491 \pm 0.178 ^{+ 0.164 }_{- 0.131 }$ \\
		&  	 	& 0.47 - 0.62 & $ 0.345 \pm 0.078 ^{+ 0.050 }_{- 0.042 }$ \\
		 &   		& 0.62 - 0.77 & $ 0.364 \pm 0.046 ^{+ 0.028 }_{- 0.034 }$ \\
		 &	   	& 0.77 - 0.92 & $ 0.440 \pm 0.042 ^{+ 0.087 }_{- 0.032 }$ \\
\hline
   $ 1.42\cdot 10^{-2}$ &  88 	& 0.32 - 0.47 & $ 0.351 \pm 0.133 ^{+ 0.171 }_{- 0.054 }$ \\
		 &  	 	& 0.47 - 0.62 & $ 0.479 \pm 0.101 ^{+ 0.169 }_{- 0.103 }$ \\
		 &  	 	& 0.62 - 0.77 & $ 0.390 \pm 0.054 ^{+ 0.098 }_{- 0.037 }$ \\
		 &  	 	& 0.77 - 0.92 & $ 0.498 \pm 0.050 ^{+ 0.042 }_{- 0.112 }$ \\
\hline
   $ 3.01\cdot 10^{-2}$ &  88 	& 0.32 - 0.52 & $ 0.449 \pm 0.211 ^{+ 0.075 }_{- 0.264 }$ \\
		 &  	 	& 0.52 - 0.72 & $ 0.330 \pm 0.063 ^{+ 0.121 }_{- 0.023 }$ \\
		 & 	  	& 0.72 - 0.92 & $ 0.440 \pm 0.056 ^{+ 0.062 }_{- 0.044 }$ \\
\hline
   $ 4.00\cdot 10^{-3}$ & 237 	& 0.32 - 0.52 & $ 0.250 \pm 0.153 ^{+ 0.115 }_{- 0.129 }$ \\
		&  		& 0.52 - 0.72 & $ 0.175 \pm 0.068 ^{+ 0.241 }_{- 0.013 }$ \\
 		& 	 	& 0.72 - 0.92 & $ 0.451 \pm 0.071 ^{+ 0.079 }_{- 0.044 }$ \\
\hline
   $ 7.52\cdot 10^{-3}$ & 237 	& 0.32 - 0.52 & $ 0.467 \pm 0.195 ^{+ 0.105 }_{- 0.086 }$ \\
		 &  		& 0.52 - 0.72 & $ 0.482 \pm 0.088 ^{+ 0.067 }_{- 0.205 }$ \\
		 & 	 	& 0.72 - 0.92 & $ 0.430 \pm 0.049 ^{+ 0.069 }_{- 0.053 }$ \\
\hline
   $ 1.47\cdot 10^{-2}$ & 237 	& 0.32 - 0.47 & $ 0.301 \pm 0.156 ^{+ 0.366 }_{- 0.040 }$ \\
 		& 	 	& 0.47 - 0.62 & $ 0.249 \pm 0.074 ^{+ 0.087 }_{- 0.037 }$ \\
 		& 		& 0.62 - 0.77 & $ 0.445 \pm 0.070 ^{+ 0.112 }_{- 0.040 }$ \\
 		& 	 	& 0.77 - 0.92 & $ 0.381 \pm 0.045 ^{+ 0.040 }_{- 0.028 }$ \\
\hline
   $ 3.25\cdot 10^{-2}$ & 237 	& 0.32 - 0.47 & $ 0.102 \pm 0.082 ^{+ 0.119 }_{- 0.011 }$ \\
 		&  		& 0.47 - 0.62 & $ 0.355 \pm 0.090 ^{+ 0.049 }_{- 0.072 }$ \\
 		& 	 	& 0.62 - 0.77 & $ 0.359 \pm 0.055 ^{+ 0.135 }_{- 0.028 }$ \\
 		& 		& 0.77 - 0.92 & $ 0.403 \pm 0.049 ^{+ 0.041 }_{- 0.064 }$ \\
 \hline \hline
\end{supertabular}
\end{center}

\clearpage


\begin{center}

\tablefirsthead{
\hline \hline 
$\langle x \rangle$  & $\langle Q^2 \rangle$  (GeV$^2$) &  $x_L$ range &  $r^{\rm LP(3)}$ \\ 
\hline \hline 
}

\tablehead{
\multicolumn{4}{l}{\tab{rlp3_0.04} (cont.)} \\
\hline  \hline
$\langle x \rangle$  & $\langle Q^2 \rangle$  (GeV$^2$) &  $x_L$ range &  $r^{\rm LP(3)}$ \\ 
\hline \hline
}

\tabletail{\hline}

\bottomcaption{
The leading proton production rate,  $r^{\rm LP(3)}$, measured as a
function of $x_L$ for protons with $p_T^2<0.04$ $GeV^2$, in bins of 
$x$ and $Q^2$, with averages $\langle x \rangle$  and $\langle Q^2 \rangle$.
Statistical uncertainties are listed first, followed by systematic uncertainties.
}

\begin{supertabular}{|c|c|c|c|}
 $  9.6\cdot 10^{-5}$ &   4.2 	 & 0.32 - 0.47  & $ 0.082\pm0.015^{+0.043}_{-0.014}$ \\
 			 &    	 & 0.47 - 0.62  & $ 0.091\pm0.012^{+0.015}_{-0.019}$ \\
 			 &    	 & 0.62 - 0.77  & $ 0.080\pm0.006^{+0.005}_{-0.026}$ \\
			 &    	 & 0.77 - 0.92  & $ 0.104\pm0.008^{+0.007}_{-0.014}$ \\

\hline
 $  1.7\cdot 10^{-4}$ &   4.2    & 0.32 - 0.47 & $ 0.103\pm0.013^{+0.006}_{-0.026}$ \\
			 &  	 & 0.47 - 0.62 & $ 0.083\pm0.008^{+0.009}_{-0.010}$ \\ 
			 &  	 & 0.62 - 0.77 & $ 0.096\pm0.005^{+0.003}_{-0.023}$ \\
			 &  	 & 0.77 - 0.92 & $ 0.089\pm0.004^{+0.003}_{-0.008}$ \\ 
\hline
 $  3.5\cdot 10^{-4}$ &   4.2 	 & 0.32 - 0.47 & $ 0.077\pm0.010^{+0.009}_{-0.009}$ \\
 			 &  	 & 0.47 - 0.62 & $ 0.087\pm0.008^{+0.012}_{-0.006}$ \\
			 &  	 & 0.62 - 0.77 & $ 0.108\pm0.006^{+0.009}_{-0.007}$ \\ 
			 &   	 & 0.77 - 0.92 & $ 0.093\pm0.004^{+0.004}_{-0.003}$ \\
\hline	
 $  6.9\cdot 10^{-4}$ &   4.2    & 0.32 - 0.47 & $ 0.077\pm0.012^{+0.013}_{-0.004}$ \\ 
			 &  	 & 0.47 - 0.62 & $ 0.085\pm0.009^{+0.007}_{-0.007}$ \\
			 &   	 & 0.62 - 0.77 & $ 0.095\pm0.006^{+0.009}_{-0.002}$ \\
			 &   	 & 0.77 - 0.92 & $ 0.087\pm0.004^{+0.006}_{-0.002}$ \\
\hline 
 $  1.46\cdot 10^{-3}$ &   4.2   & 0.32 - 0.47 & $ 0.074\pm0.016^{+0.012}_{-0.015}$ \\
			 &   	 & 0.47 - 0.62 & $ 0.046\pm0.008^{+0.011}_{-0.004}$ \\
			 &   	 & 0.62 - 0.77 & $ 0.083\pm0.006^{+0.012}_{-0.003}$ \\
			 &   	 & 0.77 - 0.92 & $ 0.086\pm0.006^{+0.005}_{-0.007}$ \\
\end{supertabular}
\label{tab-rlp3_0.04}
\end{center}

\clearpage

\begin{center}

\tablefirsthead{
\multicolumn{4}{l}{\tab{rlp3_0.04} (cont.)} \\
\hline 
$\langle x \rangle$  & $\langle Q^2 \rangle$  (GeV$^2$) &  $x_L$ range &  $r^{\rm LP(3)}$ \\ 
\hline \hline 
}

\tablehead{
\multicolumn{4}{l}{\tab{rlp3_0.04} (cont.)} \\
\hline  
$\langle x \rangle$  & $\langle Q^2 \rangle$  (GeV$^2$) &  $x_L$ range & $r^{\rm LP(3)}$ \\ 
\hline \hline
}

\tabletail{\hline}

\begin{supertabular}{|c|c|c|c|}
   $ 1.9\cdot 10^{-4}$ &   7.3  & 0.32 - 0.47 & $ 0.093\pm0.017^{+0.009}_{-0.015}$ \\
 			 &   	& 0.47 - 0.62 & $ 0.098\pm0.012^{+0.008}_{-0.027}$ \\
			 &  	& 0.62 - 0.77 & $ 0.087\pm0.007^{+0.013}_{-0.003}$ \\
			 &    	& 0.77 - 0.92 & $ 0.092\pm0.007^{+0.006}_{-0.015}$ \\
\hline
   $ 3.4\cdot 10^{-4}$ &   7.3  & 0.32 - 0.47 & $ 0.090\pm0.015^{+0.012}_{-0.019}$ \\
			 &   	& 0.47 - 0.62 & $ 0.103\pm0.011^{+0.006}_{-0.011}$ \\
			 &  	& 0.62 - 0.77 & $ 0.079\pm0.005^{+0.007}_{-0.002}$ \\
			 &   	& 0.77 - 0.92 & $ 0.088\pm0.005^{+0.004}_{-0.003}$ \\
\hline
   $ 6.9\cdot 10^{-4}$ &   7.3  & 0.32 - 0.47 & $ 0.090\pm0.015^{+0.005}_{-0.025}$ \\
		 	&  	& 0.47 - 0.62 & $ 0.093\pm0.011^{+0.004}_{-0.017}$ \\
		 	&   	& 0.62 - 0.77 & $ 0.102\pm0.007^{+0.004}_{-0.006}$ \\
			 &   	& 0.77 - 0.92 & $ 0.095\pm0.006^{+0.002}_{-0.017}$ \\
\hline
   $ 1.36\cdot 10^{-3}$ &   7.3 & 0.32 - 0.47 & $ 0.075\pm0.014^{+0.011}_{-0.006}$ \\
			 &   	& 0.47 - 0.62 & $ 0.104\pm0.013^{+0.010}_{-0.014}$ \\
			 &    	& 0.62 - 0.77 & $ 0.113\pm0.008^{+0.003}_{-0.017}$ \\
			 &    	& 0.77 - 0.92 & $ 0.098\pm0.006^{+0.006}_{-0.009}$ \\
\hline
   $ 2.67\cdot 10^{-3}$ &   7.3 & 0.32 - 0.47 & $ 0.067\pm0.025^{+0.007}_{-0.034}$ \\
		 	&    	& 0.47 - 0.62 & $ 0.112\pm0.028^{+0.023}_{-0.036}$ \\
		 	&    	& 0.62 - 0.77 & $ 0.097\pm0.015^{+0.017}_{-0.012}$ \\
		 	&    	& 0.77 - 0.92 & $ 0.095\pm0.011^{+0.003}_{-0.017}$ \\
\hline

$ 2.6\cdot 10^{-4}$ &  11 	& 0.32 - 0.47 & $ 0.098\pm0.024^{+0.006}_{-0.031}$ \\
		 	&   	& 0.47 - 0.62 & $ 0.075\pm0.013^{+0.038}_{-0.006}$ \\
		 	&   	& 0.62 - 0.77 & $ 0.129\pm0.014^{+0.005}_{-0.011}$ \\
		 	&   	& 0.77 - 0.92 & $ 0.085\pm0.008^{+0.003}_{-0.005}$ \\
\hline
  $ 4.6\cdot 10^{-4}$ &  11      & 0.32 - 0.47 & $ 0.114\pm0.018^{+0.008}_{-0.027}$ \\
 			 &  	 & 0.47 - 0.62 & $ 0.081\pm0.009^{+0.014}_{-0.004}$ \\
 			 &  	 & 0.62 - 0.77 & $ 0.101\pm0.007^{+0.003}_{-0.011}$ \\
			  &   	 & 0.77 - 0.92 & $ 0.099\pm0.006^{+0.007}_{-0.003}$ \\
\hline
   $ 9.2\cdot 10^{-4}$ &  11     & 0.32 - 0.47 & $ 0.080\pm0.013^{+0.009}_{-0.006}$ \\
			 &   	 & 0.47 - 0.62 & $ 0.110\pm0.012^{+0.019}_{-0.004}$ \\
			 & 	 & 0.62 - 0.77 & $ 0.080\pm0.006^{+0.009}_{-0.007}$ \\
			 &   	 & 0.77 - 0.92 & $ 0.091\pm0.005^{+0.002}_{-0.005}$ \\
\hline
   $ 1.83\cdot 10^{-3}$ &  11    & 0.32 - 0.47 & $ 0.082\pm0.015^{+0.008}_{-0.007}$ \\
			 &  	 & 0.47 - 0.62 & $ 0.094\pm0.011^{+0.004}_{-0.018}$ \\
			 &  	 & 0.62 - 0.77 & $ 0.100\pm0.007^{+0.004}_{-0.006}$ \\
			 &   	 & 0.77 - 0.92 & $ 0.089\pm0.005^{+0.008}_{-0.002}$ \\

\hline
   $ 3.98\cdot 10^{-3}$ &  11    & 0.32 - 0.47 & $ 0.041\pm0.015^{+0.009}_{-0.020}$ \\
			 &  	 & 0.47 - 0.62 & $ 0.118\pm0.021^{+0.010}_{-0.042}$ \\
			 & 	 & 0.62 - 0.77 & $ 0.102\pm0.010^{+0.003}_{-0.024}$ \\
			 &  	 & 0.77 - 0.92 & $ 0.107\pm0.009^{+0.006}_{-0.014}$ \\
\hline
   $ 5.1\cdot 10^{-4}$ &  22     & 0.32 - 0.47 & $ 0.098\pm0.029^{+0.013}_{-0.056}$ \\
			 & 	 & 0.47 - 0.62 & $ 0.081\pm0.014^{+0.010}_{-0.006}$ \\
			 &  	 & 0.62 - 0.77 & $ 0.114\pm0.014^{+0.021}_{-0.007}$ \\
			 & 	 & 0.77 - 0.92 & $ 0.088\pm0.009^{+0.005}_{-0.023}$ \\
\hline		
   $ 9.2\cdot 10^{-4}$ &  22     & 0.32 - 0.47 & $ 0.094\pm0.018^{+0.007}_{-0.020}$ \\
			 &	 & 0.47 - 0.62 & $ 0.094\pm0.012^{+0.018}_{-0.003}$ \\
			 &	 & 0.62 - 0.77 & $ 0.100\pm0.008^{+0.009}_{-0.008}$ \\
			 &	 & 0.77 - 0.92 & $ 0.090\pm0.006^{+0.008}_{-0.004}$ \\
\hline
   $ 1.84\cdot 10^{-3}$ &  22    & 0.32 - 0.47 & $ 0.089\pm0.017^{+0.022}_{-0.008}$ \\
			 &  	 & 0.47 - 0.62 & $ 0.080\pm0.010^{+0.007}_{-0.010}$ \\
			 &  	 & 0.62 - 0.77 & $ 0.089\pm0.007^{+0.003}_{-0.007}$ \\
 			& 	 & 0.77 - 0.92 & $ 0.089\pm0.006^{+0.010}_{-0.004}$ \\
\hline
   $ 3.66\cdot 10^{-3}$ &  22    & 0.32 - 0.47 & $ 0.078\pm0.017^{+0.026}_{-0.005}$ \\
			 & 	 & 0.47 - 0.62 & $ 0.120\pm0.019^{+0.009}_{-0.028}$ \\
			 &  	 & 0.62 - 0.77 & $ 0.114\pm0.010^{+0.007}_{-0.010}$ \\
			 & 	 & 0.77 - 0.92 & $ 0.099\pm0.007^{+0.013}_{-0.002}$ \\
\hline
   $ 7.83\cdot 10^{-3}$ &  22    & 0.32 - 0.47 & $ 0.054\pm0.019^{+0.050}_{-0.024}$ \\
			 & 	 & 0.47 - 0.62 & $ 0.087\pm0.018^{+0.011}_{-0.043}$ \\
			 & 	 & 0.62 - 0.77 & $ 0.087\pm0.010^{+0.041}_{-0.005}$ \\
			 & 	 & 0.77 - 0.92 & $ 0.116\pm0.012^{+0.033}_{-0.016}$ \\
\hline
   $ 1.03\cdot 10^{-3}$ &  44 	 & 0.32 - 0.47 & $ 0.092\pm0.031^{+0.020}_{-0.017}$ \\
			 &  	 & 0.47 - 0.62 & $ 0.086\pm0.020^{+0.048}_{-0.010}$ \\
			 &   	 & 0.62 - 0.77 & $ 0.105\pm0.016^{+0.018}_{-0.031}$ \\
			 &  	 & 0.77 - 0.92 & $ 0.108\pm0.013^{+0.015}_{-0.005}$ \\
\hline
   $ 1.86\cdot 10^{-3}$ &  44    & 0.32 - 0.47 & $ 0.096\pm0.023^{+0.024}_{-0.010}$ \\
			 &   	 & 0.47 - 0.62 & $ 0.087\pm0.016^{+0.005}_{-0.028}$ \\
			 &  	 & 0.62 - 0.77 & $ 0.130\pm0.015^{+0.004}_{-0.029}$ \\
 			&   	 & 0.77 - 0.92 & $ 0.102\pm0.010^{+0.004}_{-0.016}$ \\
\hline
   $ 3.68\cdot 10^{-3}$ &  44    & 0.32 - 0.47 & $ 0.077\pm0.023^{+0.038}_{-0.011}$ \\
			 &  	 & 0.47 - 0.62 & $ 0.079\pm0.016^{+0.004}_{-0.019}$ \\
			 &   	 & 0.62 - 0.77 & $ 0.139\pm0.017^{+0.005}_{-0.023}$ \\
			 & 	 & 0.77 - 0.92 & $ 0.117\pm0.012^{+0.003}_{-0.013}$ \\
\hline
   $ 7.33\cdot 10^{-3}$ &  44    & 0.32 - 0.47 & $ 0.124\pm0.039^{+0.021}_{-0.030}$ \\
			 & 	 & 0.47 - 0.62 & $ 0.104\pm0.023^{+0.027}_{-0.005}$ \\
			 & 	 & 0.62 - 0.77 & $ 0.107\pm0.015^{+0.022}_{-0.004}$ \\
			 & 	 & 0.77 - 0.92 & $ 0.096\pm0.011^{+0.023}_{-0.004}$ \\
\hline
   $ 1.54\cdot 10^{-2}$ &  44    & 0.32 - 0.47 & $ 0.076\pm0.036^{+0.075}_{-0.017}$ \\
		 	&   	 & 0.47 - 0.62 & $ 0.118\pm0.034^{+0.020}_{-0.040}$ \\
		 	&  	 & 0.62 - 0.77 & $ 0.108\pm0.019^{+0.036}_{-0.009}$ \\
		 	&  	 & 0.77 - 0.92 & $ 0.101\pm0.015^{+0.022}_{-0.015}$ \\
\hline
   $ 2.00\cdot 10^{-3}$ &  88    & 0.32 - 0.52 & $ 0.130\pm0.053^{+0.014}_{-0.076}$ \\
			 &  	 & 0.52 - 0.72 & $ 0.128\pm0.031^{+0.038}_{-0.049}$ \\
			 &   	 & 0.72 - 0.92 & $ 0.089\pm0.014^{+0.021}_{-0.002}$ \\
\hline
   $ 3.59\cdot 10^{-3}$ &  88    & 0.32 - 0.52 & $ 0.126\pm0.037^{+0.009}_{-0.046}$ \\
			 & 	 & 0.52 - 0.72 & $ 0.080\pm0.014^{+0.035}_{-0.004}$ \\
			 & 	 & 0.72 - 0.92 & $ 0.104\pm0.012^{+0.015}_{-0.003}$ \\
\hline
   $ 7.37\cdot 10^{-3}$ &  88    & 0.32 - 0.52 & $ 0.091\pm0.031^{+0.034}_{-0.015}$ \\
			&  	 & 0.52 - 0.72 & $ 0.077\pm0.017^{+0.029}_{-0.006}$ \\
			 &   	 & 0.72 - 0.92 & $ 0.090\pm0.011^{+0.026}_{-0.004}$ \\
\hline
   $ 1.42\cdot 10^{-2}$ &  88    & 0.32 - 0.52 & $ 0.075\pm0.024^{+0.048}_{-0.005}$ \\
			 &  	 & 0.52 - 0.72 & $ 0.118\pm0.027^{+0.051}_{-0.010}$ \\
			 &  	 & 0.72 - 0.92 & $ 0.103\pm0.014^{+0.034}_{-0.007}$ \\
\hline
   $ 3.01\cdot 10^{-2}$ &  88    & 0.32 - 0.52 & $ 0.151\pm0.075^{+0.022}_{-0.143}$ \\
			 &  	 & 0.52 - 0.72 & $ 0.081\pm0.020^{+0.015}_{-0.012}$ \\
			 & 	 & 0.72 - 0.92 & $ 0.099\pm0.017^{+0.033}_{-0.004}$ \\
\hline
   $ 4.00\cdot 10^{-3}$ & 237   & 0.32 - 0.52 & $ 0.053\pm0.036^{+0.020}_{-0.045}$ \\
			&  	& 0.52 - 0.72 & $ 0.092\pm0.033^{+0.102}_{-0.006}$ \\
 			& 	& 0.72 - 0.92 & $ 0.116\pm0.025^{+0.022}_{-0.010}$ \\
\hline
   $ 7.52\cdot 10^{-3}$ & 237   & 0.32 - 0.52 & $ 0.116\pm0.052^{+0.036}_{-0.015}$ \\
			 &  	& 0.52 - 0.72 & $ 0.108\pm0.030^{+0.016}_{-0.057}$ \\
			 & 	& 0.72 - 0.92 & $ 0.098\pm0.017^{+0.007}_{-0.024}$ \\
\hline
   $ 1.47\cdot 10^{-2}$ & 237   & 0.32 - 0.52 & $ 0.072\pm0.036^{+0.053}_{-0.007}$ \\
 			& 	& 0.52 - 0.72 & $ 0.098\pm0.025^{+0.062}_{-0.022}$ \\
 			& 	& 0.72 - 0.92 & $ 0.128\pm0.020^{+0.012}_{-0.025}$ \\
\hline
   $ 3.25\cdot 10^{-2}$ & 237   & 0.32 - 0.52 & $ 0.033\pm0.020^{+0.073}_{-0.002}$ \\
	 		&  	& 0.52 - 0.72 & $ 0.071\pm0.019^{+0.058}_{-0.013}$ \\
 			& 	& 0.72 - 0.92 & $ 0.102\pm0.017^{+0.014}_{-0.019}$ \\
 \hline \hline
\end{supertabular}
\end{center}

\clearpage


\begin{center}

\tablefirsthead{
\hline \hline 
$\langle x \rangle$  & $\langle Q^2 \rangle$  (GeV$^2$) &  $x_L$ range & $r^{\rm LP(3)}$ \\ 
\hline \hline 
}

\tablehead{
\multicolumn{4}{l}{\tab{rlp3_0.15} (cont.)} \\
\hline  \hline
$\langle x \rangle$  & $\langle Q^2 \rangle$  (GeV$^2$) &  $x_L$ range & $r^{\rm LP(3)}$ \\ 
\hline \hline
}

\tabletail{\hline}

\bottomcaption{
The leading proton production rate,  $r^{\rm LP(3)}$, measured as a
function of $x_L$ for protons with $0.04<p_T^2<0.15$ $GeV^2$, in bins of 
$x$ and $Q^2$, with averages $\langle x \rangle$  and $\langle Q^2 \rangle$.
Statistical uncertainties are listed first, followed by systematic uncertainties.
}

\begin{supertabular}{|c|c|c|c|}
 $  9.6\cdot 10^{-5}$ &   4.2 	 & 0.38 - 0.56 &$ 0.146 \pm 0.027 ^{+ 0.022 }_{- 0.039 }$\\
 			 &    	 & 0.56 - 0.74 &$ 0.125 \pm 0.012 ^{+ 0.015 }_{- 0.013 }$\\
 			 &    	 & 0.74 - 0.92 &$ 0.177 \pm 0.013 ^{+ 0.017 }_{- 0.058 }$\\
\hline
 $  1.7\cdot 10^{-4}$ &   4.2    & 0.38 - 0.56 &$ 0.133 \pm 0.017 ^{+ 0.025 }_{- 0.026 }$\\
			 &  	 & 0.56 - 0.74 &$ 0.145 \pm 0.010 ^{+ 0.006 }_{- 0.026 }$\\
			 &   	 & 0.74 - 0.92 &$ 0.137 \pm 0.006 ^{+ 0.009 }_{- 0.005 }$\\
\hline
 $  3.5\cdot 10^{-4}$ &   4.2 	 & 0.38 - 0.56 &$ 0.123 \pm 0.018 ^{+ 0.020 }_{- 0.011 }$\\
 			 &  	 & 0.56 - 0.74 &$ 0.140 \pm 0.009 ^{+ 0.009 }_{- 0.013 }$\\
			 &  	 & 0.74 - 0.92 &$ 0.141 \pm 0.007 ^{+ 0.008 }_{- 0.012 }$\\
\hline
 $  6.9\cdot 10^{-4}$ &   4.2    & 0.38 - 0.56 &$ 0.128 \pm 0.019 ^{+ 0.018 }_{- 0.028 }$\\
			 &  	 & 0.56 - 0.74 &$ 0.128 \pm 0.010 ^{+ 0.008 }_{- 0.018 }$\\
			 &   	 & 0.74 - 0.92 &$ 0.151 \pm 0.008 ^{+ 0.006 }_{- 0.020 }$\\
\hline
 $  1.46\cdot 10^{-3}$ &   4.2   & 0.38 - 0.56 &$ 0.160 \pm 0.033 ^{+ 0.035 }_{- 0.016 }$\\
			 &   	 & 0.56 - 0.74 &$ 0.156 \pm 0.018 ^{+ 0.015 }_{- 0.019 }$\\
			 &   	 & 0.74 - 0.92 &$ 0.136 \pm 0.010 ^{+ 0.017 }_{- 0.020 }$\\
\hline
   $ 1.9\cdot 10^{-4}$ &   7.3  & 0.38 - 0.56 &$ 0.126 \pm 0.023 ^{+ 0.048 }_{- 0.006 }$\\
 			 &   	& 0.56 - 0.74 &$ 0.123 \pm 0.012 ^{+ 0.009 }_{- 0.016 }$\\
			 &  	& 0.74 - 0.92 &$ 0.149 \pm 0.011 ^{+ 0.004 }_{- 0.020 }$\\
\hline
   $ 3.4\cdot 10^{-4}$ &   7.3  & 0.38 - 0.56 &$ 0.140 \pm 0.022 ^{+ 0.018 }_{- 0.028 }$\\
			 &   	& 0.56 - 0.74 &$ 0.144 \pm 0.012 ^{+ 0.006 }_{- 0.026 }$\\
			 &  	& 0.74 - 0.92 &$ 0.136 \pm 0.008 ^{+ 0.008 }_{- 0.008 }$\\
\hline
   $ 6.9\cdot 10^{-4}$ &   7.3  & 0.38 - 0.56 &$ 0.123 \pm 0.021 ^{+ 0.019 }_{- 0.022 }$\\
		 	&  	& 0.56 - 0.74 &$ 0.122 \pm 0.011 ^{+ 0.012 }_{- 0.004 }$\\
		 	&   	& 0.74 - 0.92 &$ 0.141 \pm 0.008 ^{+ 0.009 }_{- 0.014 }$\\
\hline
   $ 1.36\cdot 10^{-3}$ &   7.3 & 0.38 - 0.56 &$ 0.158 \pm 0.027 ^{+ 0.052 }_{- 0.024 }$\\
			 &   	& 0.56 - 0.74 &$ 0.140 \pm 0.013 ^{+ 0.016 }_{- 0.009 }$\\
			 &    	& 0.74 - 0.92 &$ 0.149 \pm 0.010 ^{+ 0.013 }_{- 0.006 }$\\
\hline
   $ 2.67\cdot 10^{-3}$ &   7.3 & 0.38 - 0.56 &$ 0.175 \pm 0.070 ^{+ 0.023 }_{- 0.153 }$\\
		 	&    	& 0.56 - 0.74 &$ 0.193 \pm 0.038 ^{+ 0.077 }_{- 0.054 }$\\
		 	&    	& 0.74 - 0.92 &$ 0.161 \pm 0.021 ^{+ 0.032 }_{- 0.024 }$\\
\end{supertabular}
\label{tab-rlp3_0.15}
\end{center}

\clearpage

\begin{center}

\tablefirsthead{
\multicolumn{4}{l}{\tab{rlp3_0.15} (cont.)} \\
\hline 
$\langle x \rangle$  & $\langle Q^2 \rangle$  (GeV$^2$) &  $x_L$ range &  $r^{\rm LP(3)}$ \\ 
\hline \hline 
}

\tablehead{
\multicolumn{4}{l}{\tab{rlp3_0.15} (cont.)} \\
\hline
$\langle x \rangle$  & $\langle Q^2 \rangle$  (GeV$^2$) &  $x_L$ range &  $r^{\rm LP(3)}$ \\ 
\hline \hline
}

\tabletail{\hline}

\begin{supertabular}{|c|c|c|c|}
$ 2.6\cdot 10^{-4}$ &  11 	& 0.38 - 0.56 &$ 0.084 \pm 0.021 ^{+ 0.033 }_{- 0.008 }$\\
		 	&   	& 0.56 - 0.74 &$ 0.151 \pm 0.019 ^{+ 0.010 }_{- 0.030 }$\\
		 	&   	& 0.74 - 0.92 &$ 0.140 \pm 0.012 ^{+ 0.013 }_{- 0.006 }$\\
\hline
  $ 4.6\cdot 10^{-4}$ &  11      & 0.38 - 0.56 &$ 0.134 \pm 0.020 ^{+ 0.056 }_{- 0.009 }$\\
 			 &  	 & 0.56 - 0.74 &$ 0.132 \pm 0.011 ^{+ 0.009 }_{- 0.016 }$\\
 			 &  	 & 0.74 - 0.92 &$ 0.133 \pm 0.007 ^{+ 0.032 }_{- 0.003 }$\\
\hline
   $ 9.2\cdot 10^{-4}$ &  11     & 0.38 - 0.56 &$ 0.129 \pm 0.021 ^{+ 0.008 }_{- 0.015 }$\\
			 &   	 & 0.56 - 0.74 &$ 0.140 \pm 0.011 ^{+ 0.032 }_{- 0.006 }$\\
			 & 	 & 0.74 - 0.92 &$ 0.150 \pm 0.008 ^{+ 0.003 }_{- 0.022 }$\\
\hline
   $ 1.83\cdot 10^{-3}$ &  11    & 0.38 - 0.56 &$ 0.115 \pm 0.018 ^{+ 0.026 }_{- 0.007 }$\\
			 &  	 & 0.56 - 0.74 &$ 0.165 \pm 0.013 ^{+ 0.007 }_{- 0.036 }$\\
			 &  	 & 0.74 - 0.92 &$ 0.127 \pm 0.007 ^{+ 0.022 }_{- 0.003 }$\\
\hline
   $ 3.98\cdot 10^{-3}$ &  11    & 0.38 - 0.56 &$ 0.110 \pm 0.031 ^{+ 0.099 }_{- 0.013 }$\\
			 &  	 & 0.56 - 0.74 &$ 0.129 \pm 0.018 ^{+ 0.007 }_{- 0.032 }$\\
			 & 	 & 0.74 - 0.92 &$ 0.157 \pm 0.013 ^{+ 0.018 }_{- 0.024 }$\\
\hline
   $ 5.1\cdot 10^{-4}$ &  22     & 0.38 - 0.56 &$ 0.112 \pm 0.029 ^{+ 0.079 }_{- 0.009 }$\\
			 & 	 & 0.56 - 0.74 &$ 0.133 \pm 0.017 ^{+ 0.009 }_{- 0.019 }$\\
			 &  	 & 0.74 - 0.92 &$ 0.148 \pm 0.013 ^{+ 0.037 }_{- 0.003 }$\\
\hline		
   $ 9.2\cdot 10^{-4}$ &  22     & 0.38 - 0.56 &$ 0.146 \pm 0.028 ^{+ 0.039 }_{- 0.025 }$\\
			 &	 & 0.56 - 0.74 &$ 0.153 \pm 0.015 ^{+ 0.007 }_{- 0.025 }$\\
			 &	 & 0.74 - 0.92 &$ 0.157 \pm 0.010 ^{+ 0.004 }_{- 0.016 }$\\
\hline
   $ 1.84\cdot 10^{-3}$ &  22    & 0.38 - 0.56 &$ 0.101 \pm 0.019 ^{+ 0.050 }_{- 0.006 }$\\
			 &  	 & 0.56 - 0.74 &$ 0.141 \pm 0.013 ^{+ 0.009 }_{- 0.012 }$\\
			 &  	 & 0.74 - 0.92 &$ 0.144 \pm 0.009 ^{+ 0.019 }_{- 0.006 }$\\
\hline
   $ 3.66\cdot 10^{-3}$ &  22    & 0.38 - 0.56 &$ 0.177 \pm 0.036 ^{+ 0.042 }_{- 0.018 }$\\
			 & 	 & 0.56 - 0.74 &$ 0.165 \pm 0.018 ^{+ 0.014 }_{- 0.020 }$\\
			 &  	 & 0.74 - 0.92 &$ 0.156 \pm 0.011 ^{+ 0.005 }_{- 0.020 }$\\
\hline
   $ 7.83\cdot 10^{-3}$ &  22    & 0.38 - 0.56 &$ 0.166 \pm 0.046 ^{+ 0.063 }_{- 0.014 }$\\
			 & 	 & 0.56 - 0.74 &$ 0.185 \pm 0.031 ^{+ 0.011 }_{- 0.065 }$\\
			 & 	 & 0.74 - 0.92 &$ 0.147 \pm 0.015 ^{+ 0.007 }_{- 0.023 }$\\
\hline
   $ 1.03\cdot 10^{-3}$ &  44 	 & 0.38 - 0.56 &$ 0.128 \pm 0.042 ^{+ 0.155 }_{- 0.008 }$\\
			 &  	 & 0.56 - 0.74 &$ 0.090 \pm 0.016 ^{+ 0.052 }_{- 0.006 }$\\
			 &   	 & 0.74 - 0.92 &$ 0.152 \pm 0.018 ^{+ 0.005 }_{- 0.042 }$\\
\hline
   $ 1.86\cdot 10^{-3}$ &  44    & 0.38 - 0.56 &$ 0.190 \pm 0.048 ^{+ 0.059 }_{- 0.015 }$\\
			 &   	 & 0.56 - 0.74 &$ 0.136 \pm 0.019 ^{+ 0.006 }_{- 0.029 }$\\
			 &  	 & 0.74 - 0.92 &$ 0.115 \pm 0.010 ^{+ 0.049 }_{- 0.003 }$\\
\hline
   $ 3.68\cdot 10^{-3}$ &  44    & 0.38 - 0.56 &$ 0.082 \pm 0.023 ^{+ 0.176 }_{- 0.010 }$\\
			 &  	 & 0.56 - 0.74 &$ 0.156 \pm 0.023 ^{+ 0.017 }_{- 0.031 }$\\
			 &   	 & 0.74 - 0.92 &$ 0.158 \pm 0.015 ^{+ 0.021 }_{- 0.004 }$\\
\hline
   $ 7.33\cdot 10^{-3}$ &  44    & 0.38 - 0.56 &$ 0.138 \pm 0.045 ^{+ 0.070 }_{- 0.011 }$\\
			 & 	 & 0.56 - 0.74 &$ 0.169 \pm 0.028 ^{+ 0.027 }_{- 0.022 }$\\
			 & 	 & 0.74 - 0.92 &$ 0.188 \pm 0.020 ^{+ 0.010 }_{- 0.056 }$\\
\hline
   $ 1.54\cdot 10^{-2}$ &  44    & 0.38 - 0.56 &$ 0.130 \pm 0.055 ^{+ 0.056 }_{- 0.036 }$\\
		 	&   	 & 0.56 - 0.74 &$ 0.195 \pm 0.043 ^{+ 0.042 }_{- 0.078 }$\\
		 	&  	 & 0.74 - 0.92 &$ 0.118 \pm 0.016 ^{+ 0.052 }_{- 0.004 }$\\
\hline
   $ 2.00\cdot 10^{-3}$ &  88    & 0.38 - 0.65 &$ 0.092 \pm 0.030 ^{+ 0.025 }_{- 0.018 }$\\
			 &  	 & 0.65 - 0.92 &$ 0.109 \pm 0.018 ^{+ 0.020 }_{- 0.009 }$\\
\hline
   $ 3.59\cdot 10^{-3}$ &  88    & 0.38 - 0.65 &$ 0.236 \pm 0.053 ^{+ 0.030 }_{- 0.098 }$\\
			 & 	 & 0.65 - 0.92 &$ 0.153 \pm 0.018 ^{+ 0.018 }_{- 0.006 }$\\
\hline
   $ 7.37\cdot 10^{-3}$ &  88    & 0.38 - 0.65 &$ 0.190 \pm 0.047 ^{+ 0.078 }_{- 0.030 }$\\
			&  	 & 0.65 - 0.92 &$ 0.191 \pm 0.024 ^{+ 0.022 }_{- 0.047 }$\\
\hline
   $ 1.42\cdot 10^{-2}$ &  88    & 0.38 - 0.65 &$ 0.153 \pm 0.040 ^{+ 0.072 }_{- 0.018 }$\\
			 &  	 & 0.65 - 0.92 &$ 0.179 \pm 0.026 ^{+ 0.004 }_{- 0.063 }$\\
\hline
   $ 3.01\cdot 10^{-2}$ &  88    & 0.38 - 0.65 &$ 0.137 \pm 0.053 ^{+ 0.095 }_{- 0.010 }$\\
			 &  	 & 0.65 - 0.92 &$ 0.158 \pm 0.032 ^{+ 0.030 }_{- 0.026 }$\\
\hline
   $ 4.00\cdot 10^{-3}$ & 237   & 0.38 - 0.65 &$ 0.051 \pm 0.048 ^{+ 0.119 }_{- 0.003 }$\\
			&  	& 0.65 - 0.92 &$ 0.067 \pm 0.027 ^{+ 0.069 }_{- 0.012 }$\\
\hline
   $ 7.52\cdot 10^{-3}$ & 237   & 0.38 - 0.65 &$ 0.162 \pm 0.053 ^{+ 0.023 }_{- 0.048 }$\\
			 &  	& 0.65 - 0.92 &$ 0.164 \pm 0.027 ^{+ 0.094 }_{- 0.036 }$\\
\hline
   $ 1.47\cdot 10^{-2}$ & 237   & 0.38 - 0.65 &$ 0.161 \pm 0.052 ^{+ 0.049 }_{- 0.048 }$\\
 			& 	& 0.65 - 0.92 &$ 0.095 \pm 0.015 ^{+ 0.056 }_{- 0.005 }$\\
\hline
   $ 3.25\cdot 10^{-2}$ & 237   & 0.38 - 0.65 &$ 0.179 \pm 0.054 ^{+ 0.026 }_{- 0.056 }$\\
	 		&  	& 0.65 - 0.92 &$ 0.123 \pm 0.019 ^{+ 0.017 }_{- 0.008 }$\\
 \hline \hline
\end{supertabular}
\end{center}

\clearpage


\begin{center}

\tablefirsthead{
\hline \hline 
$\langle x \rangle$  & $\langle Q^2 \rangle$  (GeV$^2$) &  $x_L$ range & $r^{\rm LP(3)}$ \\ 
\hline \hline 
}

\tablehead{
\multicolumn{4}{l}{\tab{rlp3_0.5} (cont.)} \\
\hline  \hline
$\langle x \rangle$  & $\langle Q^2 \rangle$  (GeV$^2$) &  $x_L$ range &  $r^{\rm LP(3)}$ \\ 
\hline \hline
}

\tabletail{\hline}

\bottomcaption{
The leading proton production rate,  $r^{\rm LP(3)}$, measured as a
function of $x_L$ for protons with $0.15<p_T^2<0.5$ $GeV^2$, in bins of 
$x$ and $Q^2$, with averages $\langle x \rangle$  and $\langle Q^2 \rangle$.
Statistical uncertainties are listed first, followed by systematic uncertainties.
}

\begin{supertabular}{|c|c|c|c|}
 $  9.6\cdot 10^{-5}$ &   4.2 	 & 0.62 - 0.72 & $ 0.114\pm0.027^{+0.072}_{-0.014}$ \\
 			 &    	 & 0.72 - 0.82 & $ 0.116\pm0.033^{+0.066}_{-0.026}$ \\
 			 &    	 & 0.82 - 0.92 & $ 0.121\pm0.016^{+0.054}_{-0.015}$ \\
\hline
 $  1.7\cdot 10^{-4}$ &   4.2    & 0.62 - 0.72 & $ 0.102\pm0.014^{+0.018}_{-0.026}$ \\
			 &  	 & 0.72 - 0.82 & $ 0.105\pm0.018^{+0.053}_{-0.034}$ \\
			 &   	 & 0.82 - 0.92 & $ 0.133\pm0.012^{+0.036}_{-0.011}$ \\
\hline
 $  3.5\cdot 10^{-4}$ &   4.2 	 & 0.62 - 0.72 & $ 0.114\pm0.018^{+0.010}_{-0.021}$ \\
 			 &  	 & 0.72 - 0.82 & $ 0.212\pm0.043^{+0.023}_{-0.172}$ \\
			 &  	 & 0.82 - 0.92 & $ 0.099\pm0.010^{+0.010}_{-0.011}$ \\
\hline
 $  6.9\cdot 10^{-4}$ &   4.2    & 0.62 - 0.72 & $ 0.097\pm0.015^{+0.040}_{-0.007}$ \\
			 &  	 & 0.72 - 0.82 & $ 0.166\pm0.034^{+0.036}_{-0.048}$ \\
			 &   	 & 0.82 - 0.92 & $ 0.151\pm0.016^{+0.011}_{-0.020}$ \\
\hline
 $  1.46\cdot 10^{-3}$ &   4.2   & 0.62 - 0.72 & $ 0.106\pm0.024^{+0.015}_{-0.022}$ \\
			 &   	 & 0.72 - 0.82 & $ 0.090\pm0.027^{+0.021}_{-0.039}$ \\
			 &   	 & 0.82 - 0.92 & $ 0.154\pm0.022^{+0.043}_{-0.033}$ \\
\hline
   $ 1.9\cdot 10^{-4}$ &   7.3  & 0.62 - 0.72 & $ 0.158\pm0.034^{+0.012}_{-0.069}$ \\
 			 &   	& 0.72 - 0.82 & $ 0.068\pm0.026^{+0.030}_{-0.040}$ \\
			 &  	& 0.82 - 0.92 & $ 0.100\pm0.017^{+0.019}_{-0.035}$ \\
\hline
   $ 3.4\cdot 10^{-4}$ &   7.3  & 0.62 - 0.72 & $ 0.138\pm0.025^{+0.013}_{-0.025}$ \\
			 &   	& 0.72 - 0.82 & $ 0.149\pm0.034^{+0.020}_{-0.077}$ \\
			 &  	& 0.82 - 0.92 & $ 0.110\pm0.013^{+0.012}_{-0.017}$ \\
\hline
   $ 6.9\cdot 10^{-4}$ &   7.3  & 0.62 - 0.72 & $ 0.137\pm0.025^{+0.010}_{-0.042}$ \\
		 	&  	& 0.72 - 0.82 & $ 0.098\pm0.024^{+0.009}_{-0.035}$ \\
		 	&   	& 0.82 - 0.92 & $ 0.160\pm0.018^{+0.012}_{-0.032}$ \\
\hline
   $ 1.36\cdot 10^{-3}$ &   7.3 & 0.62 - 0.72 & $ 0.127\pm0.024^{+0.038}_{-0.008}$ \\
			 &   	& 0.72 - 0.82 & $ 0.170\pm0.040^{+0.068}_{-0.066}$ \\
			 &    	& 0.82 - 0.92 & $ 0.127\pm0.016^{+0.009}_{-0.023}$ \\
\hline
   $ 2.67\cdot 10^{-3}$ &   7.3 & 0.62 - 0.77 & $ 0.245\pm0.103^{+0.119}_{-0.120}$ \\
		 	&    	& 0.77 - 0.92 & $ 0.113\pm0.027^{+0.060}_{-0.018}$ \\
\end{supertabular}
\label{tab-rlp3_0.5}
\end{center}

\clearpage

\begin{center}

\tablefirsthead{
\multicolumn{4}{l}{\tab{rlp3_0.5} (cont.)} \\
\hline 
$\langle x \rangle$  & $\langle Q^2 \rangle$  (GeV$^2$) &  $x_L$ range  & $r^{\rm LP(3)}$ \\ 
\hline \hline 
}

\tablehead{
\multicolumn{4}{l}{\tab{rlp3_0.5} (cont.)} \\
\hline
$\langle x \rangle$  & $\langle Q^2 \rangle$  (GeV$^2$) &  $x_L$ range  & $r^{\rm LP(3)}$ \\ 
\hline \hline
}

\tabletail{\hline}

\begin{supertabular}{|c|c|c|c|}
$ 2.6\cdot 10^{-4}$ &  11 	& 0.62 - 0.72 & $ 0.170\pm0.047^{+0.063}_{-0.036}$ \\
		 	&   	& 0.72 - 0.82 & $ 0.088\pm0.048^{+0.225}_{-0.012}$ \\
		 	&   	& 0.82 - 0.92 & $ 0.169\pm0.027^{+0.023}_{-0.059}$ \\
\hline
  $ 4.6\cdot 10^{-4}$ &  11      & 0.62 - 0.72 & $ 0.176\pm0.030^{+0.015}_{-0.040}$ \\
 			 &  	 & 0.72 - 0.82 & $ 0.126\pm0.030^{+0.076}_{-0.074}$ \\
 			 &  	 & 0.82 - 0.92 & $ 0.134\pm0.015^{+0.020}_{-0.027}$ \\
\hline
   $ 9.2\cdot 10^{-4}$ &  11     & 0.62 - 0.72 & $ 0.142\pm0.023^{+0.015}_{-0.039}$ \\
			 &   	 & 0.72 - 0.82 & $ 0.158\pm0.040^{+0.026}_{-0.117}$ \\
			 & 	 & 0.82 - 0.92 & $ 0.148\pm0.016^{+0.017}_{-0.019}$ \\
\hline
   $ 1.83\cdot 10^{-3}$ &  11    & 0.62 - 0.72 & $ 0.107\pm0.018^{+0.018}_{-0.012}$ \\
			 &  	 & 0.72 - 0.82 & $ 0.071\pm0.019^{+0.089}_{-0.008}$ \\
			 &  	 & 0.82 - 0.92 & $ 0.145\pm0.015^{+0.020}_{-0.017}$ \\
\hline
   $ 3.98\cdot 10^{-3}$ &  11    & 0.62 - 0.77 & $ 0.135\pm0.034^{+0.058}_{-0.014}$ \\
			 &  	 & 0.77 - 0.92 & $ 0.199\pm0.033^{+0.018}_{-0.050}$ \\
\hline
   $ 5.1\cdot 10^{-4}$ &  22     & 0.62 - 0.72 & $ 0.125\pm0.033^{+0.019}_{-0.052}$ \\
			 & 	 & 0.72 - 0.82 & $ 0.179\pm0.069^{+0.038}_{-0.121}$ \\
			 &  	 & 0.82 - 0.92 & $ 0.127\pm0.020^{+0.053}_{-0.012}$ \\
\hline		
   $ 9.2\cdot 10^{-4}$ &  22     & 0.62 - 0.72 & $ 0.127\pm0.024^{+0.057}_{-0.016}$ \\
			 &	 & 0.72 - 0.82 & $ 0.106\pm0.027^{+0.014}_{-0.019}$ \\
			 &	 & 0.82 - 0.92 & $ 0.147\pm0.018^{+0.010}_{-0.021}$ \\
\hline
   $ 1.84\cdot 10^{-3}$ &  22    & 0.62 - 0.72 & $ 0.084\pm0.017^{+0.028}_{-0.008}$ \\
			 &  	 & 0.72 - 0.82 & $ 0.158\pm0.039^{+0.017}_{-0.063}$ \\
			 &  	 & 0.82 - 0.92 & $ 0.157\pm0.021^{+0.016}_{-0.036}$ \\
\hline
   $ 3.66\cdot 10^{-3}$ &  22    & 0.62 - 0.72 & $ 0.115\pm0.029^{+0.019}_{-0.052}$ \\
			 & 	 & 0.72 - 0.82 & $ 0.112\pm0.030^{+0.097}_{-0.011}$ \\
			 &  	 & 0.82 - 0.92 & $ 0.124\pm0.020^{+0.018}_{-0.068}$ \\
\hline
   $ 7.83\cdot 10^{-3}$ &  22    & 0.62 - 0.77 & $ 0.110\pm0.037^{+0.064}_{-0.024}$ \\
			 & 	 & 0.77 - 0.92 & $ 0.223\pm0.044^{+0.015}_{-0.127}$ \\
\hline
   $ 1.03\cdot 10^{-3}$ &  44 	 & 0.62 - 0.77 & $ 0.171\pm0.054^{+0.048}_{-0.039}$ \\
			 &  	 & 0.77 - 0.92 & $ 0.100\pm0.023^{+0.014}_{-0.024}$ \\
\hline
   $ 1.86\cdot 10^{-3}$ &  44    & 0.62 - 0.77 & $ 0.124\pm0.030^{+0.013}_{-0.024}$ \\
			 &   	 & 0.77 - 0.92 & $ 0.141\pm0.022^{+0.068}_{-0.015}$ \\
\hline
   $ 3.68\cdot 10^{-3}$ &  44    & 0.62 - 0.77 & $ 0.170\pm0.049^{+0.060}_{-0.058}$ \\
			 &  	 & 0.77 - 0.92 & $ 0.117\pm0.022^{+0.026}_{-0.016}$ \\
\hline
   $ 7.33\cdot 10^{-3}$ &  44    & 0.62 - 0.77 & $ 0.185\pm0.078^{+0.044}_{-0.103}$ \\
			 & 	 & 0.77 - 0.92 & $ 0.120\pm0.024^{+0.037}_{-0.074}$ \\
\hline
   $ 1.54\cdot 10^{-2}$ &  44    & 0.62 - 0.77 & $ 0.196\pm0.078^{+0.228}_{-0.020}$ \\
		 	&   	 & 0.77 - 0.92 & $ 0.182\pm0.050^{+0.041}_{-0.079}$ \\
\hline
   $ 2.00\cdot 10^{-3}$ &  88    & 0.62 - 0.77 & $ 0.156\pm0.077^{+0.040}_{-0.081}$ \\
			 &  	 & 0.77 - 0.92 & $ 0.125\pm0.034^{+0.011}_{-0.047}$ \\
\hline
   $ 3.59\cdot 10^{-3}$ &  88    & 0.62 - 0.77 & $ 0.208\pm0.087^{+0.115}_{-0.044}$ \\
			 & 	 & 0.77 - 0.92 & $ 0.129\pm0.027^{+0.067}_{-0.019}$ \\
\hline
   $ 7.37\cdot 10^{-3}$ &  88    & 0.62 - 0.77 & $ 0.061\pm0.031^{+0.003}_{-0.022}$ \\
			&  	 & 0.77 - 0.92 & $ 0.127\pm0.035^{+0.050}_{-0.040}$ \\
\hline
   $ 1.42\cdot 10^{-2}$ &  88    & 0.62 - 0.77 & $ 0.145\pm0.056^{+0.338}_{-0.013}$ \\
			 &  	 & 0.77 - 0.92 & $ 0.268\pm0.079^{+0.030}_{-0.264}$ \\
\hline
   $ 3.01\cdot 10^{-2}$ &  88    & 0.62 - 0.77 & $ 0.106\pm0.063^{+0.200}_{-0.040}$ \\
			 &  	 & 0.77 - 0.92 & $ 0.215\pm0.107^{+0.031}_{-0.146}$ \\
\hline
   $ 4.00\cdot 10^{-3}$ & 237   & 0.62 - 0.92 & $ 0.248\pm0.107^{+0.024}_{-0.132}$ \\
\hline
   $ 7.52\cdot 10^{-3}$ & 237   & 0.62 - 0.92 & $ 0.183\pm0.049^{+0.068}_{-0.027}$ \\
\hline
   $ 1.47\cdot 10^{-2}$ & 237   & 0.62 - 0.92 & $ 0.203\pm0.057^{+0.016}_{-0.059}$ \\
\hline
   $ 3.25\cdot 10^{-2}$ & 237   & 0.62 - 0.92 & $ 0.164\pm0.045^{+0.061}_{-0.055}$ \\
 \hline \hline
\end{supertabular}
\end{center}

\clearpage


\begin{table}
\begin{small}
\begin{center}
\begin{tabular}{|c|c|c|}
\hline \hline
$\langle Q^2 \rangle$ (GeV$^2$) & $x$ &  $ r^{\rm LP(2)}$  \\
\hline \hline
4.2 & $  9.6\cdot 10^{-5} $ 	& $0.256\pm0.006^{+0.023}_{-0.022} $ \\
 & $  1.7\cdot 10^{-4} $ 	&  $0.245\pm0.004^{+0.018}_{-0.021} $ \\
 & $  3.5\cdot 10^{-4} $ 	&  $0.237\pm0.004^{+0.018}_{-0.017} $ \\
 & $  6.9\cdot 10^{-4} $ 	&  $0.234\pm0.004^{+0.018}_{-0.016} $ \\
 & $  1.46\cdot 10^{-3} $ 	& $0.214\pm0.006^{+0.020}_{-0.016} $ \\
\hline
7.3  & $  1.9\cdot 10^{-4} $ 	& $0.246\pm0.006^{+0.020}_{-0.024} $ \\
 & $  3.4\cdot 10^{-4} $ 	& $0.231\pm0.005^{+0.019}_{-0.016} $ \\
 & $  6.9\cdot 10^{-4} $ 	& $0.236\pm0.005^{+0.018}_{-0.021} $ \\
 & $  1.36\cdot 10^{-3} $ 	& $0.243\pm0.006^{+0.019}_{-0.018} $ \\
 & $  2.67\cdot 10^{-3} $ 	& $0.252\pm0.012^{+0.025}_{-0.030} $ \\
\hline
11 & $  2.6\cdot 10^{-4} $ 	& $0.253\pm0.007^{+0.020}_{-0.016} $ \\
 & $  4.6\cdot 10^{-4} $ 	& $0.241\pm0.005^{+0.020}_{-0.015} $ \\
 & $  9.2\cdot 10^{-4} $	& $0.242\pm0.005^{+0.018}_{-0.019} $ \\
 & $  1.83\cdot 10^{-3} $ 	& $0.230\pm0.005^{+0.019}_{-0.014} $ \\
 & $  3.98\cdot 10^{-3} $ 	& $0.258\pm0.008^{+0.019}_{-0.030} $ \\
\hline
22  & $  5.1\cdot 10^{-4} $ 	&  $0.242\pm0.007^{+0.020}_{-0.023} $ \\
 & $  9.2\cdot 10^{-4} $ 	&  $0.244\pm0.006^{+0.018}_{-0.018} $ \\
 & $  1.84\cdot 10^{-4} $ 	&  $0.227\pm0.005^{+0.022}_{-0.016} $ \\
 & $  3.66\cdot 10^{-3} $ 	&  $0.253\pm0.007^{+0.020}_{-0.018} $ \\
 & $  7.83\cdot 10^{-3} $ 	&  $0.254\pm0.010^{+0.025}_{-0.020} $ \\
\hline
44  & $  1.03\cdot 10^{-3} $ 	& $0.228\pm0.009^{+0.017}_{-0.015} $ \\
 & $  1.86\cdot 10^{-3} $ 	& $0.238\pm0.008^{+0.019}_{-0.016} $ \\
 & $  3.68\cdot 10^{-3} $ 	& $0.270\pm0.009^{+0.020}_{-0.021} $ \\
 & $  7.33\cdot 10^{-3} $ 	& $0.271\pm0.011^{+0.023}_{-0.017} $ \\
 & $  1.54\cdot 10^{-3} $ 	& $0.253\pm0.014^{+0.031}_{-0.016} $ \\
\hline
 88 & $  2.00\cdot 10^{-3} $ 	& $0.215\pm0.013^{+0.015}_{-0.019} $ \\
 & $  3.59\cdot 10^{-3} $ 	& $0.245\pm0.011^{+0.043}_{-0.017} $ \\
 & $  7.37\cdot 10^{-3} $ 	& $0.243\pm0.013^{+0.026}_{-0.017} $ \\
 & $  1.42\cdot 10^{-2} $ 	& $0.265\pm0.015^{+0.032}_{-0.023} $ \\
 & $  3.01\cdot 10^{-2} $ 	& $0.241\pm0.019^{+0.034}_{-0.024} $ \\
\hline
237 & $  4.00\cdot 10^{-2} $ 	& $0.242\pm0.024^{+0.069}_{-0.021} $ \\
    & $  7.52\cdot 10^{-2} $ 	& $0.264\pm0.018^{+0.029}_{-0.038} $ \\
    & $  1.47\cdot 10^{-2} $ 	& $0.227\pm0.015^{+0.029}_{-0.017} $ \\
    & $  3.25\cdot 10^{-2} $ 	& $0.232\pm0.015^{+0.021}_{-0.019} $ \\
\hline \hline
\end{tabular}
\caption{
The leading proton production rate,  $r^{\rm LP(2)}$, measured as a
function of $x$ in bins of $Q^2$, with average $\langle Q^2 \rangle$, 
for protons with $0.32<x_L<0.92$ and $p_T^2<0.5$ $GeV^2$.
Statistical uncertainties are listed first, followed by systematic 
uncertainties.
}
\label{tab-rlp2}
\end{center}
\end{small}
\end{table}


\begin{table}
\begin{center}
\begin{tabular}{|c|c|c|}
\hline \hline
$Q^2$ (GeV$^2$) &\multicolumn{2}{c|}{$\langle r^{\rm LP(2)} \rangle$}   \\
\hline
                 & $0.6<x_L<0.97$ & $0.32<x_L<0.92$ \\
\hline \hline
  5.1 &  $0.145\pm0.001^{+0.010}_{-0.010} $ & $0.238\pm0.002^{+0.018}_{-0.017} $  \\ 
 15.8 &  $0.149\pm0.001^{+0.011}_{-0.009} $ & $0.241\pm0.002^{+0.018}_{-0.015} $  \\
 81.1 &  $0.151\pm0.002^{+0.011}_{-0.009} $ & $0.245\pm0.003^{+0.019}_{-0.015} $  \\
\hline \hline
\end{tabular}
\caption{
The leading proton production rate, $\langle r^{\rm LP(2)}
\rangle$, averaged over $x$, as a function of  $Q^2$,  for protons with
$p_T^2<0.5$ $GeV^2$ in  two $x_L$ ranges as denoted in the table.
Statistical uncertainties are listed first, followed
by systematic uncertainties. 
}
\label{tab-rlpmean}
\end{center}
\end{table}


\begin{table}
\begin{small}
\begin{center}
\begin{tabular}{|c|c|c|}
\hline \hline
$\langle Q^2 \rangle$ (GeV$^2$) & $x$ &  $ F_2^{\rm LP(2)}$  \\
\hline \hline
4.2 & $  9.6\cdot 10^{-5} $ &  $0.262\pm0.006^{+0.023}_{-0.024} $ \\
    & $  1.7\cdot 10^{-4} $ &  $0.234\pm0.004^{+0.021}_{-0.018} $ \\
    & $  3.5\cdot 10^{-4} $ &  $0.195\pm0.003^{+0.015}_{-0.015} $ \\
    & $  6.9\cdot 10^{-4} $ &  $0.167\pm0.003^{+0.012}_{-0.013} $ \\
    & $  1.46\cdot 10^{-3} $ & $0.133\pm0.004^{+0.011}_{-0.013} $ \\
\hline
7.3 & $  1.9\cdot 10^{-4} $ &  $0.290\pm0.007^{+0.030}_{-0.024} $ \\
    & $  3.4\cdot 10^{-4} $ &  $0.239\pm0.005^{+0.018}_{-0.020} $ \\
    & $  6.9\cdot 10^{-4} $ &  $0.209\pm0.004^{+0.019}_{-0.016} $ \\
    & $  1.36\cdot 10^{-3} $ & $0.183\pm0.004^{+0.014}_{-0.015} $ \\
    & $  2.67\cdot 10^{-3} $ & $0.165\pm0.008^{+0.021}_{-0.018} $ \\
\hline
11 & $  2.6\cdot 10^{-4} $ &  $0.327\pm0.010^{+0.023}_{-0.027} $ \\
   & $  4.6\cdot 10^{-4} $ &  $0.281\pm0.005^{+0.018}_{-0.024} $ \\
   & $  9.2\cdot 10^{-4} $ &  $0.239\pm0.005^{+0.020}_{-0.019} $ \\
   & $  1.83\cdot 10^{-3} $ & $0.191\pm0.004^{+0.012}_{-0.017} $ \\
   & $  3.98\cdot 10^{-3} $ & $0.181\pm0.006^{+0.022}_{-0.015} $ \\
\hline
22 & $  5.1\cdot 10^{-4} $ &  $0.336\pm0.010^{+0.033}_{-0.030} $ \\
   & $  9.2\cdot 10^{-4} $ &  $0.299\pm0.007^{+0.023}_{-0.023} $ \\
   & $  1.84\cdot 10^{-3} $ & $0.230\pm0.005^{+0.017}_{-0.023} $ \\
   & $  3.66\cdot 10^{-3} $ & $0.210\pm0.006^{+0.016}_{-0.018} $ \\
   & $  7.83\cdot 10^{-3} $ & $0.176\pm0.007^{+0.015}_{-0.018} $ \\
\hline  
44 & $  1.03\cdot 10^{-4} $ & $0.322\pm0.013^{+0.025}_{-0.027} $ \\
   & $  1.86\cdot 10^{-3} $ & $0.286\pm0.009^{+0.022}_{-0.025} $ \\
   & $  3.68\cdot 10^{-3} $ & $0.265\pm0.009^{+0.023}_{-0.022} $ \\
   & $  7.33\cdot 10^{-3} $ & $0.215\pm0.008^{+0.016}_{-0.020} $ \\
   & $  1.54\cdot 10^{-2} $ & $0.164\pm0.009^{+0.014}_{-0.022} $ \\
\hline
88 & $  2.00\cdot 10^{-3} $ & $0.286\pm0.017^{+0.030}_{-0.027} $ \\
   & $  3.59\cdot 10^{-3} $ & $0.275\pm0.013^{+0.023}_{-0.050} $ \\
   & $  7.37\cdot 10^{-3} $ & $0.214\pm0.011^{+0.018}_{-0.026} $ \\
   & $  1.42\cdot 10^{-2} $ & $0.190\pm0.010^{+0.020}_{-0.025} $ \\
   & $  3.01\cdot 10^{-2} $ & $0.138\pm0.011^{+0.017}_{-0.022} $ \\
\hline
273 & $  4.00\cdot 10^{-3} $ & $0.287\pm0.028^{+0.038}_{-0.087} $ \\
    & $  7.52\cdot 10^{-3} $ & $0.260\pm0.017^{+0.041}_{-0.034} $ \\
    & $  1.47\cdot 10^{-2} $ & $0.173\pm0.012^{+0.017}_{-0.025} $ \\
    & $  3.25\cdot 10^{-2} $ & $0.139\pm0.009^{+0.015}_{-0.016} $ \\
\hline \hline
\end{tabular}
\caption{
The leading proton structure function, $F_2^{\rm LP(2)}$, measured as a
function of $x$ in bins of $Q^2$, with average $\langle Q^2 \rangle$, 
for protons with $0.32<x_L<0.92$ and $p_T^2<0.5$ $GeV^2$.
Statistical uncertainties are listed first, followed by systematic 
uncertainties.
}
\label{tab-flp2}
\end{center}
\end{small}
\end{table}

\end{small}

\clearpage


\clearpage


\clearpage


%
%
\end{document}